%% file: main.tex
\documentclass[12pt]{article}

\usepackage{iftex}
\ifPDFTeX
  \usepackage[T1]{fontenc}
  \usepackage[utf8]{inputenc}
  \usepackage{textcomp}
  \usepackage{lmodern}
\else
  \usepackage{unicode-math}
  \defaultfontfeatures{Scale=MatchLowercase}
  \defaultfontfeatures[\rmfamily]{Ligatures=TeX,Scale=1}
\fi

\usepackage[margin=1in]{geometry}
\usepackage{setspace}
\usepackage[dvipsnames,svgnames,x11names]{xcolor}
\usepackage{xspace}
\usepackage{comment}
\usepackage{multirow}
\usepackage{calc}
\usepackage{etoolbox}

\IfFileExists{upquote.sty}{\usepackage{upquote}}{}
\IfFileExists{microtype.sty}{%
  \usepackage{microtype}
  \UseMicrotypeSet[protrusion]{basicmath}
}{}
\IfFileExists{xurl.sty}{\usepackage{xurl}}{}
\makeatletter
\@ifundefined{KOMAClassName}{%
  \IfFileExists{parskip.sty}{%
    \usepackage{parskip}
  }{%
    \setlength{\parindent}{0pt}
    \setlength{\parskip}{6pt plus 2pt minus 1pt}
  }
}{%
  \KOMAoptions{parskip=half}
}
\makeatother

\makeatletter
\ifx\paragraph\undefined\else
  \let\oldparagraph\paragraph
  \renewcommand{\paragraph}{%
    \@ifstar{\oldparagraph*}{\oldparagraph}%
  }
\fi
\ifx\subparagraph\undefined\else
  \let\oldsubparagraph\subparagraph
  \renewcommand{\subparagraph}{%
    \@ifstar{\oldsubparagraph*}{\oldsubparagraph}%
  }
\fi
\makeatother

\usepackage{array}
\usepackage{booktabs}
\usepackage{longtable}
\usepackage{graphicx}
\usepackage{float}
\usepackage[font=small]{caption}
\usepackage{subcaption}

\IfFileExists{footnotehyper.sty}{\usepackage{footnotehyper}}{\usepackage{footnote}}
\makesavenoteenv{longtable}

\usepackage{booktabs,longtable,array}
\newcolumntype{L}[1]{>{\raggedright\arraybackslash}p{#1}}

\makeatletter
\patchcmd\longtable{\par}{\if@noskipsec\mbox{}\fi\par}{}{}
\makeatother

\makeatletter
\def\maxwidth{\ifdim\Gin@nat@width>\linewidth\linewidth\else\Gin@nat@width\fi}
\def\maxheight{\ifdim\Gin@nat@height>\textheight\textheight\else\Gin@nat@height\fi}
\setkeys{Gin}{width=\maxwidth,height=\maxheight,keepaspectratio}
\def\fps@figure{htbp}
\makeatother

\floatstyle{ruled}
\newfloat{codelisting}{h}{lop}
\floatname{codelisting}{Listing}

\usepackage{amsmath,amsthm,amssymb,amsfonts,amstext,bm}
\usepackage{mathtools}
\usepackage{thmtools}
\usepackage{upgreek}
\usepackage{breqn}
\allowdisplaybreaks

\newtheorem{theorem}{Theorem}
\newtheorem{corollary}[theorem]{Corollary}
\newtheorem{lemma}[theorem]{Lemma}

\newtheorem{proposition}[theorem]{Proposition}

\theoremstyle{definition}
\newtheorem{definition}{Definition}
\declaretheorem[style=definition, name=Remark, qed=$\diamond$]{remark}
\declaretheorem[style=definition, name=Example, qed=$\diamond$]{example}

\numberwithin{equation}{section}
\numberwithin{figure}{section}

\usepackage{algorithm}
\usepackage{algpseudocode}
\usepackage{tikz}

\usepackage[authoryear]{natbib}
\usepackage{etoc}

\usepackage[
  unicode,
  colorlinks=true,
  linkcolor=blue,
  filecolor=Maroon,
  citecolor=Blue,
  urlcolor=Blue,
  pdftitle={Title},
  pdfauthor={Author 1; Author 2},
  pdfkeywords={3 to 6 keywords, that do not appear in the title},
  pdfcreator={LaTeX via pdfLaTeX}
]{hyperref}

\usepackage{bookmark}
\usepackage{cleveref}

\crefname{proposition}{Proposition}{Propositions}
\Crefname{proposition}{Proposition}{Propositions}
\crefname{lemma}{Lemma}{Lemmas}
\Crefname{lemma}{Lemma}{Lemmas}
\crefname{corollary}{Corollary}{Corollaries}
\Crefname{corollary}{Corollary}{Corollaries}
\crefname{claim}{Claim}{Claims}
\Crefname{claim}{Claim}{Claims}
\crefname{appendix}{Appendix}{Appendices}
\Crefname{appendix}{Appendix}{Appendices}

\ifLuaTeX
  \usepackage{selnolig}
\fi

\AtBeginDocument{%
  \renewcommand*\contentsname{Table of contents}

}

\newcommand{\anon}{1}

\newcommand{\minimize}{\mathop{\mathrm{minimize}}}
\newcommand{\maximize}{\mathop{\mathrm{maximize}}}

\newcommand\dotp[1]{\left\langle #1 \right\rangle}

\newcommand{\Diag}{\operatorname{Diag}}

\newcommand{\var}{\operatorname{Var}}
\newcommand{\cov}{\operatorname{Cov}}

\newcommand{\calK}{{\mathcal K}}

\newcommand{\calN}{{\mathcal N}}
\newcommand{\calO}{{\mathcal O}}

\newcommand{\mat}[1]{\begin{bmatrix} #1 \end{bmatrix}}

\newcommand{\gsM}{\mat{B}}
\newcommand{\gscv}[1]{\vec{b}_{#1}}

\newcommand{\alive}{\mathcal{A}}

\newcommand{\gsvsym}{z}
\newcommand{\gsv}[1]{\vec{\gsvsym}_{#1}}
\newcommand{\gsvt}{\gsv{t}}
\newcommand{\gsve}[2]{\gsv{#1}(#2)}

\newcommand{\gsusym}{u}
\newcommand{\gsuv}[1]{\vec{\gsusym}_{#1}}
\newcommand{\gsuvt}{\gsuv{t}}

\newcommand{\gsun}[1]{\gsusym({#1})}

\newcommand{\gsdsym}{\delta}
\newcommand{\gsd}[1]{\gsdsym_{#1}}

\newcommand{\gsdn}{\gsdsym}
\newcommand{\gsdnp}{\gsdsym^+}
\newcommand{\gsdnm}{\gsdsym^-}

\newcommand{\gsdt}{\gsd{t}}

\DeclarePairedDelimiterXPP\spanv[1]{\mathrm{span}}{\lbrace}{\rbrace}{}{#1}
\DeclareMathOperator*{\argmin}{argmin}

\makeatletter
\numberwithin{equation}{section}
\numberwithin{figure}{section}

\makeatother

\begin{document}
\etocdepthtag.toc{main}
\def\spacingset#1{\renewcommand{\baselinestretch}%
{#1}\small\normalsize} \spacingset{1}
% \linenumbers
% \modulolinenumbers[1]
\if1\anon
{
\title{\bf Optimal Design under Interference, Homophily, and Robustness Trade-offs}

\author{
Vydhourie Thiyageswaran\thanks{University of Washington} \and
Alex Kokot\footnotemark[1] \and
Jennifer Brennan\thanks{Google Research} \and
Marina Meil\u{a}\thanks{University of Waterloo} \and
Christina Lee Yu\thanks{Cornell University} \and
Maryam Fazel\footnotemark[1]
}
\date{} 
\maketitle
} \fi

\if0\anon
{
  \bigskip
  \bigskip
  \bigskip
  \begin{center}
    {\LARGE\bf Optimal Design under Interference, Homophily, and Robustness Trade-offs}
\end{center}
  \medskip
} \fi

\begin{abstract}
    To minimize the mean squared error (MSE) in global average treatment effect (GATE) estimation under network interference, a popular approach is to use a cluster-randomized design. However, in the presence of homophily, which is common in social networks, cluster randomization can instead increase the MSE. We develop a novel potential outcomes model that accounts for interference, homophily, and heterogeneous variation. In this setting, we establish a framework for optimizing designs for worst-case MSE under the Horvitz-Thompson estimator. This leads to an optimization problem over the covariance matrices of the treatment assignment, trading off interference, homophily, and robustness. We frame and solve this problem using two complementary approaches. The first involves formulating a semidefinite program (SDP) and employing Gaussian rounding, in the spirit of the Goemans-Williamson approximation algorithm for MAXCUT. The second is an adaptation of the Gram-Schmidt Walk, a vector balancing algorithm which has recently received much attention. Finally, we evaluate the performance of our designs through various experiments on simulated network data and a real village network dataset.
\end{abstract}

\noindent%
\vfill

% \maketitle
\newpage
\spacingset{1.8} % DON'T change the spacing!

\section{Introduction}
 Network interference models settings in which the outcome of one unit depends on the treatment assignment of its neighbors in a network. The presence of network interference complicates the estimation of global average treatment effects (GATE), the difference between the average outcome when all units are treated and the average outcome when all units are assigned to control, as it violates the Stable Unit Treatment Value Assumption (SUTVA). In particular, in an experiment with partial treatment assignments where there is a mixture of treated units and control units, the difference in treatment assignments between neighboring units can bias each unit's outcome relative to the all-treated and all-control outcomes. Therefore, basic experimental designs such as unit-level Bernoulli randomization lead to biased GATE estimation under standard estimators.
 % such as the vanilla Horvitz-Thompson and difference-in-means estimators. 

 \begin{figure}[b]
    \centering
    \begin{subfigure}[b]{0.3\textwidth}
        \centering
        \includegraphics[width=2in]{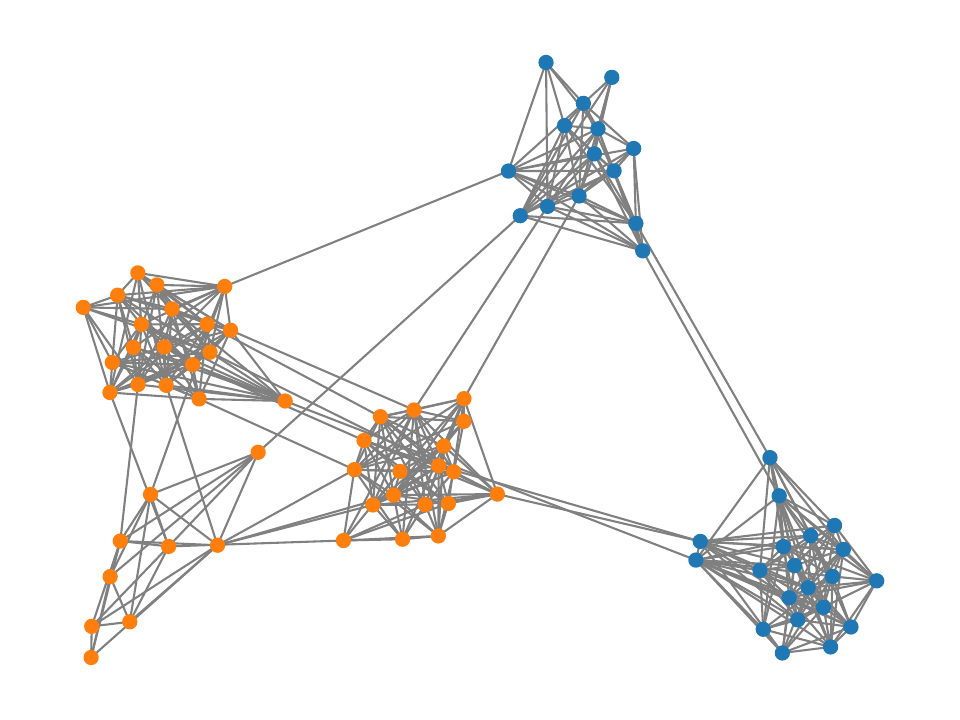}
    \end{subfigure}
    \hfill
    \begin{subfigure}[b]{0.3\textwidth}
        \centering
        \includegraphics[width=2in]{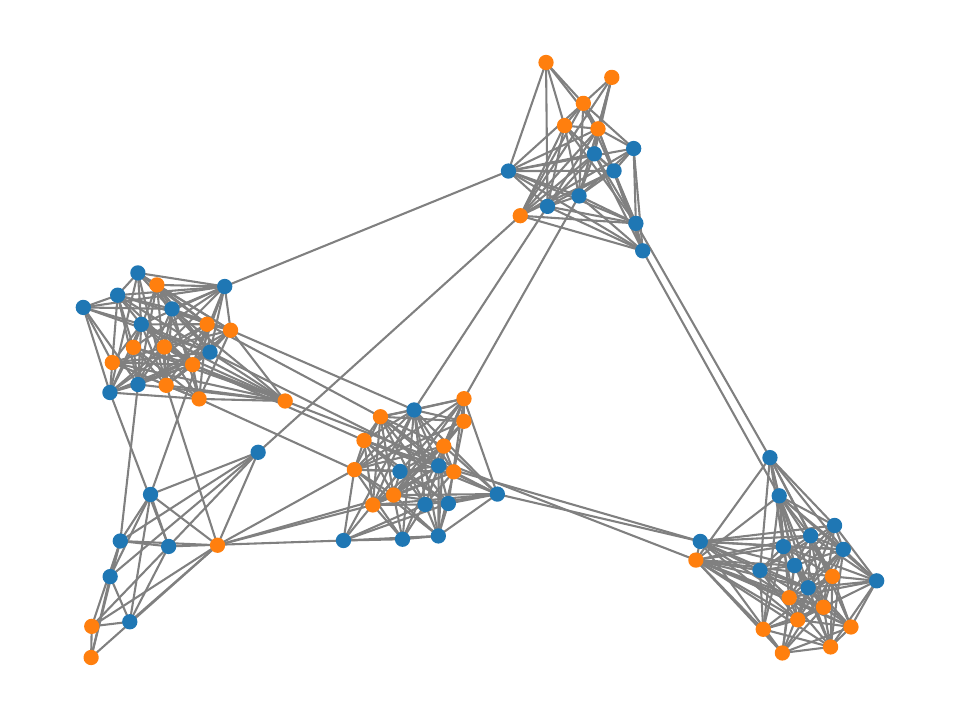}
    \end{subfigure}
    \hfill
    \begin{subfigure}[b]{0.3\textwidth}
        \centering
        \includegraphics[width=2in]{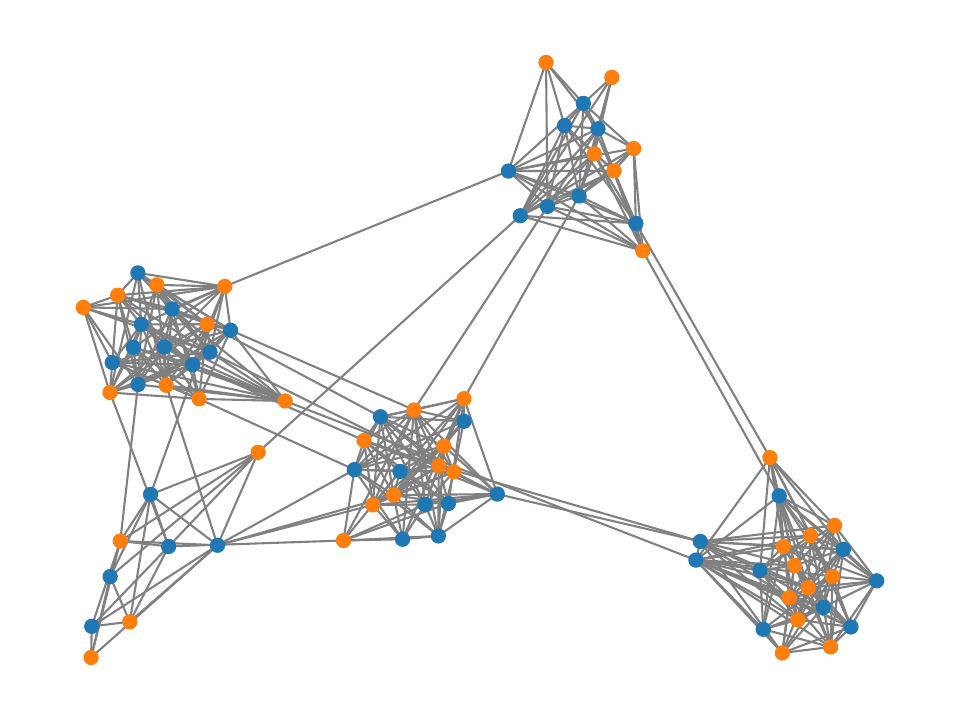}
    \end{subfigure}
    \caption{Optimized treatment assignments under interference-homophily tradeoffs on a synthetic network as homophily levels increase from left to right. See \Cref{app:more illustrations} for more details.}
    \label{fig:implementation_five_cluster_transition_h_m}
\end{figure}

Cluster randomization approaches \citep{ugander2013graph, ugander2023randomized, brennan2022cluster, viviano2023causal} are considered to be a useful way to reduce this bias. All of these methods stem from the same principle. Since bias is introduced by the conflicting treatment assignments between neighboring units, if we were to cluster together units that are strongly connected before assigning treatment at the cluster level, we would successfully reduce the bias.
Consider a social network in which the outcome of interest is each individual’s likelihood of joining a premium social media platform. If treatment consists of offering a free trial, an individual’s outcome may depend not only on their own assignment, but also on whether their friends receive free trials and subsequently join the platform.
Consequently, the outcomes from when all people in the social network are treated and from when all people are assigned to control will be different from when there is an experiment run where people are uniformly randomly assigned to treatment and control, due to the interference between friends who have different treatment assignments. Assigning tightly connected friends to the same treatment group, whether treatment or control, reduces this discrepancy between the outcomes in the experimental setting and in the full treatment and control settings.
\begin{figure}[tb]
    \centering
    \includegraphics[width=\linewidth]{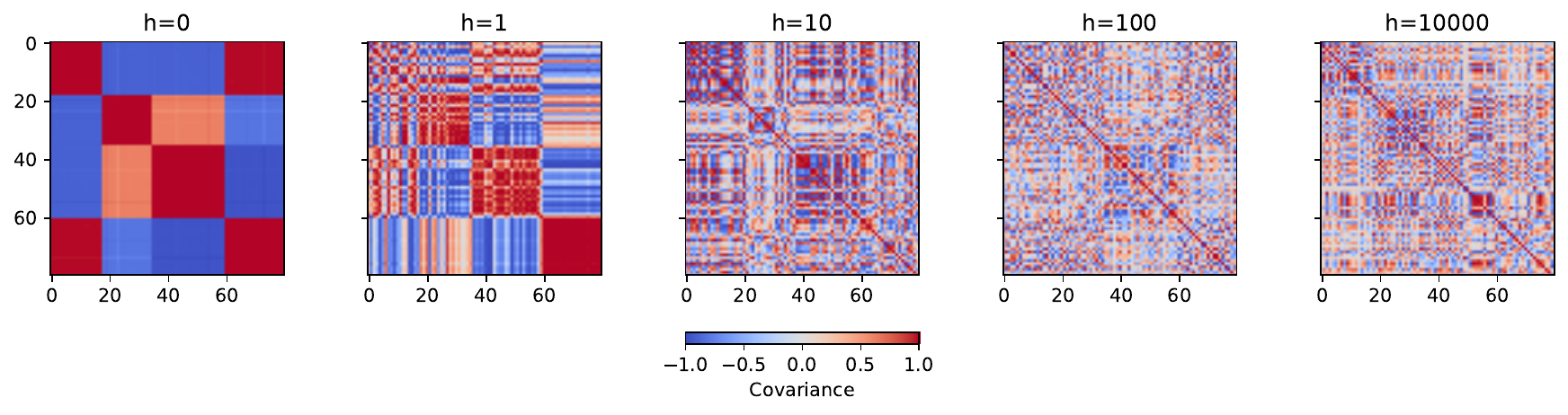}
    \caption{Optimal design covariance matrices under interference-homophily tradeoffs as homophily increases from left to right. See \Cref{app:more illustrations} for more details.}
    \label{fig:cov_mat_m_h}
\end{figure}
However, networks are often homophilous: individuals tend to form connections with others who are similar to themselves, a phenomenon first noted by \citet{lazarsfeld1954friendship} and commonly summarized as “birds of a feather flock together” \citep{mcpherson2001birds}. Similarities along socioeconomic, demographic, or behavioral dimensions \citep{jackson2023dynamics} increase the likelihood of social connection, and these same dimensions may also drive heterogeneous treatment responses. For example, younger and older individuals may respond differently to incentives for adopting new technologies. When cluster-randomized designs align with such homophilous groupings, covariate distributions can vary substantially across clusters, leading to unrepresentative treatment or control samples \citep{ugander2023randomized,cortez2024analysis,fatemi2020minimizing}. Some covariate groups may be rarely or never exposed to treatment, inflating the imprecision of the estimator. In the presence of homophily, one would perhaps, now contradictorily, be incentivized to spread treatments across the network, to more effectively sample across different social groups. 
This reflects the typical goal of covariate balancing, where estimation accuracy is improved by enforcing similarity between treatment and control groups \citep{efron1971forcing}. Finally, an effective design is one that is robust to heterogeneous variation, i.e., differences in potential outcomes poorly explained by the network structure. In \Cref{section_read_data}, we simulate experiments using network data from villages in Karnataka, India, collected by \citet{banerjee2013diffusion}, which exhibit caste-based homophily. We observe that for large ranges of interference and heterogeneous variation levels, our designs, in the worst-case, perform better than existing unit-level and cluster-level randomization designs. 
 
\subsection{Our Contributions and Overview of Results}
We formalize the trade-off between network interference and homophily, recovering well-studied designs in limiting regimes, and show how this trade-off interacts with randomization to yield a unifying three-way framework generalizing various results in the literature \citep{harshaw2024balancing, viviano2023causal}.
We focus on the Horvitz-Thompson estimator which is commonly used in practice, and operate under the potential outcomes framework. We provide methods to solve for mean squared error (MSE)-optimal designs under the worst case potential outcomes within the subclass $\chi(\eta, \gamma, \kappa)$ (made precise in \Cref{gate_estimation}), where homophily, interference, and heterogeneous variation conditions 
are separately satisfied with parameters $\eta$, $\gamma$, and $\kappa$, respectively.

We consider the GATE, $\tau = \frac{1}{n} \dotp{\mathbf{1},Y(\mathbf{1}) - Y(\mathbf{0}) }  = \frac{1}{n} \dotp{\mathbf{1}, \phi}$, where $Y(z)$ is the outcome vector observed under the treatment assignment vector $z \in \{0,1\}^n$, and $\phi$ is the vector of global treatment effects. We focus on the vanilla Horvitz-Thompson estimator $\hat{\tau} = \dotp{Y, \frac{z}{p n} - \frac{\mathbf{1}-z}{(1-p)n}}$ with uniform treatment probability $p$ across nodes, $p = \mathbb{P}(z_i = 1) = \mathbb{P}(z_j = 1)$ for all $i,j \in [n]$. Let $L$ be the graph Laplacian\footnote{Here $L := D - W$, where $D$ is the diagonal degree matrix and $W$ is the edge-weight matrix; see \Cref{notation,homophily_heterogeneous} for details.}
 underlying the interference structure, and $L^\dagger$ its pseudo-inverse. We can then characterize the homophily, interference, and robustness trade-offs with parameters $\eta$, $\gamma$, and $\kappa$, respectively.
\begin{theorem}[Informal version of \Cref{thm:mse_bound_general_HT_symmetric}] \label{thm:mse_bound_general_HT_symmetric_short}
Let $x \in \{-1, 1\}^n$ by defining $x := 2z-1$. Define $X := \mathbb{E}[xx^T]$. Then, for $q \geq 1$, 
     \begin{align*}
          \sup_{\chi(\eta, \gamma, \kappa)} \mathbb{E}[(\hat{\tau} - \tau)^2] \leq \frac{C}{n^2} \{ \operatorname{Tr}(\mathbf{1}\mathbf{1}^T X) + C_1 \eta\operatorname{Tr}(L^{\dagger} X) + C_2 \kappa \|X\|_{q} + C_3 \gamma \operatorname{Tr}(L X) \},
     \end{align*}
for known constants $C, C_1, C_2, C_3$ 
that depend on $p$.
\end{theorem}
We propose two complementary approaches. The first approach involves formulating a semidefinite program (SDP) and employing Gaussian rounding, in the spirit of the Goemans-Williamson approximation algorithm \citep{goemans1995improved} for MAXCUT. In particular, we cast the optimal design problem as an optimization problem over $X$, the covariance matrix of the design. 
Reparameterizing with $\Tilde{\gamma}$,$\Tilde{\eta}$, and $\Tilde{\kappa}$, so that
$\Tilde{\gamma} := C_3 \gamma, \Tilde{\eta} := C_1 \eta ,\text{ and } \Tilde{\kappa} := C_2  \kappa$,
we can write an SDP more explicitly in terms of the new trade-off parameters:
\begin{align} \label{sdp_informal}
    \min_{X\in\mathbb{R}^{n\times n}} \quad & \Tilde{\eta}\text{Tr}(L^\dagger X)  +   \Tilde{\gamma} \text{Tr}(L X) + \Tilde{\kappa}\|X\|_{q} + \operatorname{Tr}(\mathbf{1}\mathbf{1}^TX) \\
    \text{s.t.} \quad & X \succeq  0, \quad \text{diag}(X) = \mathbf{1}. \notag 
\end{align} 
To generate treatment assignments, we proceed with a Gaussian rounding procedure. As a second approach, we adapt the Gram-Schmidt Walk \citep{harshaw2024balancing}, a vector balancing algorithm.

We uncover a natural interpretation of interference and homophily by viewing eigenvalue-scaled Laplacian eigenvector embeddings as unit-level covariates, connecting the covariate balancing and interference literatures.

We demonstrate improvements over existing designs through various experiments on simulated network data and a real village network dataset. 

\subsection{Related Work}
\citet{harshaw2024balancing} propose the Gram-Schmidt Walk design that navigates trade-offs between covariate balancing and robustness under a potential outcomes model. Our paper is of a similar flavor as we study designs that trade off homophily, interference, and robustness to heterogeneous variation. 
We adapt the Gram-Schmidt Walk algorithm to our setting as one approach, and contrast its performance with an alternative method of solving an SDP followed by Gaussian rounding. We also note a connection to \citet{bhat2020near}, who study treatment-effect estimation, under a different potential outcomes model, in a non-network setting with the goal of maximizing precision. They derive an SDP relaxation for this maximization problem and directly apply the Goemans–Williamson rounding scheme \citep{goemans1995improved}. They further show that their precision maximization admits a covariate balancing interpretation.
Our framework is formulated instead through a quadratic-form minimization perspective, and our characterization of homophily and covariate balancing reveals a natural conceptual bridge between these approaches, with connections to \Cref{alg:gaussian_round,alg:gramschmidt}.

Optimization-based designs for network interference settings are well studied. \citet{viviano2023causal} propose an SDP to construct clusters for cluster-randomized designs by minimizing the MSE of a difference-in-means estimator, but do not consider the effects of homophily.
The work most closely related to ours is \citet{chen2024optimized}, who also minimize an MSE objective over the treatment assignment covariance matrix. Their focus, however, is restricted to cluster randomization and does not address homophily. They consider a non-convex formulation arising from Grothendieck’s identity and propose a projected gradient method to find stationary points. In contrast, we adopt a convex SDP formulation and apply a Gaussian rounding procedure that generalizes the rounding scheme for the MAXCUT approximation of \citet{goemans1995improved}. In \Cref{other_related_work}, we discuss other related works.

\subsection{Finite-population setup and network structure}\label{setup}\label{notation}

We consider a finite population of $n$ units indexed by $V=[n]$. To represent possible pathways of interference, as well as patterns of homophily among the units, we equip this population with an undirected weighted network 
$G=(V,E,W)$,
where $E \subseteq V\times V$ is the set of edges and $W\in[0,1]^{n\times n}$ is a symmetric weight matrix. We take $(i,j)\in E$ if and only if $W_{ij}>0$,
so that an edge between units $i$ and $j$ indicates that they are linked through a relationship relevant to interference or homophily, while the magnitude of $W_{ij}$ reflects the strength of that relationship. We write
$d_i := \sum_{j=1}^n W_{ij}$, $D := \Diag(d)$, $L := D-W$,
where $d_i$ is the weighted degree of unit $i$, $D$ is the diagonal degree matrix, and $L$ is the graph Laplacian. We let $L^\dagger$ denote the Moore--Penrose pseudoinverse of $L$.

Throughout most of the paper, for clarity of exposition, we assume that the relevant interference/homophily network is observed without error. That is, the analyst observes a network
$G_{\mathrm{obs}}=(V,E_{\mathrm{obs}},W_{\text{obs}})$
that coincides with the underlying network
$G=(V,E,W)$
governing the dependence structure of interest.

In many applications, however, the relevant network is not directly observed and must instead be constructed from a proxy. Examples include self-reported ties, administrative records, spatial or infrastructural adjacency, digital interaction logs, and exposure records. Such proxies may be subject to missingness, measurement error, thresholding, temporal mismatch, or other distortions, so that the observed network need not coincide with the latent one. Within the network interference literature, a small body of work has begun to examine the consequences of network and exposure misspecification \citep{aronow2017estimating,savje2024causal,weinstein2026causal}. In this spirit, \Cref{robustness} develops our robustness framework for settings in which the observed network is a noisy proxy for the underlying interference/homophily structure. In particular, we study a natural model in which the observed edge weights are unbiased for the latent edge weights, $\mathbb{E}[{(W_{\mathrm{obs}})}_{ij}] = W_{ij}$. 
% $\mathbb{E}[{W_{\mathrm{obs}}}_{ij}] = W_{ij}$. 
We also discuss in \Cref{example:noisy_interference} the setting in which the observed network is a first-order approximation to a richer interference structure, for example when interference extends beyond immediate neighbors and decays with distance. 

For ease of reference, a glossary containing a subset of key notation is provided in \Cref{tab:glossary} in the Appendix.

\section{Homophily and Heterogeneous Variation} \label{homophily_heterogeneous}

Network interference models how the outcome of units depend on the treatment assignments of neighboring units in the network. With regards to the GATE, the bias arising from network interference can be written as the component of the potential outcome model accounting for disparate treatments among units. As such, the bias notably vanishes in a graph where all units are treated (respectively controlled). In this section, we focus on the terms that are left over, the baseline $\alpha$, and treatment effect $\phi$. We discuss the interference terms in full detail in \Cref{section:interference}. 
Denote the outcomes by $Y:=Y(z)$, with treatment assignment $z \in \{0, 1\}^n$. We consider outcomes associated with $\mathbf{0}$ and $\mathbf{1}$, the treatment assignment vectors of all 0s and 1s, respectively. We can express, for each instance of the sampled network,
\begin{align} \label{extremes_ate}
    \alpha:=Y(\mathbf{0}),\quad 
    \alpha + \phi:= Y(\mathbf{0}) + [Y(\mathbf{1}) - Y(\mathbf{0})] = Y(\mathbf{1}). 
\end{align}

We decompose these vectors $\alpha = h_{\alpha} + \varepsilon_{\alpha}$, and $\phi = h_{\phi} + \varepsilon_{\phi}$, where $h_{\alpha}, h_{\phi}$ are homophilous components, capturing smooth variation of the potential outcome coefficients with the graph structure. We formalize this smoothness via a quadratic constraint, and define this formally in \Cref{m-homophily}. The $\varepsilon_{\alpha}, \varepsilon_{\phi}$ terms are heterogeneous variation, components not explained by homophily and interference, formalized by moment constraints. These are nothing but the residuals or model misspecifications, terms more commonly seen in other literature, if we model the true generative (potential) outcomes by just interference and homophily. 

\begin{remark} \label{remark:mean_decomposition_nonunique}
    It is equivalent to write $\alpha = \bar{\alpha} + h_\alpha + \varepsilon_\alpha$, and $\phi = \bar{\phi} + h_\phi + \varepsilon_\phi$, where $h_\alpha, h_\phi, \varepsilon_\alpha, \varepsilon_\phi$ are mean-zero, and $\bar{\alpha}, \bar{\phi}$ are constant vectors.
    We note that the decompositions into $h_\alpha + \varepsilon_\alpha$ and $h_\phi + \varepsilon_\phi$ are generally not unique. Our results below hold for any admissible decomposition satisfying the stated homophily, interference, and heterogeneous-variation conditions. In \Cref{gate_estimation}, we distinguish a particular admissible decomposition, which we call the \textit{min-min-max} decomposition.
\end{remark}

\subsection{Homophily}
We formalize the definition of homophily we consider. To our knowledge, the only other work that has used a similar definition is \citet{ugander2023randomized}. 
Informally, homophily captures the extent to which similar nodes are strongly connected and dissimilar nodes are weakly connected.
The quantity $\sum_{(i,j) \in E} W_{ij}(h_i - h_j)^2$ encapsulates this idea by aggregating the dissimilarities in potential outcomes coefficients $h_i$
across edges. In a connected graph that is well-clusterable, high levels of homophily could result in clusters of nodes that are very similar while being significantly dissimilar across clusters. Large dissimilarities across clusters would result in a large value of $\sum_{(i,j) \in E} W_{ij}(h_i - h_j)^2$. 

\begin{definition}[$\eta$-homophily] \label{m-homophily}
We say that the vector $h$ satisfies ($\eta, L$)-homophily, or is ($\eta, L$)-homophilous, if $h^TLh \leq \eta$, where $L:= D- W$ is  the graph Laplacian. That is, $\sum_{(i,j) \in E} W_{ij}(h_i - h_j)^2 \leq \eta$.
\end{definition}

\begin{figure}
    \centering
    \includegraphics[width=0.5\linewidth]{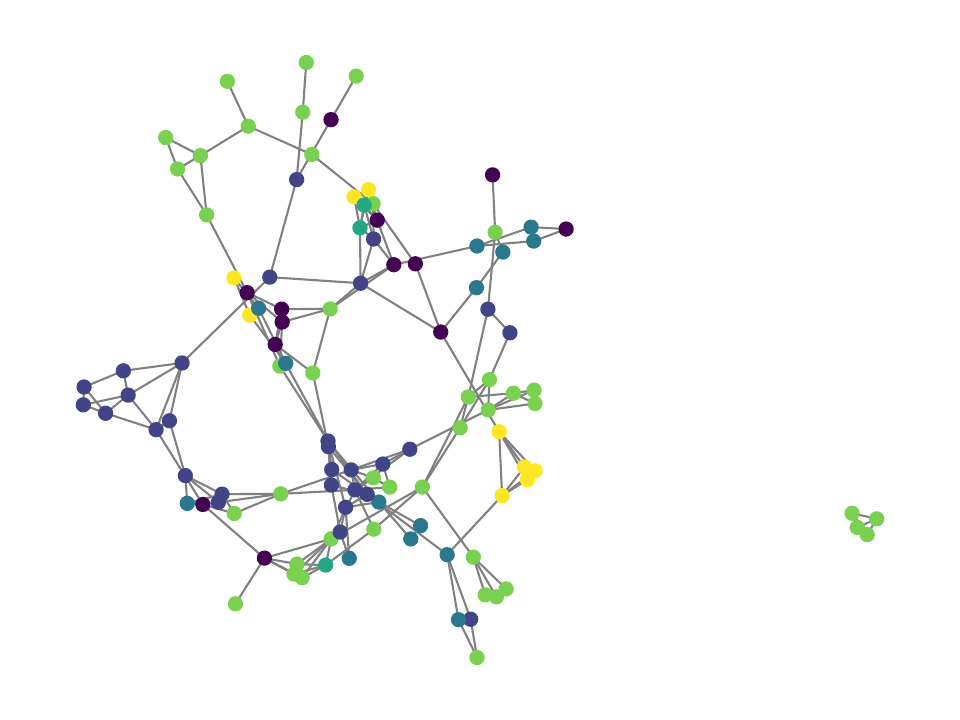}
    \caption{A real example of homophily: network of home visits between members of a village (no.6), using data from \cite{banerjee2013diffusion}. Node colors indicate the caste of each village member.}
    \label{fig:house_visits_village}
\end{figure}

In such a setting, one natural design might be to have a spread out treatment assignment, to diversify treatment samples across the natural clusters. 
In \Cref{section_mse_bound}, we demonstrate how this formalization leads to a design that distributes treatment assignments across the underlying graph clusters, aligning with intuition.

\begin{remark}\label{remark_homophily}
Unless otherwise specified, we use $\eta$-homophily to refer throughout to homophily defined with respect to the graph Laplacian $L$, and we suppress this dependence in the notation for brevity. Other kernels may be more appropriate in certain applications; see, for example, \Cref{remark_homophily_lapsym}. More generally, \Cref{general_misspecifications_sim_kernels} treats $(\eta,\mathcal{K})$-homophily for an arbitrary similarity kernel $\mathcal{K}$, allowing the notion of homophily to capture, for example, spatial proximity or similarity in other covariates.
\end{remark}

\subsection{Robustness to heterogeneous variation} \label{robustness}

Our model accounts for heterogeneous variation via the $\varepsilon$ components, which we only constrain by certain moment conditions. In its simplest form, this corresponds to an $\ell_2$ constraint on $\varepsilon$ which leads to an operator norm as previously realized in \citet{harshaw2024balancing}. Indeed, let $X = \mathbb{E}[xx^T]$, $x=2z-1$, then
$    \sup_{\|\varepsilon_f\|_2^2 \leq \kappa} \operatorname{Tr}(\varepsilon_f^T X \varepsilon_f) =  \kappa \|X\|_{\text{op}},$
% \begin{align*}
%     \sup_{\|\varepsilon_f\|^2 \leq \kappa} \operatorname{Tr}(\varepsilon_f^T X \varepsilon_f) = \|X\|_{\text{op}} \kappa,
% \end{align*}
for $f = \alpha, \phi.$

Our framework encompasses heterogeneous variation arising both from unobserved subject covariates and from unobserved dependence between units independent of treatment assignment.
In works such as \cite{li2021causal}, the setting of interference under noisy network observations has previously been considered, resulting in stochasticity of the potential outcome model.
The observed network is a sample from a super-population of possible dependencies between the finite population of experimental units.
In \Cref{example:noisy_interference}, we discuss how this noisy interference is encapsulated by our generalized notion of network interference. 
Our notion of robustness in the current section addresses the resulting correlation in the baseline $\alpha$ and treatment effect $\phi$ in addition to that induced by unobserved subject covariates and homophily with respect to the observed network.
In particular, we allow for $\varepsilon_f$, $f= \alpha, \phi$, to act as a random effect, drawing particular inspiration from the spatial statistics literature \citep{gaetan2010spatial}. 
In our potential outcomes framework, we place constraints on $\Sigma_f := \mathbb{E}[\varepsilon_f \varepsilon_f^T]$, %\mf{or just norm of $\varepsilon$?}
and we discuss how different penalties relate to inter-unit dependencies.
This encompasses relationships between the outcomes that are not restricted to the observed network encoded by homophily. We employ duality to relate this back to the treatment design.

\begin{example}\label{example:noisy_misspecification}
    A prominent example of stochastic heterogeneous variation is through the lens of noisy network observations. Consider the setting where $\alpha$ (and similarly $\phi$) is a fixed function of the units in our finite population that is $\eta$-homophilous with respect to the graph $G$ with Laplacian $L$. Thus, we can express $\alpha = L^{\dagger/2} \beta$ for a coefficient vector $\beta$ with length $\sqrt{\eta}$. Now consider the observed network $G_{\text{obs}}$ with corresponding Laplacian $L_{\text{obs}}$ (see, for example, \Cref{fig:noisy_netw_obs}). We can decompose
    $\alpha = L^{\dagger/2} \beta = h + \varepsilon, h:= L_{\text{obs}}^{\dagger/2} \beta, \text{ and } \varepsilon := (L^{\dagger/2} -L_{\text{obs}}^{\dagger/2}) \beta$.
    % \[
    % \alpha = L^{\dagger/2} \beta = h + \varepsilon,\quad h:= L_{\text{obs}}^{\dagger/2} \beta,\quad \varepsilon := (L^{\dagger/2} -L_{\text{obs}}^{\dagger/2}) \beta.
    % \]
    In this decomposition, $h$ is $\eta$-homophilous with respect to the observed graph, and $\varepsilon$ is a function of the randomly observed graph structure. Imposing conditions on $\Sigma := \mathbb{E}[\varepsilon \varepsilon^T]$ imposes minimal moment constraints on the extent of heterogeneous variation.
\end{example}

We can characterize worst-case bounds in terms of Schatten norms, $\sup_{\|\Sigma_f\|_{q^\star} \leq \kappa} \operatorname{Tr}(X\Sigma_f) = \kappa \|X\|_{q}.$
% \begin{align*}
% \sup_{\|\Sigma_f\|_{q^\star} \leq \kappa} \operatorname{Tr}(X\Sigma_f) = \|X\|_{q} \kappa.
% \end{align*}
We write this out formally in \Cref{section_mse_bound}. When $q^* = 1,\, q= \infty$, we recover the previous operator-norm bound.
When $q^*=q=2$, we get a bound in terms of the Frobenius norm of the covariance matrix, $\kappa \| X \|_F$. This leads to a novel connection to \citet{viviano2023causal}. 

Concretely, these are realized as penalizations that encourage randomization in our designs, with the extreme case being unit-level Bernoulli randomization.
In this sense, randomization yields robust designs that ensure a controlled worst-case MSE under minimal moment assumptions. This underscores the interpolation of the two extremes of randomization and determinism. In principle, we uniformly independently randomize when we have no information and assumptions at all. 
Conversely, if we had all the necessary information, i.e., all potential outcomes were encoded through our notions of homophily and interference, then we can deterministically solve for an optimal design. 

The importance of randomization for robustness has a long history, and here we mention only a few representative references most closely related to our perspective. \citet{fisher1925statistical, fisher1935design} first emphasized this importance through the role of randomization in yielding unbiased estimates of experimental error, which are vital for valid testing procedures. \citet{wu1981robustness} further studied robustness from the perspective of the worst-case mean squared error; showing that the Bernoulli randomized design is minimax optimal over invariant sets of model violations, as is also reflected in our framework. The paper also describes how efficiency can be gained with structured assignments aligned with particular patterns of violations. More recently, \citet{kallus2018optimal} showed that, without structural information on how the outcomes depend on the particular covariates, Bernoulli randomization is minimax optimal with respect to variance.

\section{Potential Outcomes Model and Network Interference} \label{section:interference}
%\subsection{Potential Outcomes Model}
Let $z \in \{0, 1\}^n$ denote the binary treatment assignment vector. We consider the following potential outcome model
\begin{align} \label{potential_outcome_model}
    Y_i(z) = \alpha_i + \phi_i z_i + {s_{i}}(z),
\end{align} with $z_i \in \{0, 1\}$, for all $i \in [n]$. The term $s(z)$ denotes interference that vanishes under constant treatment, i.e., $z \in \{\mathbf{1}, \mathbf{0}\}$, with $\alpha$ denoting the baseline, and $\phi$ the treatment effect (see \Cref{extremes_ate}). Together with \Cref{extremes_ate}, $s(z)$ is uniquely determined.

\begin{definition}[$\gamma$-interference] \label{gamma-interference}
We say that the outcome component $s$ satisfies ($\gamma, L$)-interference, or is ($\gamma, L$)-interferent, if $\sup_{z \in \{0,1\}^n} s(z)^TL^{\dagger}s(z) \leq \gamma$, where $L^{\dagger}$ is the pseudo-inverse graph Laplacian.
\end{definition}

\begin{remark}
Similar to our definition of homophily (see \Cref{remark_homophily}), unless otherwise specified, we use $\gamma$-interference to refer to interference with respect to $L$, suppressing this dependence in the notation for brevity. For a similar level of interference with respect to a different kernel $\mathcal{K}$, we say that such a vector $s$ is $(\gamma, \mathcal{K})$-interferent.
\end{remark}

\begin{example}[Neighborhood Interference] \label{neighborhood_interference}
A form of $s(z)$ that is commonly studied in the literature, such as in \cite{chin2019regression, eckles2017design, harshaw2023design, yu2022estimating}, is the linear neighborhood interference term, i.e., $s(z) \propto D^{-1}Lz$. This corresponds to a neighborhood interference setting, where a unit’s outcome depends only on its own and its neighbors’ treatment assignments, not on assignments beyond its neighborhood. Let the normalized graph Laplacian $\Tilde{L} := I-D^{-1}W$, i.e., $\Tilde{L} = D^{-1}L$. A simple outcome model with such a ($\gamma, \Tilde{L}$)-interferent $s$ component with a $\sqrt{n}$-scaling reminiscent of the local asymptotic framework \citep{viviano2023causal, hirano2009asymptotics} is
$Y_i = \alpha_i + \beta_i z_i + \frac{\gamma}{\sqrt{n}} \sum_{j} \frac{W_{ij}}{\sum_j W_{ij}} z_j$. Here, $s(z) = -  \frac{\gamma}{\sqrt{n}} \Tilde{L} z$, i.e., $s$ is $(\gamma, \Tilde{L})$-interferent. Additionally, the normalized Laplacian term appropriately vanishes when the full population is treated or assigned to control. We describe this in more detail in \Cref{neighborhood_interference_full}.
\end{example}

To capture settings with stochastically misspecified interference, we cannot use the approach from \Cref{example:noisy_misspecification}, since the resulting misspecification would be correlated with treatment assignment, complicating our later analysis. Instead, we rely on the intrinsic robustness of our constraint, which allows generalization beyond neighborhood interference models. 

\begin{example}[Full Network Interference] \label{example:noisy_interference}
Our interference characterization allows for ``perturbations'' of the interference structure. Let $\Tilde{L}$ be the graph Laplacian of the perturbed network, and let $L$ be the graph Laplacian underlying a neighborhood interference structure satisfying $\gamma$-interference. If both the perturbed and unperturbed networks are connected, then the column and row spaces of $L$ and $\Tilde{L}$ are equal, and hence $\Tilde{L}^{\dagger/2} L^{1/2} L^{\dagger/2} = \Tilde{L}^{\dagger/2}$, and $L^{\dagger/2} L^{1/2} \Tilde{L}^{\dagger/2} = \Tilde{L}^{\dagger/2}$. Thus, if $\|\Tilde{L}^{\dagger} L \|_{\text{op}}\leq c$, and both the perturbed and unperturbed networks are connected, we can write ${s(z)}^T \Tilde{L}^\dagger s(z) = {s(z)}^T (\Tilde{L}^{\dagger/2} L^{1/2}) L^{\dagger} (L^{1/2} \Tilde{L}^{\dagger/2}) s(z) \leq c  {s(z)}^T L^\dagger s(z) \leq c \gamma$ for all $z \in \{0,1\}^n$.
In other words, if $L$ and $\tilde{L}$ are close in spectrum, i.e., $\|\tilde{L}^\dagger L \|_{\text{op}} \leq c$, then $s$ being $(\gamma, L)$-interferent implies that $s$ is $ (c\gamma, \tilde{L})$-interferent. For instance, if $\Tilde{L}$ encapsulates a full-interference structure with decaying interference (for example, \citet{leung2022causal}) across nodes with longer distances such that $\Tilde{L}$ is well approximated by the neighborhood interference Laplacian $L$,
% $L_{\calN}$ \textcolor{magenta}{was this defined?},
we can write $c = 1 + O(\psi)$ for some $\psi \ll 1$. We describe this in more detail in \Cref{example:noisy_interference_full}.
\end{example}

\section{Estimation and Trade-offs} \label{section_mse_bound}

\subsection{GATE estimation}  \label{gate_estimation}

\begin{definition}[Global Average Treatment Effect (GATE)] \label{def:gate}
The Global Average Treatment Effect (GATE) is defined to be
\begin{align*}
    \tau = \frac{1}{n} \dotp{\mathbf{1},Y(\mathbf{1}) - Y(\mathbf{0}) }  = \frac{1}{n} \dotp{\mathbf{1}, \phi}.
\end{align*}
\end{definition}

Our primary focus is on the Horvitz-Thompson estimator with uniform treatment probability $p$ across nodes, $p = \mathbb{P}(z_i = 1) = \mathbb{P}(z_j = 1)$ for all $i,j \in [n]$:
\begin{align*}
    \hat{\tau} = \dotp{Y, \frac{z}{p n} - \frac{\mathbf{1}-z}{(1-p)n}}.
\end{align*}
For ease of exposition, we present our framework primarily in the symmetric setting, i.e., $p =1/2$. The general case is given in \Cref{section:non_symmetric}.

We note here that the particular Horvitz-Thompson estimator we focus on throughout the paper is the vanilla one. In the literature on interference and cluster randomized designs, it is common to pair Horvitz-Thompson estimators with neighborhood exposure mappings  \citep{aronow2017estimating, ugander2013graph, leung2022rate, thiyageswaran2024data, eichhorn2024low}.

Under a correctly specified exposure mapping under network interference, the Horvitz-Thompson estimator remains unbiased, albeit with high variance. In contrast, the vanilla Horvitz-Thompson estimator is biased in the presence of interference, though it typically exhibits lower variance. 

We focus on the vanilla estimator because it is much simpler, and our primary goal is to study the tradeoffs that occur under varying levels of interference, homophily and heterogeneous variation, for various designs. This includes settings without interference, where using a Horvitz-Thompson estimator paired with a neighborhood exposure mapping would introduce unnecessary variance. Moreover, applying exposure mappings requires specifying them correctly, and under more general forms of interference beyond neighborhood interference, characterizing the correct exposure mapping (and the imprecision that arises from a misspecified exposure mapping) becomes much more complicated. In \Cref{section_dim} we consider our framework with respect to the difference-in-means estimator. In fact, our framework applies to more general estimators of the form $\hat{\tau} = \dotp{Y, w}$, with general weights $w$.

The measure of precision of the estimator we focus on is the worst-case mean squared error (MSE) $\sup_{h_f, \Sigma_f, s \in \chi(\eta, \gamma, \kappa)} \mathbb{E}[(\hat{\tau} - \tau)^2]$, with respect to the potential outcomes within the class $\chi(\eta, \gamma, \kappa) = \chi_{q^*}(\eta, \gamma, \kappa) = \{h_f, \varepsilon_f, s: \|L^{1/2} h_f\|^2 \leq \eta, \|\Sigma_f\|_{q^*}\leq \kappa, \|L^{\dagger/2} s(z)\|^2 \leq \gamma$ for all $z\in \{0,1\}^n$, for $f = \alpha, \phi \}$. Here $q^*$ denotes the Schatten exponent used to connstrain the exponent $\Sigma_f$, $q$ denotes its Hölder conjugate exponent, defined by $1/{q^*} + 1/q = 1$. Consequently, the dual penalty appearing in the MSE bound is $\| X \|_q$.

When $\tau$ and $\hat{\tau}$ incorporate stochastic network misspecification, this expectation is understood to be the iterated expectation $\mathbb{E}[\mathbb{E}[(\hat{\tau} - \tau)^2 \mid \varepsilon_f]]$, $f=\alpha, \phi$, over the network misspecification residual and the design, respectively. 

We call $\chi(\eta, \gamma, \kappa)$ the \textit{min-min-max} model class to reflect $(\eta,\gamma,\kappa)$ chosen to minimize the resulting worst-case MSE upper bound in \Cref{eq: MSE bound} among all admissible decompositions of $\alpha$ and $\phi$. This parameterization, arising from the corresponding trade-off-admissible \textit{min-min-max} decomposition of $\alpha$ and $\phi$, yields the sharpest worst-case MSE control within our framework and is unique up to alternative optimal tuples.  
Specifically, the components $h_\alpha, h_\phi$ and $\varepsilon_\alpha, \varepsilon_\phi$ optimally capture, relative to the underlying graph structure, the homophily and heterogeneous variation present in $\alpha$ and $\phi$, by effectively separating graph-aligned low-Rayleigh-quotient variation from residual heterogeneous variation; see also \Cref{remark_homophily_clustering}. 

\begin{theorem}[MSE bound] \label{thm:mse_bound_general_HT_symmetric}
Let $p:= \mathbb{P}(z_i = 1) = 1/2$ for all $i \in [n]$, $a = \langle \alpha, \mathbf{1}/n\rangle$, and $b = \langle \phi, \mathbf{1}/n\rangle$. Let $|\langle s(z), \mathbf{1}/n\rangle| \leq \sqrt{\delta}$ for all $z$. Let also $x \in \{-1, 1\}^n$ by defining $x := 2z-1$. Define $X := \mathbb{E}[xx^T]$ and $\Delta := (a + b/2)^2 + \delta$. Then, for $q \geq 1$,
     \begin{align}\label{eq: MSE bound}
         \sup_{\chi(\eta, \gamma, \kappa)} \mathbb{E}[(\hat{\tau} - \tau)^2] \leq \frac{28}{n^2} \{ \Delta \operatorname{Tr}(\mathbf{1}\mathbf{1}^T X) + \frac{5\eta}{4}\operatorname{Tr}(L^{\dagger} X) + \frac{5\kappa}{4}\|X\|_{q} + \gamma \operatorname{Tr}(L X) \}.
     \end{align}
\end{theorem}

Thus, to control the worst-case MSE it suffices to minimize the right-hand side in \Cref{eq: MSE bound} over the covariance matrix $X$. The proof of \Cref{thm:mse_bound_general_HT_symmetric} is in \Cref{proof_mse_bound_general_HT}. We assume that the interference term is bounded in its center, i.e., $\sup_z \dotp{s({z}), \mathbf{1}/n} \leq \sqrt{\delta}$, and this is implicitly adopted in later MSE bounds. 

In \Cref{alternative_interference}, we discuss an alternative where we constrain the quadratic form $s(z)^T (\delta \mathbf{1}\mathbf{1}^T +L)^{-1} s(z)$. If instead we were to constrain the quadratic form $s(z)^T (\lambda I +L)^{-1} s(z)$, we would arrive at the same bound up to a constant additive factor without necessitating this assumption. There are many alternatives that lead to fundamentally equivalent optimization problems, with another alternative provided in \Cref{exp: quad interference}.

\subsection{Trade-offs} \label{model_class_tradeoffs}

\subsubsection*{Trade-off parameters} \label{assumptions}
Our design flexibly navigates homophily, robustness, and interference trade-offs parameterized by $\eta, \kappa, \gamma \in \mathbb{R}$:
\begin{enumerate} 
    \item ${h_{\alpha}}^T L h_{\alpha} \leq \eta$, and ${h_{\phi}}^T L h_{\phi} \leq \eta$ (see also \Cref{remark_homophily} for a measure of homophily including degree similarity)
    \item $\| \Sigma_\alpha \|_{q^*} \leq \kappa$,  and $\|\Sigma_\phi\|_{q^*} \leq \kappa$
    \item $s(z)^T L^{\dagger} s(z) \leq \gamma$ for all $z \in \{ 0,1\}^n$
\end{enumerate}

As is typical in conducting power analyses, our method depends on baseline and treatment effect size, encapsulated by $\Delta$. We dissect the three components in the trade-offs above. 

\subsubsection{Cut Minimization}
If we fix the treatment vector $x$, then $X = xx^T$. The term $\operatorname{Tr}(LX) = x^T L x$ in the theorem statement above is the same as the objective function in the cut minimization problem. Indeed, for $L = D- W$, and $x \in \{-1, 1 \}^n,$ $\frac{1}{4} x^T L x =  \sum_{i \in S, j\in S^c}  W_{ij}$,
% \begin{align*}
%      \frac{1}{4} x^T L x =  \sum_{i \in S, j\in S^c}  W_{ij}, 
% \end{align*} 
where $S$ and $S^c$ make up the cut partition. Thus, this $x^T L x$ term alone encourages a more ``clustered'' design. The term $\sum_{i \in S, j\in S^c}  W_{ij}$ has previously appeared in \citet{brennan2022cluster, viviano2023causal}.

\subsubsection{Effective Resistance}
Once again, if we fix the treatment vector $x$, then $X = xx^T$. The term $\operatorname{Tr}(L^\dagger X) = x^TL^\dagger x$ can be expressed in terms of the effective resistances. Let $R$ be the matrix of effective resistances across edges, so that the $(i,j)$-th entry is the effective resistance $R_{ij}$ across edge $(i,j).$ Then, for
$x \in \{-1,1\}^n$, $x^T L^\dagger x =  - \frac{1}{2} x^T \Pi^T R \Pi x$, where $\Pi$ is the projection matrix to the subspace orthogonal to $\mathbf{1}$ (see \citet{fontan2021properties}).
% \begin{align*}
%      x^T L^\dagger x &=  - \frac{1}{2} x^T \Pi^T R \Pi x, 
% \end{align*} 
In the graph sparsification literature, \citet{spielman2008graph} use effective resistances to sample network edges.
Selecting a cut to minimize $x^TL^\dagger x$ corresponds to selecting a treatment design ``spread-out'' in the graph.
We  see in our MSE bounds that this provides a counterbalance to the $x^TLx$ term above that incentivizes clustered designs, and thus improves our estimation procedure in the presence of homophilous outcomes. 

\begin{figure}
    \centering
    \begin{subfigure}[b]{0.3\textwidth}
        \centering
        \includegraphics[width=2in]{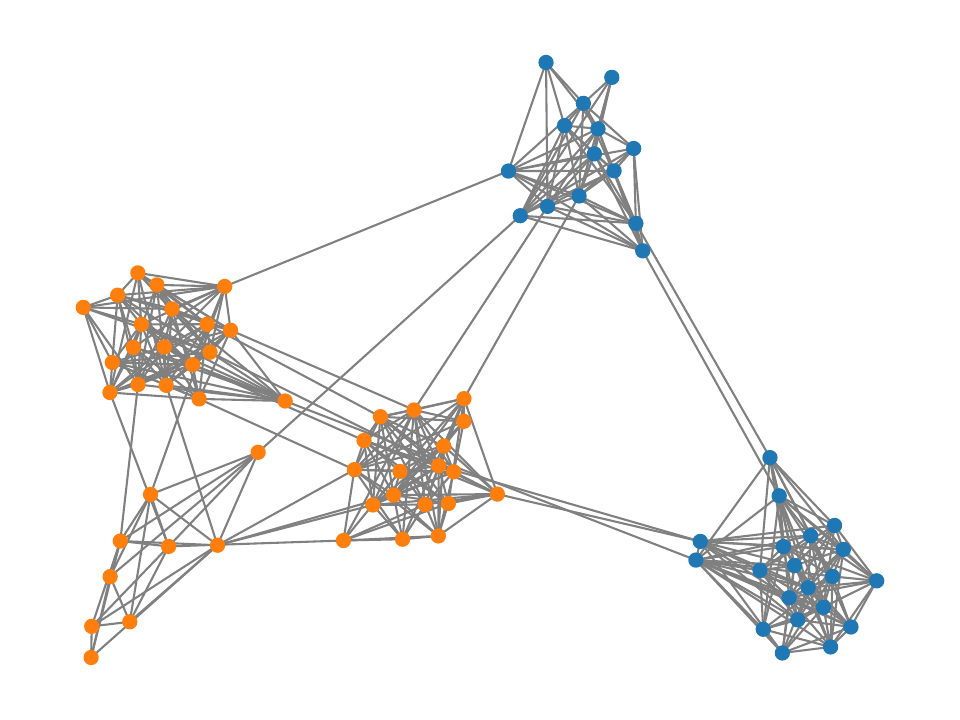}
    \end{subfigure}
    \hfill
    \begin{subfigure}[b]{0.3\textwidth}
        \centering
        \includegraphics[width=2in]{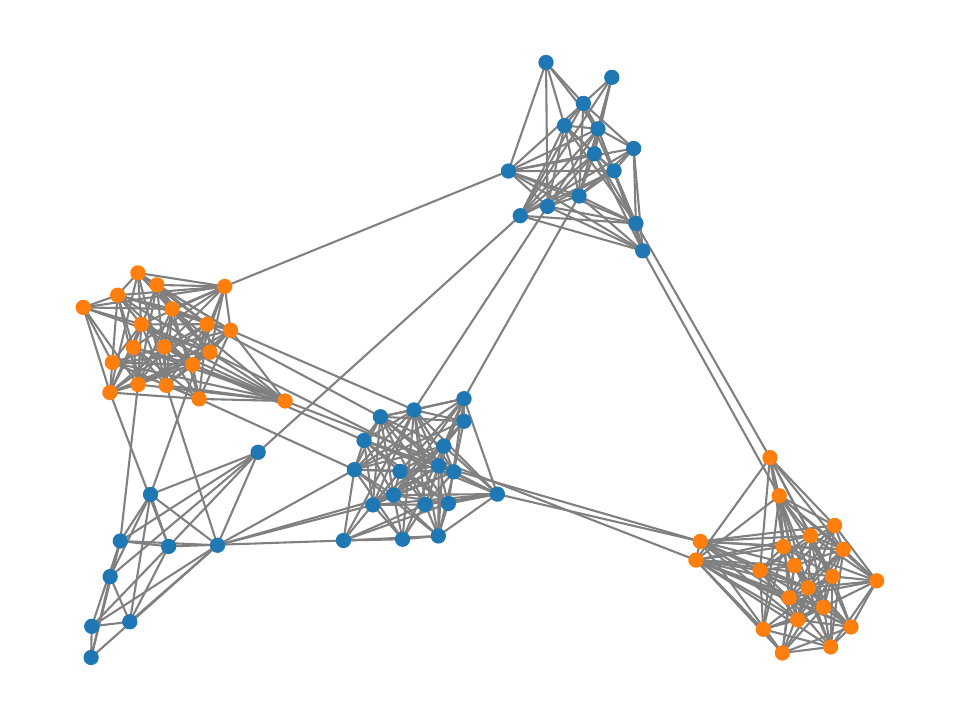}
    \end{subfigure}
    \hfill
    \begin{subfigure}[b]{0.3\textwidth}
        \centering
        \includegraphics[width=2in]{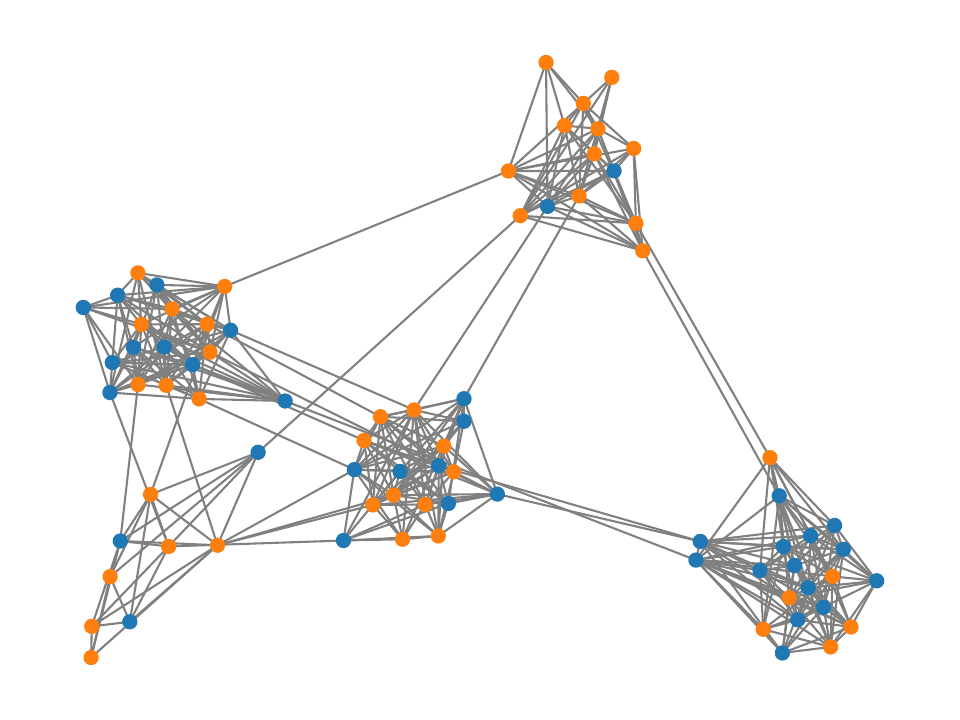}
    \end{subfigure}
    \caption{Optimized treatment assignments under interference-robustness tradeoffs on a synthetic network as heterogeneous variation levels increase from left to right. See \Cref{app:more illustrations} for more details.}
    \label{fig:implementation_five_cluster_transition_h_k}
\end{figure}

\begin{figure}[tb]
    \centering
    \includegraphics[width=\linewidth]{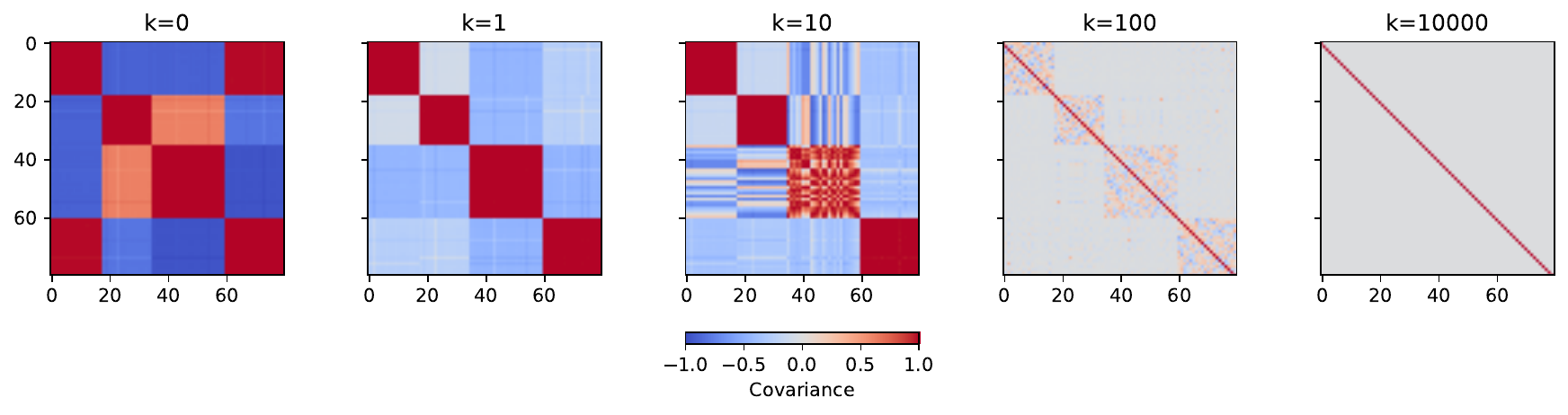}
    % \caption{Solutions of the SDPs described in \Cref{section_sdp} on the graph in  \cref{fig:sbm_four_clusters} trading-off $10\operatorname{Tr}(LX) + k \| X\|_{F} + 10 \operatorname{Tr}(\mathbf{1}\mathbf{1}^T X)$. $k$ ranges between $(0,10^0, 10, 10^2, 10^4)$ from left to right.}
    \caption{Optimal design covariance matrices under interference-robustness tradeoffs as heterogeneous variation levels increase from left to right. See \Cref{app:more illustrations}  for more details.}
    \label{fig:cov_mat_h_k}
\end{figure}

\subsubsection{Randomization} \label{section:randomization}
We use \Cref{prop:det_schatten_bound} to bound the contribution of the heterogeneous variation $\varepsilon_f$ to the MSE of our estimator.
Note that for deterministic $\varepsilon_\alpha, \varepsilon_\phi$, $\Sigma_f =\varepsilon_f \varepsilon_f^T$, and $\|\Sigma_f \|_{q^*} = \| \varepsilon_f \|_{\ell_2}^2,$ $f = \alpha, \phi$.

\begin{proposition}[Worst-case bound over deterministic heterogeneous variation] \label{prop:det_schatten_bound}
For PSD matrices $X \in \mathbb{R}^{n\times n}$, $\Sigma = \varepsilon_f \varepsilon_f^T \in \mathbb{R}^{n \times n}$, $f = \alpha, \phi$ and $\kappa \in \mathbb{R}$,
$\sup_{\|\varepsilon_f\|_{\ell_2}^2 \leq \kappa} \operatorname{Tr}(X\Sigma) = \sup_{\|\varepsilon_f\|_{\ell_2}^2 \leq \kappa} \varepsilon_f^T X \varepsilon_f = \kappa \lambda_{\text{max}} (X)$, where $\lambda_{\text{max}} (X)$ is the maximum eigenvalue of $X$.
\end{proposition}

This recovers the operator-norm bound that is used as a robustness measure in \citet{harshaw2024balancing}. We can generalize this further for the (uncentered) covariance matrix of the stochastic network misspecification.

\begin{proposition}[{\citealp[Exercise IV.2.12]{bhatia2013matrix}}] \label{prop:general_schatten_p_bound}
Let $\| \cdot \|_{q}$ denote the Schatten $q$-norm of a matrix. For symmetric positive semidefinite (PSD) matrices $X \in \mathbb{R}^{n\times n}$ and $\Sigma \in \mathbb{R}^{n \times n}$, and $1/{q^*} + 1/q = 1$, \(\sup_{\|\Sigma\|_{q^\star} \leq \kappa} \operatorname{Tr}(X\Sigma) = \kappa \|X\|_{q} .\)  
\end{proposition}
Particular choices of $q^*$ relate to existing trade-offs in the literature.

\begin{corollary}[Worst-case bound over the marginal variances] \label{variance_schatten_bound}
For PSD matrices $X \in \mathbb{R}^{n\times n}$ and $\Sigma \in \mathbb{R}^{n \times n}$, $q^\star = 1,\, q= \infty$, and $\kappa \in \mathbb{R}$, $\sup_{\|\Sigma\|_{1} \leq \kappa} \operatorname{Tr}(X\Sigma) = \kappa \lambda_{\text{max}} (X) .$
\end{corollary}

As a practical consideration, it is worth noting that the above supremization places little constraint on the off-diagonal of $\Sigma$. 
Indeed, the supremum is realized by a dense $\Sigma$, and as $\Sigma$ is PSD, the Schatten-1 norm corresponds to the trace, hence the marginal variances of $\varepsilon$ are directly constrained. 
However, in many circumstances, one may expect the coordinates of $\varepsilon$ to be somewhat independent, making $q^*=1$ an unrealistic modeling assumption.

In the other extreme $\Sigma$ could be a diagonal matrix corresponding to independent heterogeneous variation, with the off-diagonal entries being completely nullified. Such a $\Sigma$ is achieved if we were to instead supremize over the constraint $\|\Sigma\|_{\infty} \leq \kappa$, and the resulting error would be $\kappa\operatorname{Tr}(X) = \kappa n$, which is constant. That is, if the (edge) misspecification was, for example, i.i.d. across the nodes, then additional randomization to account for network misspecification would be unnecessary, as this value $\kappa n$ is constant, independent of the chosen design. As a reasonable compromise between these two extremes, we highlight the Frobenius norm, $q^*=2$.

\begin{corollary}[Worst-case bound over the covariance structure] \label{covariance_schatten_bound}
For PSD matrices $X \in \mathbb{R}^{n\times n}$  and $\Sigma \in \mathbb{R}^{n \times n}$, $q^*=q=2$, $\sup_{\|\Sigma\|_{2} \leq \kappa } \operatorname{Tr}(X\Sigma) = \kappa \| X\|_{2} = \kappa\sqrt{\operatorname{Tr}(X^2)} = \kappa \sqrt{\sum_{i=1}^n \lambda_{i}^2(X)}$.
Additionally, $\sup_{\|\Sigma\|_{2} \leq \kappa \| X\|_{2}} \operatorname{Tr}(X\Sigma) = \kappa\operatorname{Tr}(X^2) = \kappa \sum_{i=1}^n \lambda_{i}^2(X)$.
\end{corollary}

In the bounds in \Cref{covariance_schatten_bound}, the total covariance structure of $({\varepsilon}_i)_{i=1}^n$ is constrained, rather than just the marginal variances. By constraining the Frobenius norm, we bound the sum of the squared entries of all of the elements of $\Sigma$, putting equal importance on on- and off-diagonal elements. The distinction between the two bounds in \Cref{covariance_schatten_bound} is that the latter allows the norm of $\Sigma$ to grow with $X$ (and therefore $n$).

Under a cluster-level Bernoulli randomization design, the second part of \Cref{covariance_schatten_bound} corresponds to the dominant term in the variance presented in \citet{viviano2023causal}. Indeed, consider $X$ such that $X_{ij} = 1$ if nodes $i,j$ are in the same cluster, and 0 otherwise. This matrix $X$ is the assignment covariance matrix for cluster-level Bernoulli randomization, where each cluster is independently assigned a treatment with probability $p$. The eigenvalues of this matrix are given by the sizes of the clusters, $|\mathcal{C}_i|$, hence $\operatorname{Tr}(X^2) = \sum_i |\mathcal{C}_i|^2$. This is exactly the dominating term in the variance observed in \citet[Theorem 3.2]{viviano2023causal}, hence the squared Frobenius norm generalizes this variance to generic designs. Further, as it appears as a convex penalty in our objective, it allows for easy computation. 
As unit-level clustering results in the smallest value of the Schatten-$q$ penalty, increases to the penalization parameter $\kappa$ will force the solution towards this covariance structure.

\begin{proposition}\label{prop:X_norm_covariance_min}
Let $x \in \{-1,1\}^n$ denote the treatment assignment under a Bernoulli randomization design. Then, for fixed $p= \mathbb{P}(x_i = 1)$, $i \in [n]$, and $q>1$, $\arg \min_{\substack{X \succeq 0; \\ X_{ii}=1}} \|X\|_{q} = \frac{1}{4p(1-p)}\operatorname{Cov}(x)$.
    % \begin{align*}
    %     \arg \min_{\substack{X \succeq 0; \\ X_{ii}=1}} \|X\|_{q} = \operatorname{Cov}(r).
    % \end{align*}
\end{proposition}

We note that the analogous result holds if $x \in \{-1,1\}^n$ is the treatment assignment under a complete randomization design and $n_1 = n_0$, with the optimization constraint $X \mathbf{1} = \mathbf{0}$. We present this formally in \Cref{proof:trade-offs}.

Thus, large $\kappa$ permits a more robust design. Indeed, $\kappa$ penalizes the Schatten norm of $X$, driving the design toward a Bernoulli randomization as $\kappa\to \infty$, as formalized in \Cref{prop:X_norm_covariance_min}. We illustrate this behavior in \Cref{fig:cov_mat_h_k,fig:cov_mat_m_k}, as the trade-off parameters in the SDP, described in \Cref{section_sdp}, are varied. 

\begin{remark} \label{randomization_levels}
    We use the term randomization in a specific sense: by increased randomization, we refer to designs with denser covariance matrices, which lead to more robustness to heterogeneous variation. This stands in contrast to trivial randomizations, such as randomly flipping between fixed treatment and control assignment vectors, that preserve the covariance matrices and lie on level sets of the objective function. 
    % We do not refer to this latter scenario.
\end{remark}

To further demonstrate the trade-offs in our framework, we generate designs for a variety of trade-off parameters. \Cref{fig:cov_mat_m_h} illustrates the trade-offs of the minimum cut and minimum effective resistance components against each other. It depicts a ``clustered'' block structure in the covariance matrices on the left-end of the spectrum when $\gamma > \eta$, and a more ``diffusive'' covariance structure on the right-end of the spectrum when $\gamma < \eta.$ \Cref{fig:implementation_five_cluster_transition_h_k,fig:implementation_five_cluster_transition_h_m} depict the trade-offs on a graph generated from an SBM with equal cluster membership probability across five clusters. \Cref{fig:implementation_five_cluster_transition_h_k} isolates the trade-offs between clustering and randomization. As the ratio $\kappa/\gamma$ increases, the designs interpolate from a clustered design to a design that resembles unit-level Bernoulli randomization. \Cref{fig:implementation_five_cluster_transition_h_m} isolates the trade-offs between clustering and effective resistance minimization. As the ratio $\eta/\gamma$ increases, the designs interpolate from a clustered design to a design that is maximally spread-out.

\begin{remark}
The term $\operatorname{Tr}(\mathbf{1}\mathbf{1}^TX)$ in the bound encourages solutions that are orthogonal to $\mathbf{1}\mathbf{1}^T$ as $\Delta \to \infty$. Since the constant vector lies in the kernel of both the graph Laplacian and its pseudo-inverse, this term encourages non-degenerate solutions.
\end{remark}

\section{Experimental Design}
We describe the two design approaches we propose. To keep ideas simple and clear, in what follows we write the robustness trade-off terms only with general Schatten norms, rather than particular special cases.

\subsection{MSE-minimizing SDP relaxation and Gaussian Rounding} \label{section_sdp}
Let $\Delta = (a + b/2)^2 + \delta.$ Reparameterizing the right-hand side of \Cref{thm:mse_bound_general_HT_symmetric} with $\Tilde{\gamma}$,$\Tilde{\eta}$, and $\Tilde{\kappa}$, so that
$\Tilde{\gamma} := \frac{\gamma}{ \Delta}, \Tilde{\eta} := \frac{5\eta}{4\Delta} ,\text{ and } \Tilde{\kappa} := \frac{5\kappa}{4\Delta}$,
we can write an SDP more explicitly in terms of the new trade-off parameters:
\begin{align} \label{sdp_relaxation_mse_p_norm_symmetric}
    \min_{X\in\mathbb{R}^{n\times n}} \quad & \Tilde{\eta}\text{Tr}(L^\dagger X)  +   \Tilde{\gamma} \text{Tr}(L X) + \Tilde{\kappa}\|X\|_{q} + \operatorname{Tr}(\mathbf{1}\mathbf{1}^TX) \\
    \text{s.t.} \quad & X \succeq  0 \notag\\
    & \text{diag}(X) = \mathbf{1} \notag 
\end{align}
The non-symmetric setting is presented in \Cref{section:non_symmetric}.
We generate our treatment assignment vector $\zeta$ using \Cref{alg:gaussian_round}, drawing ideas from  \citet{goemans1995improved}. 

\begin{algorithm}
\caption{\textsc{GaussianRounding}}
\label{alg:gaussian_round}
\begin{algorithmic}[1]
\Require Parameters: $q$, $\Tilde{\gamma}$, $\Tilde{\eta}$,$\Tilde{\kappa}$, and marginal assignment probability $p$ 
\Comment{$\mathbb{P}(x_i = 1) =p$ for all $i \in [n]$}
\State Solve the SDP in Equation~\eqref{sdp_relaxation_mse_p_norm_symmetric} with parameters $q$, $\Tilde{\gamma}$,$\Tilde{\eta}$, and $\Tilde{\kappa}$; let the solution be $X_{\mathrm{SDP}}$
\State Sample $\xi \sim \mathcal{N}(0, X_{\mathrm{SDP}})$ \Comment{Multivariate Gaussian with covariance matrix $X_{\mathrm{SDP}}$}
\For{each $i \in [n]$}
    \State $x_i' \gets \text{sgn}(\xi_i)$
    \If{$x_i' = 1$}
        \State Draw $\zeta_i \sim \text{Rademacher}(2p)$ \Comment{$\mathbb{P}(\zeta_i = 1 \mid x_i' = 1) = 2p$, $\mathbb{P}(\zeta_i = -1 \mid x_i' = 1) = 1 - 2p$}
    \Else
        \State $\zeta_i \gets -1$
    \EndIf
\EndFor
\State \Return $\zeta$
\end{algorithmic}
\end{algorithm}

\subsection{Adapted Gram-Schmidt Walk algorithm} \label{covariate_balancing_comparisons}

As a second approach, we adapt the Gram-Schmidt Walk algorithm \citep{harshaw2024balancing}, a vector balancing procedure, to our problem. In particular, by specializing the robustness hyperparameter $\lambda$, we provide an alternative algorithm that yields a guarantee for the worst-case MSE.

Focusing for simplicity on the homophily term $x^T L^\dagger x$ in our MSE expression, we can write 
$x^TL^{\dagger}x = x^T A^T A x$,
with $A=\sqrt{\eta} L^{\dagger/2}$, or $A= \sqrt{\eta}\Lambda^{1/2}V^T$, where $\Lambda$ is the matrix of eigenvalues of $L^{\dagger}$, $V$ is the matrix of the corresponding eigenvectors. We then use the Gram-Schmidt Walk algorithm \citep{harshaw2024balancing} on the augmented covariate matrix $B$ with columns $B_i= [e_i \quad A_i]^T$ where $e_i$ are standard basis vectors, and randomization parameter $\lambda = \frac{\sqrt{\kappa n^{1/q}}}{ \sqrt{\omega_A^2n} + \sqrt{\kappa n^{1/q}}}$ in \Cref{alg:gramschmidt}, where $\omega_A = \max_{i \in [n]} \| A_i \|$.

\subsection{Comparing the two approaches} \label{comparison_two_approaches}
Both proposed methods extend to more general kernels, since the final optimization objective function retains the same form after replacing the Laplacian (see \Cref{general_misspecifications_sim_kernels}).

In the adapted Gram-Schmidt Walk algorithm, the bounds grow with $n$ (see also discussion after \Cref{thm:gsw}), since this procedure is tailored to the minimization of the operator norm. By contrast, we can directly specify the Schatten-$q$ norm penalties, via an SDP for common choices of $q$ (e.g.\ $q \in \{1, 2, \infty\}$), or via convex optimization more broadly, leading to finer control.
However, the computational efficiency of the SDP significantly drops for larger $n$, making vector balancing a practical alternative. We incorporate our trade-off formulation into the Gram-Schmidt Walk algorithm \citep{harshaw2024balancing} in \Cref{alg:gramschmidt}, line~\ref{line:adapt}. 

Our application of the Gram-Schmidt Walk algorithm reduces to controlling $\operatorname{Tr}(A^TA X)$ and $\Tilde{\kappa}\| X\|_q$, where $A^TA:= \Tilde{\eta}L^{\dagger} + \Tilde{\gamma} L+ \mathbf{1} \mathbf{1}^T$. This casts worst-case MSE control as a covariate balancing problem, where the relevant covariates are induced by a choice of factor $A$ of $A^TA$. One natural choice is the spectral factor arising from an eigendecomposition of $A^TA$. Other valid choices include $A:= (\Tilde{\eta}L^{\dagger} + \Tilde{\gamma} L+ \mathbf{1} \mathbf{1}^T)^{1/2}$, or $A^T:= [ \Tilde{\eta}^{1/2} L^{\dagger/2} \quad \Tilde{\gamma}^{1/2}L^{1/2} \quad  {\mathbf{1} \mathbf{1}^T}^{1/2}]$.
This framing would allow one to think about clustering as part of the covariate balancing problem, the original focus of \citet{harshaw2024balancing}.  

The SDP followed by Gaussian rounding described in \Cref{section_sdp} is specialized to uniform treatment probabilities, whereas the Gram-Schmidt Walk approach allows direct specification of non-uniform treatment probabilities: let $\mathbf{p} = (p_1, p_2, \ldots, p_n)$ denote the vector of treatment probabilities, and set the initial vector $z_1 \leftarrow 2\mathbf{p}-1$ in \Cref{alg:gramschmidt}, line~\ref{line:initialize}. On the other hand, the SDP formulation makes it straightforward to add additional convex constraints (see also \Cref{section_dim}).

\section{Theoretical guarantees}
\Cref{corr:approximation_bound} provides worst-case MSE bounds for the SDP solution paired with the Gaussian Rounding procedure. In \Cref{thm:gsw}, we provide worst-case MSE bounds for the Gram-Schmidt Walk algorithm by applying results from \citet{harshaw2024balancing}. The proofs are presented in \Cref{proof:thm_approximation_error}.

\subsection{SDP followed by Gaussian Rounding}
Our proof of \Cref{corr:approximation_bound} utilizes the seminal ideas underlying the approximation bound for MAXCUT, as introduced in \citet{goemans1995improved} (see \Cref{maxcut_context} for more details). Our problem is similar in flavour, except we look at the minimization problem instead of maximization, and as such we would like an inequality in the opposite direction for a meaningful approximation bound.

We derive geometric identities associated with \Cref{alg:gaussian_round}.
When $p:= \mathbb{P}(\zeta_i = 1) =1/2$, we have Grothendieck's identity, as derived in \citet{goemans1995improved}. That is, 
$\mathbb{E}[\zeta_i \zeta_j] = \frac{2}{\pi} \arcsin{\dotp{v_i, v_j}}$,
where $v_i$ is the $i$-th column vector of $X_{\text{SDP}}^{1/2}$, and $X_{\text{SDP}}$ is the solution to the SDP in \Cref{section_sdp}.
Indeed, when $p=1/2$, it is equivalent to setting $\zeta_i = 1$ whenever $\dotp{\xi, v_i} > 0$, $i \in [n]$. This is because $\mathbb{P}(\dotp{v_i, \xi} > 0) = 1/2.$ 

\begin{theorem}[Approximation bound for SDP + Gaussian Rounding] \label{corr:approximation_bound}
Let $\Xi := \cov(\zeta)$ be the covariance matrix of the associated Rademacher random variables. Let $X^*$ be the optimal achievable covariance matrix of the design in \Cref{thm:mse_bound_general_HT}. 
Let also $X_{\text{SDP}}$ denote the SDP solution matrix. We assume that $X_{\text{SDP}}$ is non-degenerate, i.e., $X_{\text{SDP}} \succeq \delta_\kappa I$, $\delta_\kappa > 0$. For $c_\kappa =\frac{1}{4p(1-p)} \left( \frac{8p^2}{\pi}  + \frac{8p^2 \psi_\kappa}{\pi \delta_\kappa} + \frac{4p(1-p) - 8p^2/\pi}{\delta_\kappa}  \right)$, $\psi_\kappa= n \alpha_\kappa^{3},$ where $\alpha_\kappa$ is the maximum off-diagonal entry of $X_{\text{SDP}}$ in absolute value, $\operatorname{Tr}(L\Xi) + \operatorname{Tr}(L^\dagger\Xi ) +  \|\Xi \|_{q} + \operatorname{Tr}(\mathbf{1} \mathbf{1}^T \Xi) \leq c_\kappa \left( \operatorname{Tr}(LX^*) +  \operatorname{Tr}(L^\dagger X^*)+  \| X^*\|_{q}  + \operatorname{Tr}(\mathbf{1} \mathbf{1}^T X^*)\right)$.
\end{theorem}
% \begin{proof}
% We see that $L, L^{\dagger}, \mathbf{1} \mathbf{1}^T$ are PSD. It is enough to notice that for PSD matrices $A, B, C$, with $B \succeq C$, $\operatorname{Tr}(A(B-C)) \geq 0$. Finally, notice that for $X = 4p(1-p) X_{\operatorname{SDP}}$, where $X_{\operatorname{SDP}}$ is the SDP solution, and for $X^*$ is the optimal achievable covariance matrix of a design, $\operatorname{Tr}(LX) +  \operatorname{Tr}(L^\dagger X)+  \| X\|_{q}  + \operatorname{Tr}(\mathbf{1} \mathbf{1}^T X) \leq \operatorname{Tr}(LX^*) +  \operatorname{Tr}(L^\dagger X^*)+  \| X^*\|_{q}  + \operatorname{Tr}(\mathbf{1} \mathbf{1}^T X^*).$   
% \end{proof}
% The proof is in \Cref{proof:thm_approximation_error}. Our results show that the rounding procedure provides good approximation to the optimal covariance matrix of the design when $\kappa$ is large. Under a  Bernoulli randomized design, corresponding to an extreme $\kappa$, $\alpha_\kappa = 0$. 
% The proof is given in \Cref{proof:thm_approximation_error}. 
In the proof, we make use of the Taylor expansion of $\arcsin{(X)}$, following \citet{nesterov1997quality} in the MAXCUT problem. The resulting bounds depend on the tradeoff parameter $\kappa$: as $\kappa \to \infty$, the factor $\psi_\kappa \to 0$ while $\delta_\kappa \to 1$. To encourage non-degeneracy of $X$, the design parameters must be appropriately balanced. In particular, for clusterable graphs, the small Laplacian eigenvalues can lead to heavy penalization of the corresponding inverse Laplacian eigenvectors. 

When $\kappa$ is large, the rounding procedure closely approximates the optimal design covariance; in the limiting Bernoulli randomized design, $\alpha_\kappa = 0$.
For a complete randomized design, $\alpha_\kappa = 1/(n-1)$. Our results allow for $\alpha_\kappa = \Theta(1/ n^{1/3})$, far more correlation than Bernoulli or complete randomizations, while maintaining a constant level of error. In contrast, when $\kappa$ is very small, we can optimally solve the problem deterministically.

\subsection{Adapted Gram-Schmidt Walk algorithm}

We leverage \citet[Theorem 6.4]{harshaw2024balancing} (see also \Cref{thm_gsw}) and its proof. In particular, we use the fact that the Gram-Schmidt Walk algorithm yields treatment assignments $x$ that satisfy (1) $\operatorname{Cov}(x) \preceq \lambda^{-1}I$, and (2) $\operatorname{Cov}(Ax) \preceq \omega_A^2 (1-\lambda)^{-1}I$. 
We state the results when $p= 1/2$ formally below. The general version (incorporating non-symmetric designs) is presented in \Cref{pf_thm_gsw} (\Cref{thm:gsw_nonsym}).

\begin{theorem} [Worst-case error bound from adapted Gram-Schmidt Walk algorithm] \label{thm:gsw}
Let $A:=(\Tilde{\eta}L^{\dagger} +\Tilde{\gamma} L + \Delta \mathbf{1} \mathbf{1}^T)^{1/2}$, where $\Tilde{\eta} = 5\eta/n^2$, $\Tilde{\gamma} = 4\gamma/n^2$, $\Tilde{\kappa} = 5\kappa/n^2$, and $\Delta = (a + b /2)^2$, $a = \langle \alpha, \mathbf{1}/n\rangle$, $b = \langle \phi, \mathbf{1}/n\rangle$.
The Gram-Schmidt Walk algorithm, with randomization parameter $\lambda$, returns an assignment $x$ such that $\|\operatorname{Cov}(Ax)\|_{{p'}} \leq \omega_A^2/(1-\lambda)\|I \|_{{p'}}$, and $\|\operatorname{Cov}(x)\|_{q}\leq \|I\|_{q}/\lambda=n^{1/q}/\lambda$, for $\omega_A := \max_{i\in n} \|A_i\|$, $A_i$ the column $i$ of the matrix $A$. Thus, for $q \geq 1$,
\[
\sup_{\chi(\eta, \gamma, \kappa)} \mathbb{E}[(\tau - \hat{\tau})^2] \leq 7 \left[\mathbb{E}[\|A(x-\mu)\|^2] + \Tilde{\kappa}\|X\|_{q} \right] \leq 7 \left[\frac{\omega_A^2 n}{1-\lambda} + \Tilde{\kappa}\frac{n^{1/q}}{\lambda}\right],
\] where $\mu_i = \mathbb{E}[x_i]$, for all $i \in [n].$
This is minimized for $\lambda = \frac{\sqrt{\Tilde{\kappa} n^{1/q}}}{ \sqrt{\omega_A^2n} + \sqrt{\Tilde{\kappa}n^{1/q}}}$. 
\end{theorem}

The asymptotic results in \citet{harshaw2024balancing} require the trade-off parameter $\lambda \to 1$. That is, their asymptotic results require that $\kappa/\eta, \kappa/\gamma, \kappa/\Delta \to \infty$, mirroring the improvement in the approximation error bounds in \Cref{corr:approximation_bound}.

The results in \Cref{corr:approximation_bound} and \Cref{thm:gsw} are notably different.
In \Cref{corr:approximation_bound}, we prove that rounding yields comparable MSE control to the optimal SDP solution. In contrast, in \Cref{thm:gsw} we directly bound the worst-case MSE resulting from the adapted Gram-Schmidt Walk design.
See also \Cref{comparison_two_approaches} where we discuss how the design from the Gram-Schmidt Walk algorithm compares to the design from the SDP followed by Gaussian rounding. In theory, we expect the bounds from the Gram-Schmidt Walk algorithm to be loose when $n$ is larger, and $q$ is smaller.
\section{Discussion}
We develop a novel framework to understand the three-way trade-off between homophily, network interference, and robustness to heterogeneous variation, recovering existing designs in extreme cases within the trade-off. Along the way, we generalize several existing results and designs in related literature.
We demonstrated the complementary behaviors of our two proposed approaches, solving an SDP followed by Gaussian Rounding, and the adapted Gram-Schmidt Walk algorithm. In proving approximation guarantees via the SDP approach, we leveraged and generalized ideas utilized in proving approximation guarantees for the MAXCUT problem. This generalization may be of independent interest. 

While we focus on the Horvitz–Thompson estimator for GATE estimation, our framework naturally extends to more general estimators and estimands, which we leave for future work. In this paper, we study optimal designs under known bounds $\eta, \kappa$, and $\gamma$. Such bounds may be informed by domain expertise or learned from prior experiments and historical network data, and our approach makes explicit how this information can be incorporated. An important direction for future work is extending the framework to online settings, where treatments are assigned sequentially and the parameters 
$\eta, \kappa$, and $\gamma$ are learned adaptively. Finally, for large-scale graphs,
algorithms exploiting problem-specific structure could be used to accelerate the SDP, though this lies beyond the scope of the present work.
\section{Acknowledgements}
We thank Thomas Rothvoss for insightful discussions and for pointing us to \citet{raghavendra2012approximating}, Arun Chandrasekhar for pointing us to the village network dataset, Davide Viviano for sharing the causal clustering code for verification purposes, and Seth Temple for helpful feedback on an earlier draft.

\section{Disclosure Statement}
The authors report there are no competing interests to declare.

\section{Supplementary Materials and Data Availability}
\textbf{Data Availability Statement:} The data that support the findings of this study were derived from the following resource available in the public domain: \href{https://doi.org/10.7910/DVN/U3BIHX}{DOI: 10.7910/DVN/U3BIHX}

The supplementary materials contain proofs for all formal statements in the main paper, additional theoretical and experimental results. 
% They also contain code to replicate the experiments.

\bibliography{ref.bib}
\newpage
\appendix
\etocdepthtag.toc{app}
\crefalias{section}{appendix}
\crefalias{subsection}{appendix} % optional

\begin{center}
{\large\bf Supplement to ``Optimal Design under Interference, Homophily, and Robustness Trade-offs''} \\
% {\large\bf APPENDIX}

\end{center}

\section*{Appendix: Contents}
% \addcontentsline{toc}{section}{Appendix: Contents}

% \begingroup
% \setcounter{tocdepth}{2} % 1=sections, 2=subsections, 3=subsubsections
% \tableofcontents
% \endgroup

\begingroup
    \renewcommand{\contentsname}{} % suppress "Table of contents"
  \etocsettagdepth{main}{none}        % hide everything before appendix tag
  \etocsettagdepth{app}{subsection}   % show appendix sections/subsections
  \tableofcontents
\endgroup

\clearpage

\begin{center}
\small
\setlength{\LTleft}{0pt}
\setlength{\LTright}{0pt}
\renewcommand{\arraystretch}{1.15}

\begin{longtable}{L{0.24\textwidth}L{0.71\textwidth}}
\caption{Selected glossary of notation used throughout the paper.}
\label{tab:glossary}\\
\toprule
\textbf{Term / symbol} & \textbf{Meaning} \\
\midrule
\endfirsthead

\toprule
\textbf{Term / symbol} & \textbf{Meaning} \\
\midrule
\endhead

\bottomrule
\endfoot

\multicolumn{2}{l}{\textbf{Population and network setup}}\\

$L=D-W$ & Graph Laplacian.\\
$L^\dagger$ & Moore--Penrose pseudoinverse of the graph Laplacian.\\
$\widetilde L = I-D^{-1}W$ & Normalized graph Laplacian used in neighborhood-interference examples.\\

\addlinespace
\multicolumn{2}{l}{\textbf{Potential outcomes}}\\
$x=2z-\mathbf 1$ & Rademacher recoding of the treatment assignment, so $x\in\{-1,1\}^n$.\\
$\alpha$ & Baseline vector, defined by $\alpha=Y(0)$.\\
$\varphi$ & Treatment-effect vector, defined by $\alpha+\varphi=Y(1)$, equivalently $\varphi=Y(1)-Y(0)$.\\
$s(z)$ & Interference component in the model $Y_i(z)=\alpha_i+\varphi_i z_i+s_i(z)$; it vanishes under constant treatment.\\
$h_\alpha,h_\varphi$ & Homophilous components of $\alpha$ and $\varphi$.\\
$\varepsilon_\alpha,\varepsilon_\varphi$ & Heterogeneous-variation / residual components of $\alpha$ and $\varphi$.\\

\addlinespace
\multicolumn{2}{l}{\textbf{Homophily, interference, robustness}}\\
$\eta$ & Homophily trade-off parameter.\\
$\gamma$ & Interference trade-off parameter.\\
$\kappa$ & Robustness trade-off parameter.\\
$\Sigma_f=\mathbb E[\varepsilon_f \varepsilon_f^\top]$ & Covariance matrix of heterogeneous variation, for $f\in\{\alpha,\varphi\}$.\\
$q^\ast$ & Schatten exponent used to constrain $\Sigma_f$.\\
$q$ & Hölder-conjugate exponent to $q^\ast$, defined by $1/q^\ast+1/q=1$; it appears in the penalty $\|X\|_q$.\\
$\|\cdot\|_q$ & Schatten-$q$ norm of a matrix.\\
$\chi(\eta,\gamma,\kappa)=\chi_{q^\ast}(\eta,\gamma,\kappa)$ & Model class of potential outcomes satisfying the homophily, interference, and heterogeneous-variation constraints.\\
$\widetilde\gamma$ & Rescaling; $\gamma/\Delta$ in symmetric-case and $\gamma\beta_1^2$ in general\\
$\widetilde\eta$ & Rescaling; $5\eta/(4\Delta)$ in symmetric-case and $\eta(\beta_1^2+\beta_2^2)$ in general\\
$\widetilde\kappa$ & Rescaling; $5\kappa/(4\Delta)$ in symmetric-case and $\kappa(\beta_1^2+\beta_2^2)$ in general\\

\addlinespace
\multicolumn{2}{l}{\textbf{Estimand, estimator, MSE quantities and design covariance}}\\
$p$ & Marginal treatment probability, $p=\mathbb P(z_i=1)$, taken uniform across nodes in the main development.\\
$a$ & Average baseline level, $a=\langle \alpha,\mathbf 1/n\rangle$.\\
$b$ & Average treatment-effect level, $b=\langle \varphi,\mathbf 1/n\rangle$.\\
$\delta$ & Quantity controlling the average / centered interference contribution, via $|\langle s(z),\mathbf 1/n\rangle|\le \sqrt{\delta}$.\\
$\Delta$ & Combined scale parameter: in the symmetric case, $\Delta=(a+b/2)^2+\delta$; in general, $\Delta = (a\beta_1 + b\beta_2)^2 + \beta_1^2 \delta$\\
$X = \cov(x)$ & Design covariance matrix of the assignment.\\
$X^\ast$ & Optimal achievable covariance matrix of a design.\\
$X_{\mathrm{SDP}}$ & SDP solution matrix used in the Gaussian-rounding algorithm.\\
$\beta_1$ & Constant in the general MSE bound, $\beta_1=1/(2np(1-p))$.\\
$\beta_2$ & Constant in the general MSE bound, $\beta_2=1/(2np)$.\\

\addlinespace
\multicolumn{2}{l}{\textbf{Gaussian rounding and adapted Gram--Schmidt Walk notation}}\\
$\xi$  & Gaussian sample drawn as $\xi\sim \mathcal N(0,X^\ast)$.\\
$\zeta$ & Final rounded $\{\pm 1\}^n$ assignment returned by Gaussian rounding.\\
$\Xi=\operatorname{Cov}(\zeta)$ & Covariance matrix of the rounded assignment.\\
$\lambda$ & Randomization parameter in GSW balancing robustness against covariate balance.\\
$\omega_A$ & In the GSW analysis, $\omega_A =\max_i \|A_i\|$, i.e.\ the maximum column norm of $A$.\\
$\mu$ & Mean assignment vector, with entries $\mu_i=\mathbb E[x_i]$.\\
\end{longtable}
\end{center}
%\section{Appendix}
\section{More illustrations}\label{app:more illustrations}

\begin{figure}[h!]
    \centering
    \begin{subfigure}[b]{0.49\textwidth}
    \centering
    \includegraphics[width=\linewidth]{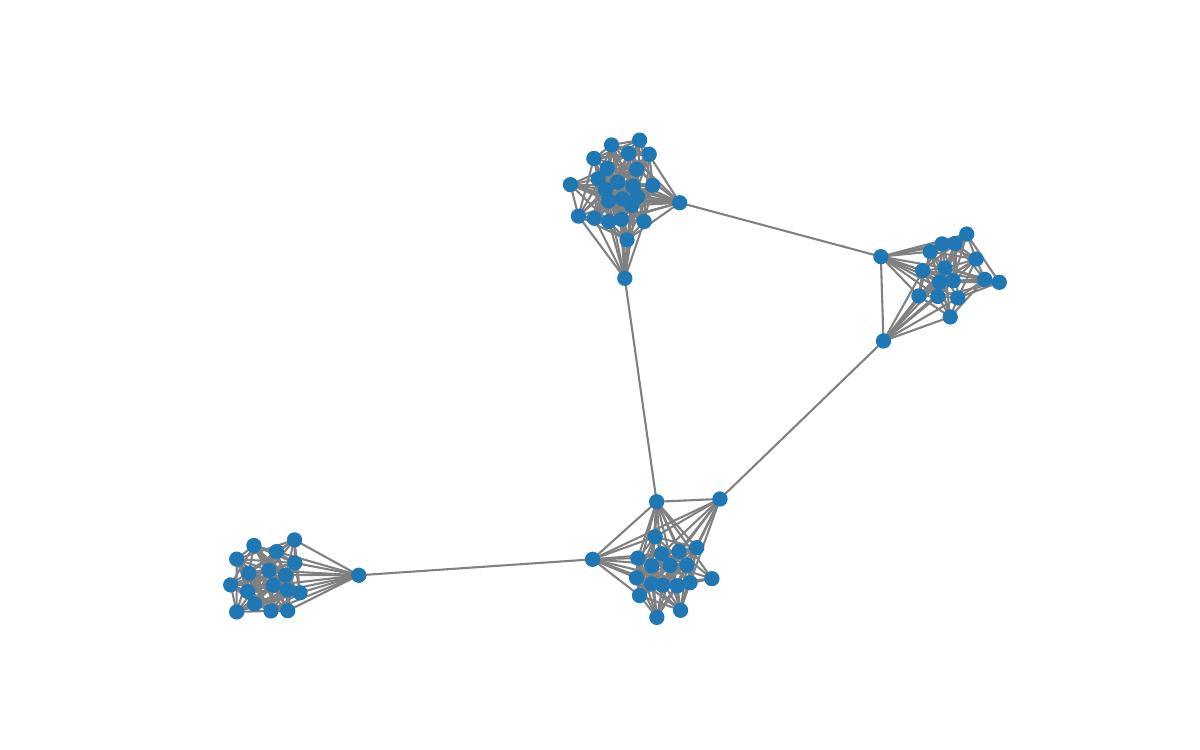}
    \end{subfigure}
    \hfill
    \begin{subfigure}[b]{0.49\textwidth}
    \centering
        \includegraphics[width=0.5\linewidth]{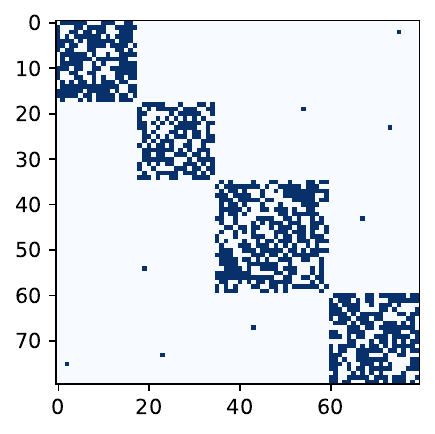}
    \end{subfigure}
    \caption{An SBM graph with four underlying clusters, and its adjacency matrix (with dark blue representing the presence of an edge, and light blue representing absence of an edge).}
    \label{fig:sbm_four_clusters}
\end{figure}

\begin{figure}[H]
    \centering
    \includegraphics[width=\linewidth]{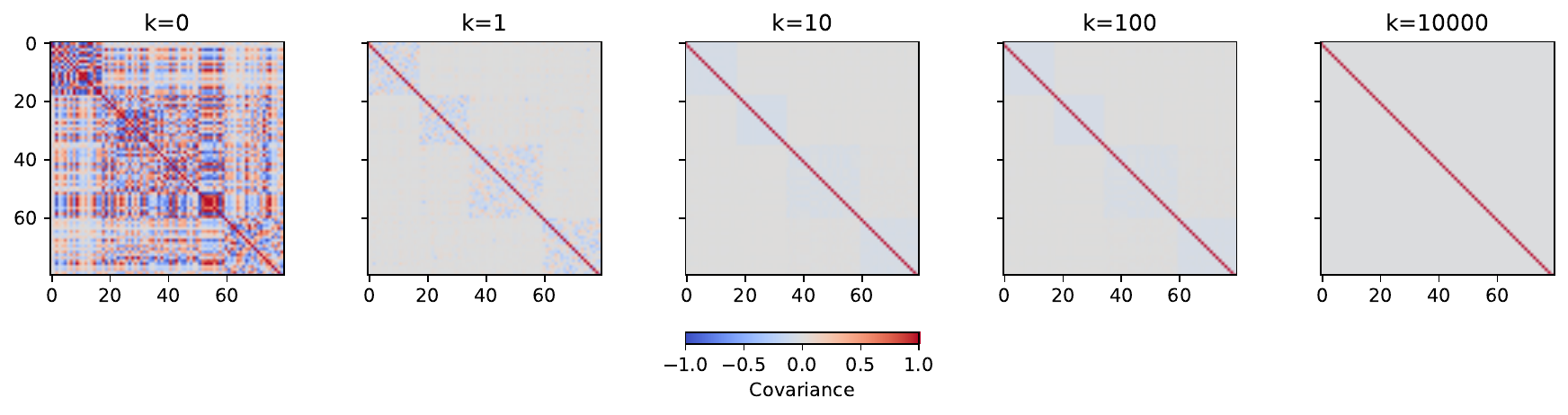}
    \caption{Solutions of the SDPs described in \Cref{section_sdp} on the graph in  \cref{fig:sbm_four_clusters} trading-off $10\operatorname{Tr}(L^{\dagger} X) + k \| X\|_{F} +10 \operatorname{Tr}(\mathbf{1}\mathbf{1}^T X)$. $k$ ranges between $(0,10^0, 10, 10^2, 10^4)$ from left to right.}
    \label{fig:cov_mat_m_k}
\end{figure}

In \Cref{fig:implementation_five_cluster_transition_h_m}, the treatment assignments are generated by Gaussian rounding. The panels vary the trade-off parameter ratio $\eta/\gamma$: Left: $\eta= 10, \gamma =100, \kappa = 1$, $q=2$; Center: $\eta= 1000, \gamma =100, \kappa = 1$, $q=2$; Right: $\eta= 10000, \gamma =10, \kappa = 1$, $q=2$.

Similarly, in \Cref{fig:implementation_five_cluster_transition_h_k}, the treatment assignments are generated by Gaussian rounding. The panels vary the trade-off parameter ratio $\kappa/\gamma$: Left: $\eta= 0, \gamma =1000, \kappa = 1$, $q=2$; Center: $\eta= 0, \gamma =1, \kappa = 1$,  $q=2$; Right: $\eta= 0, \gamma =1, \kappa = 1000$,  $q=2$.

In \Cref{fig:cov_mat_m_h}, we display solutions of the SDPs described in \Cref{section_sdp}, on the graph in \Cref{fig:sbm_four_clusters}, trading-off $10\operatorname{Tr}(L X) + h \operatorname{Tr}(L^{\dagger} X) + 10 \operatorname{Tr}(\mathbf{1}\mathbf{1}^T X)$. Here, $h$ ranges between $(0,10^0, 10, 10^2, 10^4)$ from left to right.

Similarly, in \Cref{fig:cov_mat_m_k}, we display solutions of the SDPs described in \Cref{section_sdp}, on the graph in  \Cref{fig:sbm_four_clusters}. This formulation explicitly trades off homophily and robustness by minimizing the objective $10\operatorname{Tr}(L^\dagger X) + k \| X\|_{F} + 10 \operatorname{Tr}(\mathbf{1}\mathbf{1}^T X)$. Here, $k$ ranges between $(0,10^0, 10, 10^2, 10^4)$ from left to right. 

In \Cref{fig:cov_mat_h_k}, we display solutions of the SDPs described in \Cref{section_sdp}, on the graph in  \Cref{fig:sbm_four_clusters}. This time, the formulation explicitly trades off interference and robustness by minimizing the objective $10\operatorname{Tr}(LX) + k \| X\|_{F} + 10 \operatorname{Tr}(\mathbf{1}\mathbf{1}^T X)$. Here, $k$ ranges between $(0,10^0, 10, 10^2, 10^4)$ from left to right. 

\section{Non-symmetric Case} \label{section:non_symmetric}

We now consider the non-symmetric case, i.e., when $\mathbb{P}(z_i = 1) \neq \mathbb{P}(z_i = 0)$. We assume that the treatment probability $p$ is uniform across nodes, i.e., $p = \mathbb{P}(z_i = 1) = \mathbb{P}(z_j = 1)$ for all $i,j \in [n]$. As before, our goal is to relax this problem to the minimization of a quadratic form $x^T Q x$ plus a penalization term to encourage randomization, which can then be optimized via an SDP or the adapted Gram-Schmidt Walk vector balancing approach.

\begin{theorem}[Non-symmetric MSE bound] \label{thm:mse_bound_general_HT}
Let $p:= \mathbb{P}(z_i = 1)$ for all $i \in [n]$, $\beta_1 := 1/(2np(1-p))$, $\beta_2 := 1/(2np)$, $a = \langle \alpha, \mathbf{1}/n\rangle$, $b = \langle \phi, \mathbf{1}/n\rangle$, and $\langle s(z), \mathbf{1}/n\rangle \leq \sqrt{\delta}$ for all $z$. Let also $x \in \{-1, 1\}^n$ by defining $x := 2z-1$. Define  $X:=\operatorname{Cov}(x)$. Then, for $q \geq 1$,
     \begin{align*}
          &\mathbb{E}[(\hat{\tau} -\tau)^2] \leq\\
          &\qquad 7 \{ \operatorname{Tr}((\gamma \beta_1^2 L + \eta[\beta_1^2 + \beta_2^2] L^\dagger) X) + \kappa[\beta_1^2 + \beta_2^2] \|X\|_{q} + ([a\beta_1 + b\beta_2]^2 + \beta_1^2 \delta)\operatorname{Tr}(\mathbf{1}\mathbf{1}^TX)\}.
     \end{align*}
\end{theorem}
We display the proof in \Cref{proof_mse_bound_general_HT}.

Let $\Delta = (a\beta_1 + b\beta_2)^2 + \beta_1^2 \delta$. Reparameterizing with $\Tilde{\gamma}$,$\Tilde{\eta}$, and $\Tilde{\kappa}$ so that $\Tilde{\gamma} := \frac{\gamma \beta_1^2}{\Delta}, \Tilde{\eta} := \frac{\eta[\beta_1^2 + \beta_2^2]}{\Delta}, \text{ and } \Tilde{\kappa} := \frac{\kappa [\beta_1^2 + \beta_2^2]}{\Delta}$,
we can write an SDP more explicitly in terms of the new trade-off parameters:
\begin{align} \label{sdp_relaxation_mse_p_norm}
    \min_{X\in\mathbb{R}^{n\times n}} \quad & \Tilde{\eta}\text{Tr}(L^\dagger X)  +   \Tilde{\gamma} \text{Tr}(L X) + \Tilde{\kappa} \|X\|_{q} + \operatorname{Tr}(\mathbf{1}\mathbf{1}^TX) \\
    \text{s.t.} \quad & X \succeq  0 \notag,\\
    & \text{diag}(X) = \mathbf{1}. \notag
\end{align}

\begin{figure}[H]
    \centering
    \includegraphics[width=\linewidth]{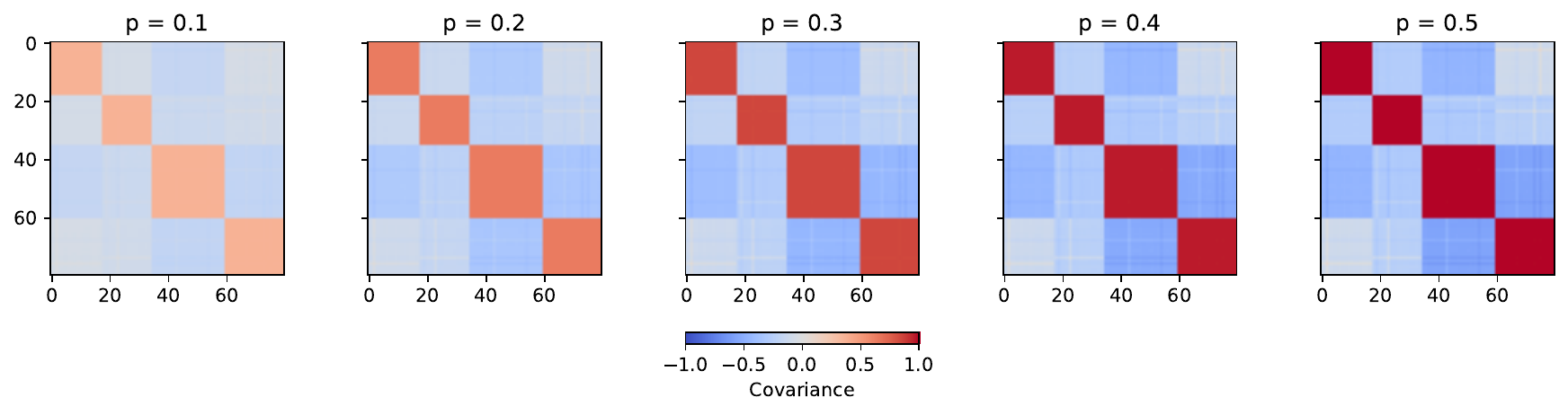}
    \caption{SDP solutions on the graph from \cref{fig:sbm_four_clusters}, with trade-off parameters $\gamma = 10,\eta= 1, \kappa= 1$, and $a= b = 1$, across varying treatment assignment probabilities $p$. Here $p$ ranges between $(0.1, 0.2, 0.3, 0.4, 0.5)$ from left to right. The probabilities $p$ scale each trade-off component of the SDP via $\beta_1$ and $\beta_2$, and thus the SDP solutions exhibit only slight variation across values of $p$.}
    \label{fig:pre_round_cov_mat_gamma_k}
\end{figure}

\begin{figure}[H]
    \centering
    \includegraphics[width=\linewidth]{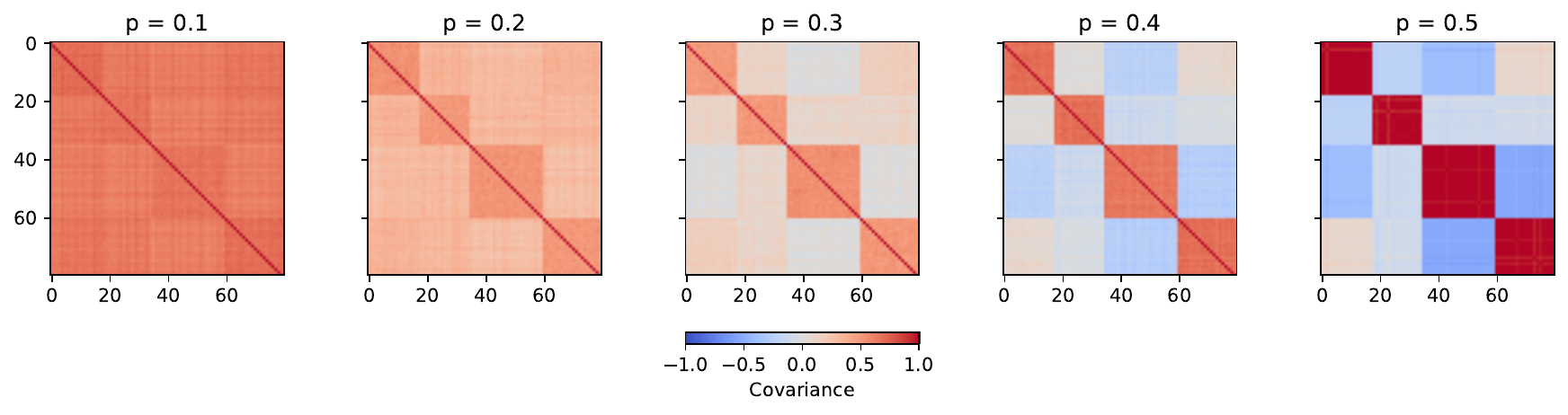}
    \caption{Covariance matrices of our designs by applying \Cref{alg:gaussian_round} to the SDP solutions in \Cref{fig:pre_round_cov_mat_gamma_k}, across treatment assignment probabilities $p$ ranging between $(0.1,0.2, 0.3, 0.4, 0.5)$ from left to right.}
    \label{fig:round_cov_mat_gamma_k}
\end{figure}

\Cref{fig:round_cov_mat_gamma_k} presents the covariance matrices of the design across different treatment assignment probabilities $p$, generated from applying \Cref{alg:gaussian_round} to the corresponding SDP solutions in \Cref{fig:pre_round_cov_mat_gamma_k}. 

In the adapted Gram-Schmidt Walk, we can take $A:= (\Tilde{\eta} L^{\dagger} + \Tilde{\gamma} L + \mathbf{1} \mathbf{1}^T)^{1/2}$. Then, we can take the robustness parameter to be $\lambda = \frac{\sqrt{\Tilde{\kappa} n^{1/q}}}{ \sqrt{\omega_A^2n} + \sqrt{\Tilde{\kappa} n^{1/q}}}$, where $\omega_A := \max_{i\in n} \|A_i\|.$
Alternatively, one could append the vectors so $A_i^T = [ \Tilde{\eta}^{1/2} L^{\dagger/2}_i \quad \Tilde{\gamma}^{1/2} L^{1/2}_i \quad {\mathbf{1} \mathbf{1}^T}^{1/2}_i].$

\section{More examples}

\subsection{Homophily}

\begin{figure}[!h]
    \centering
    \includegraphics[width=0.4\linewidth]{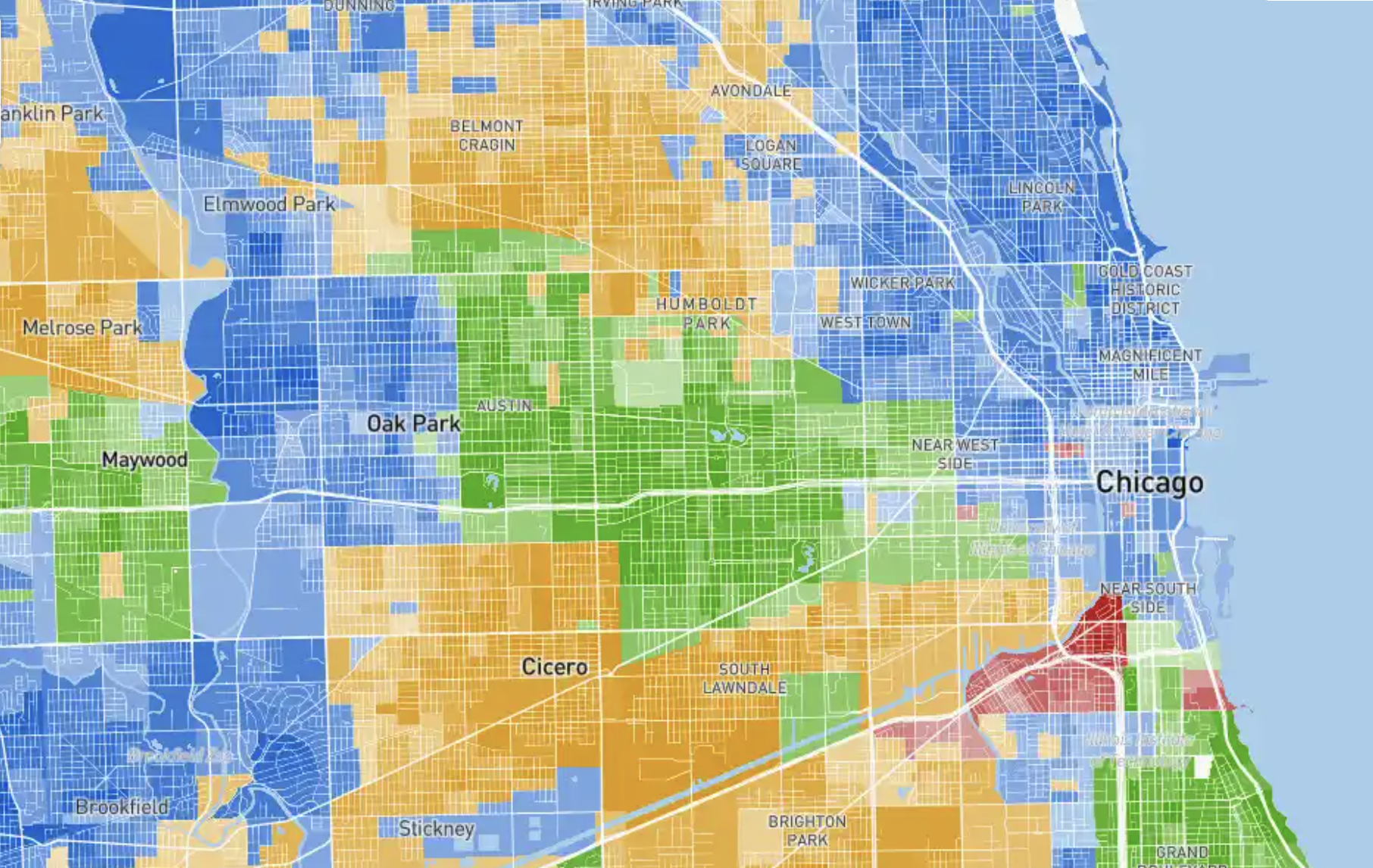}
\caption{A real example of homophily. The map from \cite{bestneighborhood} depicts majority race by geographic area in the Chicago region, based on self-identification on the US census. Darker shades indicate a larger racial majority. This spatial map can be modeled as a network, where nodes represent geographic areas and edge weights are inversely proportional to the distance between them. This network exhibits a high level of homophily, as racial composition varies smoothly across neighboring areas in the graph.}
    \label{fig:homophily_example}
\end{figure}

\Cref{fig:homophily_example} depicts a concrete real example of homophily.

\begin{example}
    One natural interpretation of homophily is via the latent space framework \citep{athreya2018statistical, hoff2002latent}, where we identify a homophilous vector $h$ with a smooth function of common latent variables that determine the graph weights $W_{ij}$.
    As a concrete example, each node may be associated to a latent vector $u_i$, and for $U:= [u_1, \dots, u_n]^T$, there is some $\beta \in \mathbb{R}^d$ such that $h =  U \beta$, and
    $W_{ij} \propto \exp(-\|u_i - u_j\|^2).$
\end{example}

\begin{remark} \label{remark_hom_gen}
    The quadratic form $g^T L g = \langle g, Lg\rangle_{\ell_2}$ directly quantifies the smoothness of the function $g$ with respect to the graph structure. Its continuous analogue, replacing the discrete graph with a manifold or a domain equipped with a measure $\mu$, is the quadratic form $\langle g, \Delta_\mu g\rangle_{L^2(\mu)} = \int g \Delta_{\mu} g\, d\mu = \int \|\nabla g\|^2\, d\mu$, where $\Delta_{\mu}$ is the Laplacian operator with respect to $\mu$, and $\nabla g$ is the gradient of a sufficiently smooth function $g$ supported on $\mu$. 
\end{remark}

\begin{remark} \label{remark_homophily_lapsym}

To define homophily with respect to similarity in node degrees $d_i$, a natural approach is to use the symmetric normalized Laplacian $L_{\text{sym}} := I - D^{-1/2}WD^{-1/2}$ and say that the vector $h$ is $(\eta, L_{\text{sym}})$-homophilous if $h^TL_{\text{sym}}h \leq \eta$. 
This formulation incorporates an appropriate normalization of the edge weights by the square roots of the degrees of the corresponding nodes. Explicitly, $h^TL_{\text{sym}}h = \sum_{(i,j) \in E} W_{ij} \left( \frac{h_i}{\sqrt{d_i}} - \frac{h_j}{\sqrt{d_j}}\right)^2 \leq \eta$ captures the smoothness of $h$ when degree similarity is taken into account. 
\end{remark}

\begin{remark} \label{remark_homophily_clustering}
Our notion of homophily is consistent with the classical view that homophily is associated with cluster formation in networks \citep{mcpherson2001birds, currarini2009economic,jackson2023dynamics, jackson2025inequality}. In our setting, this is captured by the Rayleigh quotient of the graph Laplacian $h^T L h/\|h\|^2$, the ratio $\eta'/\kappa'$ of the homophily and heterogeneity of a centered vector $h$. When this ratio is small, the potential outcomes $h$ have little discrepancy between neighbors in the graph compared to the variation 
of the vector. Equivalently, the variations  in $h$ are primarily inter-cluster rather than intra-cluster, and this can be characterized spectrally as $h$ lying predominantly in the span of the eigenvectors associated with the smallest eigenvalues of the graph Laplacian. This is most insightful in our decomposition of $\alpha$ and $\phi$ into homophilous and heterogeneous components $h_f$ and $\varepsilon_f$, where $f \in \{\alpha, \phi \}$, respectively. As will become apparent in bounds such as \Cref{thm:mse_bound_general_HT_symmetric}, this decomposition will be particularly effective numerically when $h_f$ is selected so that the corresponding ratio $\eta'/\kappa'$ is minimal, and vice versa for $\varepsilon_f$, $f \in \{\alpha, \phi\}$, thus aligning these two notions of homophily. In other words, if the outcomes associated with all the units in the population are highly heterogeneous and not driven by the underlying network structure, then a larger share of the variation in the outcomes can be attributed to $\varepsilon$ (and hence $\kappa$) rather than $h$ (and hence $\eta$).
\end{remark}

\subsection{Interference}

\begin{figure}
    \centering
    \begin{subfigure}[b]{0.45\linewidth}
    \centering
    \includegraphics[width=\linewidth, height = 0.75\linewidth]{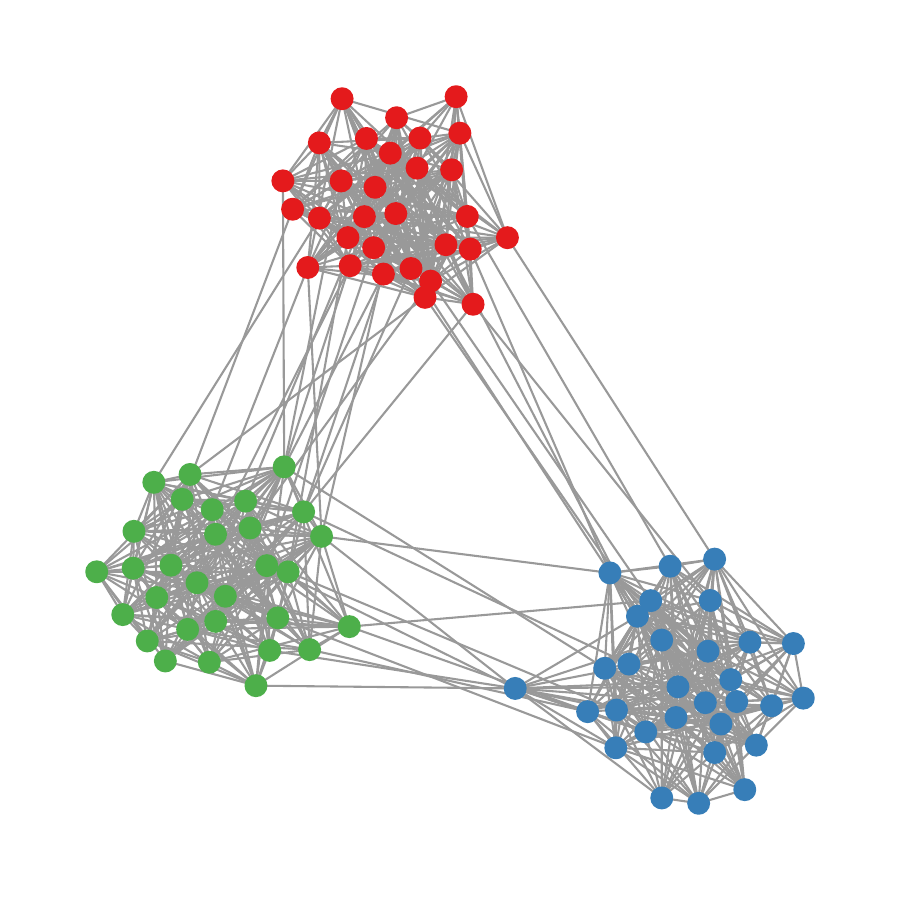}
    \end{subfigure}
    \hfill
    \begin{subfigure}[b]{0.45\linewidth}
    \centering
    \includegraphics[width=\linewidth, height = 0.75\linewidth]{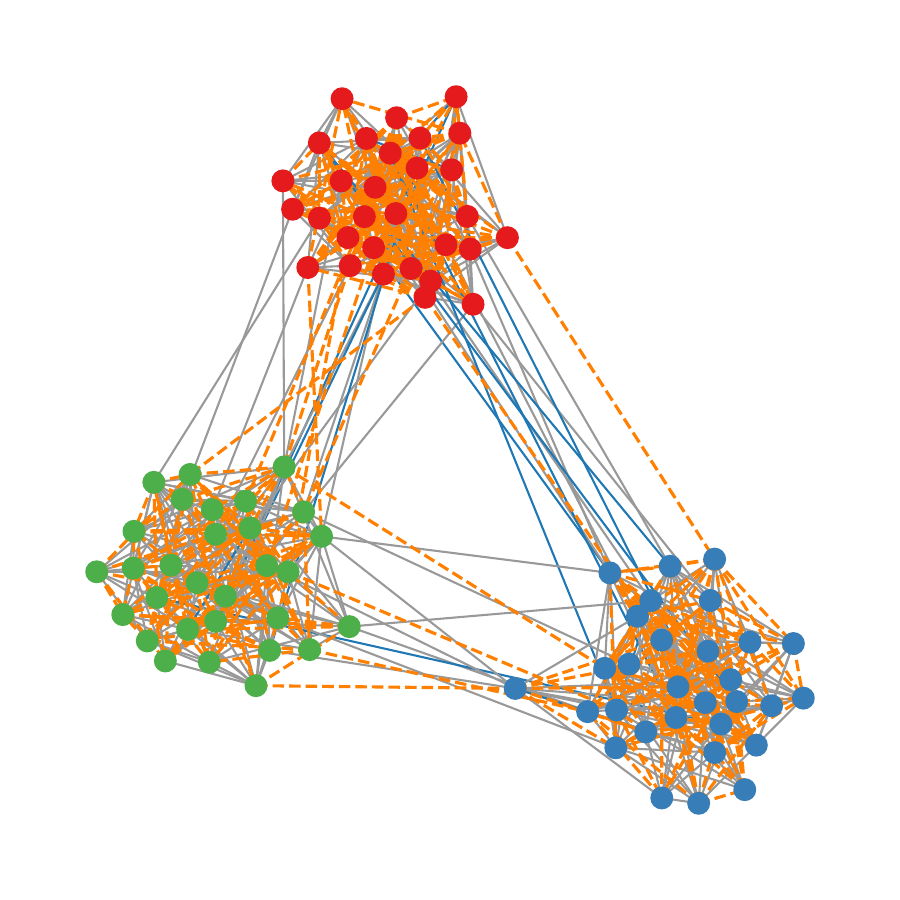}
    \end{subfigure}
    \caption{Example of noisy network observation. Left panel: True network, $G$. Right panel: Noisy observation $G_{\text{obs}}$ of the network with grey edges preserved and newly added dark blue edges. The orange edges depict missing edges.}
    \label{fig:noisy_netw_obs}
\end{figure}

\begin{example}[Neighborhood Interference] \label{neighborhood_interference_full}
A simple outcome model with a linear neighborhood interference term with a ($\gamma, \Tilde{L}$)-interferent $s$ component paired with a $\sqrt{n}$-scaling reminiscent of the local asymptotic framework \citep{viviano2023causal, hirano2009asymptotics} is
$Y_i = \alpha_i + \beta_i Z_i + \frac{\gamma}{\sqrt{n}} \sum_{j} \frac{W_{ij}}{\sum_j W_{ij}} z_j$.
Indeed, the equation above can be rewritten as
$Y_i = \alpha_i + \phi_i z_i -  \frac{\gamma}{\sqrt{n}} e_i^T (I- D^{-1}W) z$,
where $\phi_i := \beta_i + \gamma/\sqrt{n}$, 
%and $L' := I- D^{-1}W$,
$D = \text{diag}(d)$ for the vector $d\in\mathbb{R}^n$ of degrees of the nodes $d_i = \sum_{j \in N(i)} W_{ij}$, $W \in \mathbb{R}^{n \times n}$ is the matrix of the edge weights $\{W_{ij}\}_{i,j}$, and $\alpha, \phi$ are vectors with entries $\alpha_i, \phi_i$, respectively, for $i=1,2,...,n$. Let the normalized graph Laplacian $\Tilde{L} := I-D^{-1}W$, i.e., $\Tilde{L} = D^{-1}L$, and $\Tilde{L}^\dagger$ to be the pseudo-inverse of the normalized Laplacian. In matrix-vector form, 
$Y = \alpha + \Diag(\phi) z -  \frac{\gamma}{\sqrt{n}} \Tilde{L} $z. Here, $s(z) = -  \frac{\gamma}{\sqrt{n}} \Tilde{L} z$, i.e., $s$ is $(\gamma, \Tilde{L})$-interferent. The normalized Laplacian term appropriately vanishes when the full population is treated or assigned to control. 
\end{example}

\begin{figure}
    \centering
    \begin{subfigure}[b]{0.45\linewidth}
    \centering
    \includegraphics[width=\linewidth, height = 0.75\linewidth]{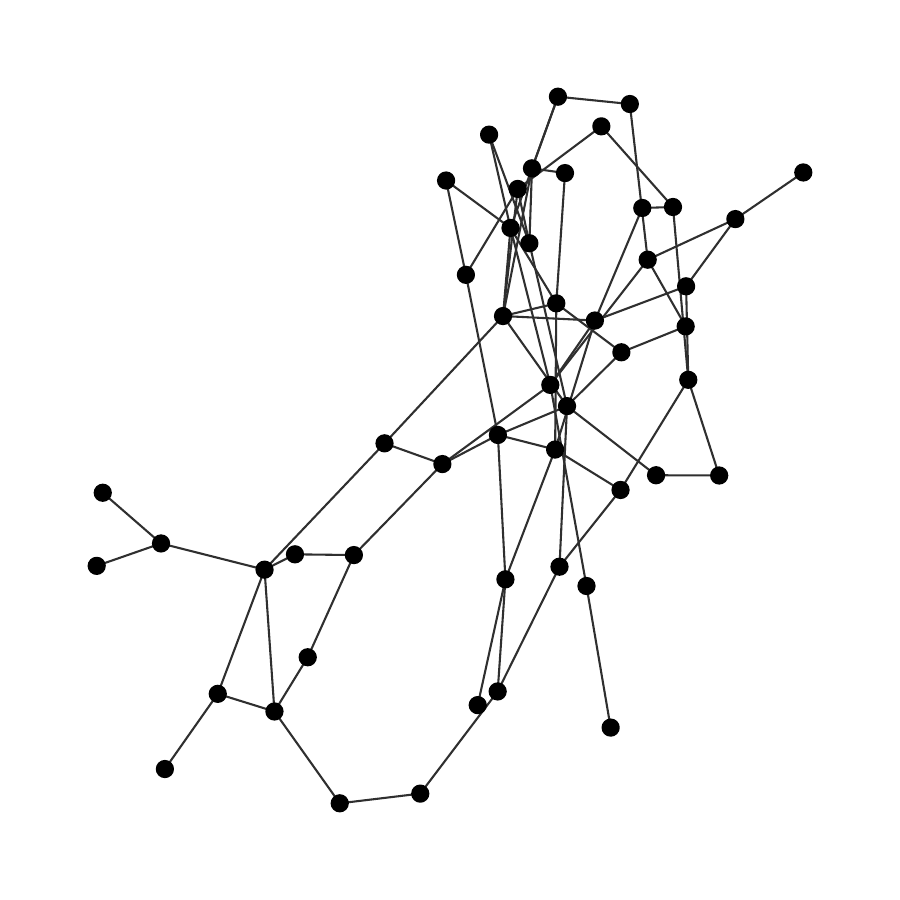}
    \end{subfigure}
    \hfill
    \begin{subfigure}[b]{0.45\linewidth}
    \centering
    \includegraphics[width=\linewidth, height = 0.75\linewidth]{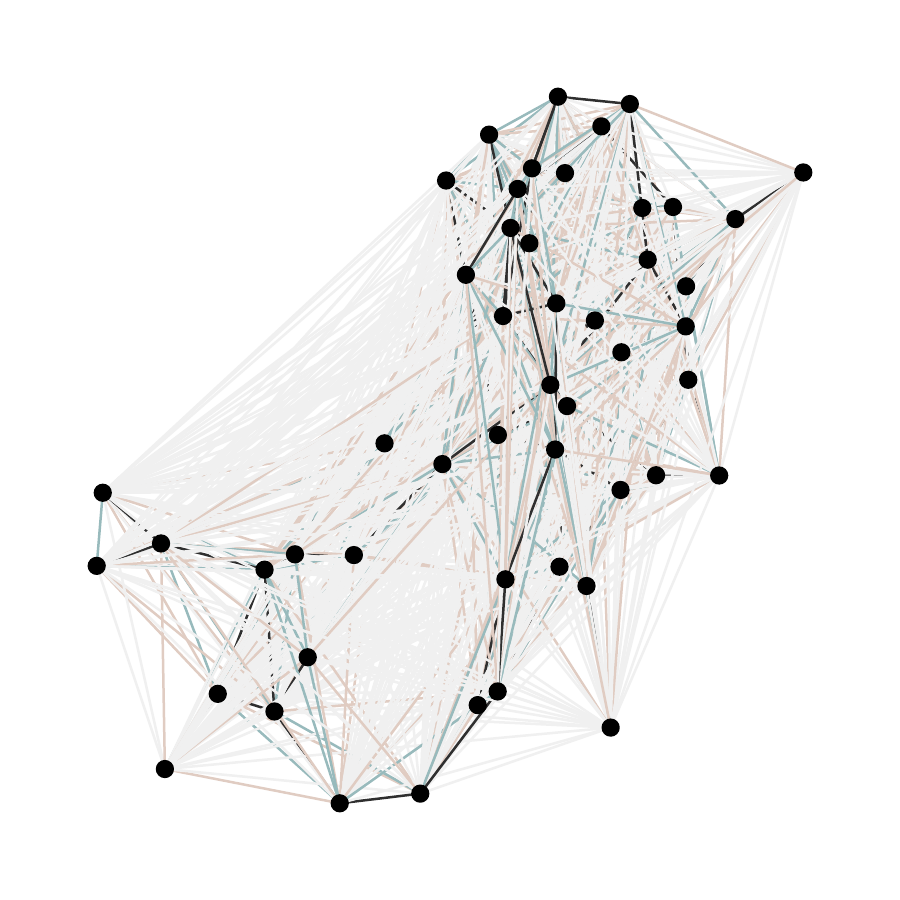}
    \end{subfigure}
    \caption{Full interference example. Left panel: First-order (neighborhood) interference, where outcomes depend only on treatments of immediate neighbors. Right panel: Full network interference induced by the same underlying graph, allowing interference to propagate along paths of increasing length, with strength decaying in path length. Darker edges indicate stronger (shorter-path) interference effects.}
    \label{fig:full_interference_example}
\end{figure}

\begin{example}[Full Network Interference] \label{example:noisy_interference_full}
Take for example an instance of the weak-dependence model \citep{leung2022causal} adapting $\upalpha$-mixing spatial processes (see, for example, \citet{jenish2009central}) to path lengths $\rho_n$ in the network, where $\sum_{m=2}^n M_m^\partial \Tilde{\theta}_{m} = o(\psi_1)$, for some $\psi_1 \ll 1$, $M_m^\partial := \frac{1}{n} \sum_{i=1}^n |\calN^\partial(i,m)|$, $\calN^\partial(i,m) :=\{j \in [n]: \rho_n(i,j) = m\}$, and correlation between outcomes $\Tilde{\theta}_m \to 0$, as $m \to \infty$. That is, the correlation $\Tilde{\theta}$ between outcomes of units decay over (path length) distances $\rho_n$. Here $\Tilde{L}$ can be taken to be the graph Laplacian of a complete graph, modeling a full interference structure, with edge weights $\Tilde{\theta}$ corresponding to the correlation under path lengths in the original path length weak-dependence structure (see, for example, \Cref{fig:full_interference_example}). Under the setting with such a $\Tilde{\theta}$ decaying quickly, we have that the full degree matrix $\Tilde{D} = D + \text{diag}(\delta)$, and full adjacency matrix $\Tilde{W} = W + \epsilon$, where $\delta_{i} = \sum_{j \in [n]} \epsilon_{ij}$, for all $i= 1, 2, \ldots, n.$ Then, writing $L_{\text{res}} := \text{diag}(\delta) - \epsilon$, $\Tilde{L} = L + L_{\text{res}}$. The spectrum of $L_{\text{res}}$ are upper-bounded by some $\psi_2 \ll 1$, such that $\|\Tilde{L}^{\dagger} L \|_{\text{op}} \leq 1 + O(\psi)$, for $\psi \ll 1$. Then, our worst-case MSE bound would consist of the perturbed bound, $\gamma(1 + O(\psi))$ instead of just $\gamma$.
\end{example}

\begin{example}
We note the distinction between our model of interference and a generalized version of the previous example when the interference term $s(z)$ is $ \frac{\bm{\upgamma}}{\sqrt{n}} \Tilde{L} z$, where $\bm{\upgamma}_i, \bm{\upgamma}_j$ are not necessarily equal for any $(i,j)$ pair. A somewhat similar interference structure encapsulated by our model is one where
$s(z) = L^{1/2} \Diag(\bm{\upgamma} / \sqrt{2 d_{\text{max}} n}) L^{1/2} z$, where $d_{\text{max}}$ is the maximum degree of the nodes in the graph. Then, if $\|\bm{\upgamma}\|_\infty^2 \leq \gamma$, indeed we have
\begin{align*}
    s(z)^T L^\dagger s(z) &= z^T L^{1/2} \Diag(\bm{\upgamma} / \sqrt{n}) L^{1/2} L^\dagger L^{1/2} \Diag(\bm{\upgamma} / \sqrt{n}) L^{1/2} z \\
    &= z^T L^{1/2} \Diag(\bm{\upgamma} / \sqrt{2 d_{\text{max}} n}) \mathcal{P} \Diag(\bm{\upgamma} / \sqrt{2 d_{\text{max}} n}) L^{1/2} z \\
    &\leq \frac{z^T L z}{2 d_{\text{max}} n} \| \bm{\upgamma} \|_{\infty}^2 \leq \gamma,
\end{align*} where $\mathcal{P}$ is the orthogonal projection matrix onto the column space of $L$. Thus, $s$ is $(\gamma, L)$-interferent. 
\end{example}

\section{More on the experimental designs}

\subsection{Gaussian Rounding on the SDP solutions}

\Cref{alg:gaussian_round} converts the SDP solution \(X_{\text{SDP}}\) into a treatment assignment by first sampling a Gaussian vector with covariance $X_{\text{SDP}}$. The entries of this vector are then rounded coordinatewise to obtain a binary assignment vector. In the symmetric case \(p=1/2\), this reduces to sign rounding; more generally, the rounding step is calibrated so that each unit has marginal treatment probability \(p\). Thus, Gaussian rounding provides a practical way to pass from the continuous SDP relaxation to an implementable randomized binary design while retaining much of the covariance structure selected by the SDP.

\subsection{Adapted Gram-Schmidt Walk algorithm}
We include the Gram-Schmidt Walk algorithm adapted to our trade-off problem in \Cref{alg:gramschmidt}. \Cref{alg:gramschmidt} applies the Gram-Schmidt Walk to the augmented covariate matrix \(B\), constructed from a factor \(A\) satisfying
\[
A^\top A = \widetilde{\eta}L^\dagger+\widetilde{\gamma}L+\mathbf{1}\mathbf{1}^\top,
\]
together with scaled standard basis vectors used to incorporate the robustness term. The algorithm is initialized at \(z = 2p-1\), ensuring the target marginal treatment probabilities $p$ are satisfied. At each iteration, it computes an update direction that seeks to preserve balance in the span of the active columns of \(B\), and then moves along this direction until one or more active coordinates hit \(\pm 1\). These coordinates are subsequently fixed, and the procedure repeats on the remaining active coordinates. When the algorithm terminates, all entries lie in \(\{\pm 1\}\), yielding a treatment assignment that balances the induced covariates while preserving the desired level of randomization.

\begin{algorithm}[!h]
    \caption{(Adapted) Gram-Schmidt Walk (adaptation of \citet{harshaw2024balancing}; see also \Cref{covariate_balancing_comparisons})}
    \label{alg:gramschmidt}
    \begin{algorithmic}[1] % [1] adds line numbers
        \Require $q, \eta, \gamma, \kappa, \mathbf{p}$
        \State Define $A_i$ as the $i$-th column of $\Lambda^{1/2} V^\top$, where $V \Lambda V^\top = \Tilde{\eta} L^\dagger + \Tilde{\gamma}L + \mathbf{1}\mathbf{1}^\top$
        \Statex \hfill (Alternatively, use the appended matrix $A$; see discussion above)
        \State Define $\omega_A \gets \max_{i \in [n]} \|A_i\|$
        \State Set $\lambda \gets \frac{\kappa n^{1/{q}}} {\eta\omega_A^2 n + \kappa n^{1/q}}$ \label{line:adapt}
        \State Define $\mat{B}_i \gets \gscv{i} \gets 
            \begin{bmatrix}
                \sqrt{\lambda} \, e_i \\
                \omega_A^{-1} \sqrt{1 - \lambda} \, A_i
            \end{bmatrix}$
        \State Initialize iteration counter $t \gets 1$
        \State Initialize assignment vector $\gsv{1} \gets 2\mathbf{p} -1$ \label{line:initialize}
        \State Select pivot index $i_{\text{piv}} \sim \mathrm{Uniform}([n])$
        
        \While{$\gsvt \notin \{\pm 1\}^n$}
            \State Define active set $\alive \gets \{ i \in [n] : |\gsve{t}{i}| < 1 \}$
            \If{$i_{\text{piv}} \notin \alive$}
                \State Resample pivot $i_{\text{piv}} \sim \mathrm{Uniform}(\alive)$
            \EndIf
            \State Compute step direction:
            \[
            \gsuvt \gets \argmin_{\gsuv{} \in \mathbb{R}^n} \| \gsM \gsuv{} \|^2 \quad
            \text{s.t.} \quad \gsun{i_{\text{piv}}} = 1,\ \gsun{i} = 0\ \forall i \notin \alive
            \]
            \State Define feasible step sizes:
            \[
                \Delta \gets \{ \gsdn \in \mathbb{R} : \gsvt + \gsdn \gsuvt \in [-1, 1]^n \}
            \]
            \State Set $\gsdnp \gets |\max \Delta|$, $\gsdnm \gets |\min \Delta|$
            \State Randomly sample step size:
            \[
            \gsdt \gets 
            \begin{cases}
                \phantom{-}\gsdnp & \text{with probability } \frac{\gsdnm}{\gsdnp + \gsdnm} \\
                -\gsdnm & \text{with probability } \frac{\gsdnp}{\gsdnp + \gsdnm}
            \end{cases}
            \]
            \State Update assignment: $\gsv{t+1} \gets \gsvt + \gsdt \gsuvt$
            \State Increment $t \gets t + 1$
        \EndWhile
        
        \State \Return Final assignment vector $\gsvt \in \{\pm 1\}^n$
    \end{algorithmic}
\end{algorithm}

\section{Other related works} \label{other_related_work}

\citet{bhat2020near} study treatment-effect estimation and show that maximizing precision is related to a covariate balancing objective. They derive an SDP relaxation for precision maximization and apply the Goemans–Williamson rounding scheme, as in the MAXCUT approximation algorithm \citep{goemans1995improved}.
Our quadratic-form characterization of homophily/covariate-balancing terms provides a bridge between this and vector balancing approaches such as \citet{harshaw2024balancing}, with natural connections to our \Cref{alg:gaussian_round,alg:gramschmidt}.

A common design approach with the goal of balancing covariates within a randomization scheme is re-randomization \citep{morgan2012rerandomization,li2018asymptotic}. While we do not directly compare our methods to re-randomization procedures, \citet{harshaw2024balancing} show superior trade-offs of their method, which ours resemble under reparameterization, in comparison to re-randomization. \citet{basse2018model} study the re-randomization approach under homophily settings alone. \citet{ugander2023randomized} introduce randomized graph cluster randomization (RGCR) to address exponentially small exposure probabilities arising in graph cluster randomization designs \citep{ugander2013graph}. They show that RGCR still exhibits large variance when the nodes in a ring graph are clustered into a small number of large randomization units under a potential outcomes model with a homophily drift, a phenomenon also observed more generally in simulations.
Related observations on the tension between homophily and cluster randomization appear in \citet{ugander2012network, cortez2024analysis}.

Another strategy is a matched-pair design, where randomization is restricted to pre-specified pairs of similar units. \citet{fatemi2020minimizing} propose CMatch, a two-stage procedure that clusters units via random walks on the graph and then matches clusters using a similarity measure to assign treatments. Our framework does not rely on local assumptions about inherent similarities between pairs of randomization units. We introduce a global optimization approach to address homophily, interference, and robustness simultaneously.

\citet{jagadeesan2020designs} study estimation of direct effects by constructing treatment assignments that balance interference-related features between treated and control units. They introduce perfect quasi-colorings, which eliminate interference bias by matching units with identical treated and control neighborhood structure, and analyze the bias and variance of restricted randomization schemes that approximate such colorings when they do not exist. Their design can be viewed as stratified randomization based on node similarity, and they characterize the resulting MSE under various settings, including models with homophily and network interference, showing improvements over Bernoulli and complete randomization in some regimes. 

Optimization-based designs in this setting include the approach of
\citet{fatemi2023network}, where the authors formulate a mixed-integer program to identify maximally independent node sets for direct effect estimation. In a different application, \citet{doudchenko2021synthetic} use a mixed-integer program to jointly optimize design and weights in synthetic control settings. 
Solving mixed-integer programs to optimality is not polynomial-time and does not scale; solving an SDP, as in our method, is polynomial-time and admits efficient solvers that can be sped up by exploiting problem structure.

Bipartite networks arising from preference-based matching processes, such as those studied by \citet{gale1962college}, naturally induce homophily. Our framework therefore extends to bipartite settings, including marketplaces, generalizing existing bias–variance and graph-cut characterizations \citep{brennan2022cluster, holtz2025reducing, pouget2018optimizing, harshaw2023design}.

\section{Generalizations and a special case} \label{generalization}

\subsection{General Similarity Kernels and Stochastic Misspecification}
\label{general_misspecifications_sim_kernels}

We use the term ``homophily'' in a broad sense. Although it technically refers to similarity in social ties, we adopt it here to encompass similarities that may arise in contexts beyond social networks. 
If the practitioner, for instance, were to include some other likeness measure such as spatial positioning or geography, then a different similarity kernel, rather than or in addition to the Laplacian $L$, can be used. For example, under a similarity kernel $\mathcal{K}$, we can write our homophily formulation as $g^T \mathcal{K} g \leq \eta$. That is, $g$ is $(\eta, \mathcal{K})$-homophilous.

This leads to a more general potential outcomes model,
\begin{align} \label{general_interference_potential_outcome}
    Y  = \sum_{i=1}^{N_\alpha} h_\alpha^i + \sum_{i=1}^{N_\alpha'} \varepsilon_\alpha^i + z \cdot \left( \sum_{i=1}^{N_{\phi}} \text{diag}(h_\phi^i) + \sum_{i=1}^{N_{\phi}'} \text{diag}(\varepsilon_\phi^i)\right),
\end{align} 
where $h_f^i$, $f \in \{\alpha, \phi\}$, satisfy quadratic constraints ${h_f^i}^T\mathcal{K}_f^i h_f^i \leq \eta_f^i$, $\mathcal{K}_f^i$ a kernel/psd matrix,
and the misspecifications $\varepsilon_f^i$, $f \in \{\alpha, \phi\}$, satisfy Schatten-norm constraints $\|\Sigma_f^i\|_{{q_f^i}^\star} := \|\mathbb{E}[\varepsilon_f^i {\varepsilon_f^i}^T\|_{{q_f^i}^\star} \leq \kappa_f^i$.
Defining $\bm{\eta} := [\eta_\alpha, \eta_\phi],\ \bm{\kappa} := [\kappa_\alpha, \kappa_\phi]$ for $\eta_f:=(\eta_f^i)_{i=1}^{N_f}$, $\kappa_f:= (\kappa_f^i)_{i=1}^{N_f'}$, where $f \in \{ \alpha, \phi\}$, and $N:=N_\alpha + N_\phi + N_\alpha' + N_\phi'$, this is summarized by the constraint set
\[
\chi(\bm{\eta}, \bm{\kappa}):= \{h_f^i, \varepsilon_f^i:\  {h_f^i}^T\mathcal{K}_f^i h_f^i \leq \eta_f^i,\ \|\Sigma_f^i\|_{{q_f^i}^*}\leq \kappa_f^i\}.
\]

Our framework extends to the following bound.
\begin{theorem}(General MSE)
\label{thm:mse_bound_general_HT_network_kernels}
Let $p:= \mathbb{P}(z_i = 1)$ for all $i \in [n]$, $a = \langle \alpha, \mathbf{1}/n\rangle$, $b = \langle \phi, \mathbf{1}/n\rangle$, and $\beta_1 := 1/(2np(1-p))$, $\beta_2 := 1/(2np)$. Let also $x \in \{-1, 1\}^n$ by defining $x := 2z-1$, $X := \operatorname{Cov}(x)$. 
Then, for $q_f^i \geq 1$, $i \in [N_f']$,
\begin{align*}
          &\sup_{\chi(\bm{\eta}, \bm{\gamma}, \bm{\kappa})} \mathbb{E}[(\hat{\tau} - \tau)^2] \leq  (N+ 1) \{ (a \beta_1 + b \beta_2)^2 \operatorname{Tr}(\mathbf{1}\mathbf{1}^T X) + \sum_{f\in \{\alpha,\phi\}} \beta_f^2\sum_{i=1}^{N_f} \eta_f^i  \operatorname{Tr}((\mathcal{K}_f^{i})^\dagger X) + \sum_{i=1}^{N_f'} \kappa_f^i \|X\|_{q_f^i} \}.
\end{align*}
\end{theorem}

Before exploring the utility of this framework, let's first see how this corresponds to the basic structure established previously in the paper.
\begin{example}
    If we set $\calK_\alpha^1 = \calK_\phi^1 = L^\dagger$, $\calK_\alpha^2= L$, $\eta_\alpha = [\eta, \gamma]$, $\eta_\phi = \eta$, $q_\alpha = q_\phi = q$, and $\kappa_\alpha = \kappa_\phi = \kappa$, then we arrive at the constraint set $\chi(\eta,\gamma, \kappa)$. Indeed, this gives us a potential outcome model
    \[
    Y = h_\alpha^1 + h_\alpha^2 + \varepsilon_\alpha^1 + z \cdot \left(h_\phi^1 + \varepsilon_\phi^1\right) = (h_\alpha^1 + \varepsilon_\alpha^1) + z \cdot \left(h_\phi^1 + \varepsilon_\phi^1\right) + h_\alpha^2 =: \alpha + z\cdot \phi + s(z).
    \]
    We note that $s(z)=h_\alpha^2$ may or may not explicitly depend on the assignment $z$ in this framework, as we only specify that it satisfies a quadratic constraint. Adversarially selecting this term to maximize the MSE bound results in a typical graph interference setting.
\end{example}
Our extended framework provides a convenient means to incorporate information from other sources besides the graph and correlation motivated objectives discussed earlier. 

\begin{example}
    Of key importance in practice is the incorporation of covariates into the potential outcome model beyond the graph adjacency. In its most basic form, this is a covariate balancing problem, where one would like to penalize the squared discrepancy $x^T CC^T x$, for $C$ the design matrix/matrix of additional covariates. In our framework, this corresponds to a potential outcome component satisfying a quadratic constraint with respect to $(CC^T)^{-1}$, with the typical example being a linear model $C \beta$. More general potential outcomes can be captured by kernel gram matrices $\calK := [k(c_i, c_j)]_{i, j=1}^n$, $c_i, c_j$ the covariates of units $i$ and $j$. This provides particular utility for designs on disconnected graphs, which we discuss in more detail in \Cref{sec:graph comps}.
\end{example}

Another point of theoretical importance is the disaggregation of the heterogeneous variation components in the model.

\begin{example}
    For the majority of the paper, we streamlined our presentation by presenting different heterogeneous variation, such as deterministic residuals and stochastic network effects, as mutually exclusive modeling choices made by the practitioner. 
    This is of course unnecessary, as our general MSE demonstrates, as both of these terms can be separately included. 
    This is particularly important for large graphs, as the Schatten-$q$, $q < \infty$, norm of $X$ may grow rapidly with the sample size. Thus, it may be desirable to assume a constant level of deterministic residuals, only penalizing the operator norm of $X$, and a vanishing stochastic network misspecification. In other words, as the graph grows, we expect the homophily and interference terms to capture the majority of the correlation structure.
\end{example}

\subsection{Disconnected Components in the Graph}\label{sec:graph comps}
We have assumed, so far, that the graph is connected. In the following, we consider the worst-case bounds if the graph had $T$ connected components instead.

Let $\Pi$ be the projection to the null-space of the Laplacian matrix. 

\begin{theorem}[Disconnected MSE bound] \label{thm:mse_bound_general_projection}
Let $p:= \mathbb{P}(z_i = 1)$ for all $i \in [n]$, $\beta_1 := 1/(2np(1-p))$, $\beta_2 := 1/(2np)$. Let also $x \in \{-1, 1\}^n$ by defining $x := 2z-1$, and $X:= \operatorname{Cov}(X)$. Then, 
     \begin{align*}
          \mathbb{E}[(\hat{\tau} -\tau)^2] \leq 6 \{ \operatorname{Tr}((\gamma \beta_1^2 L + \eta[\beta_1^2 + \beta_2^2] L^\dagger) X) + \kappa[\beta_1^2 + \beta_2^2] \|X\|_{q} + \|\Pi(\beta_1 \alpha + \beta_2 \phi)\|^2 \operatorname{Tr}(\Pi X)\}.
     \end{align*}
\end{theorem}

\begin{proof}
    We decompose $\alpha = \Pi \alpha + (I - \Pi)\alpha,\ \phi = \Pi \phi + (I - \Pi) \phi$. For the latter terms orthogonal to the kernel of $L$, the same analysis as applied to the mean 0 residuals in \Cref{thm:mse_bound_general_HT} is applicable, hence it suffices to bound the projected terms.
    Following the proof of \Cref{thm:mse_bound_general_HT}, we compute
    \begin{align*}
         &\mathbb{E}[(\langle \Pi \alpha, z/(np) - (1-z)/(n(1-p))  \rangle + \langle \Pi \phi, z/(np) - \mathbf{1}/n \rangle)^2] \\
         &= \operatorname{Var}(\langle \Pi \alpha, z/(np) - (1-z)/(n(1-p))\rangle + \langle \Pi \phi, z/(np) - \mathbf{1}/n \rangle )\\
         &= \operatorname{Var}(\langle \Pi\alpha, \beta_1 (x-\mu) \rangle + \langle \Pi \phi, \beta_2 (x-\mu) \rangle)\\
         &=\operatorname{Tr}(\Pi(\beta_1 \alpha + \beta_2 \phi)(\beta_1 \alpha + \beta_2 \phi)^T\Pi X)\\
         &\leq \|\Pi(\beta_1 \alpha + \beta_2 \phi)\|^2 \operatorname{Tr}(\Pi X).
     \end{align*}
\end{proof}

\subsection{Quadratic Penalization of $X$}

So far in this paper, we have restricted our MSE optimization to first order terms linear in the design covariance matrix $X$, either via a trace inner-product $\operatorname{Tr}(AX)$ or norm penalties. However, there are many natural examples where higher-order dependence is a more appropriate model. 

\begin{example}
    A simple and practical model for the interference is $s(z) = Lz$. We can easily compute
    \[
    (Lz)^T L^\dagger (Lz) = z^T L z,
    \]
    thus $Lz$ is $z^T L z$-interferent. In the MSE, this results in an objective $\propto (x^T L x)^2$ for the Rademacher assignment variables $x$. This quadratic dependence on the cut is exactly because both the level of interference and the extent to which interference impedes the quality of our estimator are both linear $x^T L x$.
\end{example}

Thus, assuming $x^T L x$ interference requires optimization of $\mathbb{E}[(x^T L x)^2]$, and this may similarly be of interest for other kernels $\calK$. One strategy to deal with this is to uniformly upper-bound $x^T L x$, reducing this to a linear optimization. However, this is quite loose, and fails to capture the desired behavior of the assumed outcome model.

Within the framework of Gaussian Rounding, \Cref{alg:gaussian_round}, we relax the Rademacher assignment vectors $x$ to a multivariate Gaussian distribution. This allows us to apply Isserlis's theorem (also commonly known as Wick's probability theorem) \citep{342f2b77-510b-3d97-9992-746cb8a2d27e}, to yield
\[
\mathbb{E}[(x^T L x)^2] = \operatorname{Tr}(LX)^2 + 2 \operatorname{Tr}([LX]^2) = \operatorname{Tr}(LX)^2 + 2 \|L^{1/2}XL^{1/2}\|_F^2.
\]
This is convex in $X$, and thus amenable to our optimization routine.

\begin{example}\label{exp: quad interference}
Quadratic dependencies also naturally appear when considering alternative models for the interference. As a simple example, we may consider a typical setting where $s_i(z)$ is linear in the exposure  fraction of treated neighbors adjacent to the vertex $i$ (see, for example, \citet{thiyageswaran2024data}; see also \Cref{neighborhood_interference}). In other words, $s(\mathbf{1}) - s(z) = L D{-1}W [\mathbf{1} - z] = L/2W[x] + L/2 \mathbf{1}$ where $D$ is the degree matrix and $W$ is the graph adjacency. When considering the treatment effect, we may decompose
\begin{align*}
&\langle \mathbf{1}/n, s(\mathbf{1})\rangle - \langle z/np - (\mathbf{1} - z)/np, s(z)\rangle = \\
&\quad\quad \langle \mathbf{1}/n - [z/np - (\mathbf{1} - z)/np], s(\mathbf{1})\rangle - \langle z/np - (\mathbf{1} - z)/np, s(z) - s(\mathbf{1})\rangle,
\end{align*}
so that the first term can be bounded by our typical error analysis, with some moment restrictions placed on the vector $s(\mathbf{1})$. For the remaining term, its contribution can be bounded by
\[
\langle z/np - (\mathbf{1} - z)/np, s(z) - s(\mathbf{1})\rangle   = L/2\beta_1 x^T D^{-1}W x + L/2 x^T \mathbf{1} + L/2 c_1 n.
\]
In other words, in general settings, we can evaluate the quantity
\begin{align*}
\langle z/np - (\mathbf{1} - z)/np, s(z) - s(\mathbf{1})\rangle &= \langle \beta_1 x + c_1 \mathbf{1}, s(z) - s(\mathbf{1})\rangle\\
&= 2\beta_1\langle z-\mathbf{1}, s(z) - s(\mathbf{1})\rangle + (c_1 - \beta_1)\langle \mathbf{1}, s(z) - s(\mathbf{1})\rangle
\end{align*}
and similar control of the MSE is achieved if we assume a quadratic domination $|\langle z-\mathbf{1}, s(z) - s(\mathbf{1})\rangle| \leq (z-\mathbf{1})^TA(z-\mathbf{1})$, as well as a linear bound $\langle \mathbf{1}, s(z) - s(\mathbf{1})\rangle \leq |v^T (\mathbf{1} - z)|$, for some matrix $A$ and vector $v$. A similar analysis reduces the baseline $s(\mathbf{0})$ into suitable terms for our SDP, falling within the framework of \Cref{thm:mse_bound_general_HT_network_kernels}.
\end{example}

\section{An alternative approach to the interference term} \label{alternative_interference}
Instead of using the bound $\| \langle s(z) , \mathbf{1} \rangle \| \leq \sqrt{\delta},$ one could also treat the interference term in the following manner. Suppose that $s$ is $(\gamma, L + \delta \mathbf{1}\mathbf{1}^T)$-interferent, where $\delta > 0$. Then,
\begin{align*}
   \sup_{\dotp{s(z), (L + \delta \mathbf{1}\mathbf{1}^T )^\dagger s(z)} \leq \gamma} \dotp{x, s(z)} &= \sup_{\dotp{s(z), (L + \delta \mathbf{1}\mathbf{1}^T )^\dagger s(z)} \leq \gamma} \dotp{x,(L + \delta \mathbf{1}\mathbf{1}^T)^{1/2} (L+ \delta \mathbf{1}\mathbf{1}^T)^{\dagger/2} s(z)} \\
   &=  \sup_{\dotp{s(z), (L + \delta \mathbf{1}\mathbf{1}^T )^\dagger s(z)} \leq \gamma} \dotp{ (L+ \delta \mathbf{1}\mathbf{1}^T )^{1/2}x, (L + \delta \mathbf{1}\mathbf{1}^T)^{\dagger/2}s(z)}.
\end{align*}

Therefore, by Cauchy-Schwarz, 
\begin{align*}
   \sup_{\dotp{s(z), (L + \delta \mathbf{1}\mathbf{1}^T)^{\dagger} s(z)} \leq \gamma}  \mathbb{E}[\dotp{x, s(z)}^2] = \gamma \operatorname{Tr}((L+ \delta \mathbf{1}\mathbf{1}^T ) X),
\end{align*}
and the results follow.

\section{A practical consideration on trade-off parameters}

Our framework allows practitioners with domain expertise on the experimental units and the complex mechanisms governing their interactions to navigate the trade-offs between homophily $\eta$, robustness $\kappa$, and spillover effects $\gamma$, in addition to $\Delta$, in order to identify Pareto-optimal experimental designs.

Much like in traditional power calculations, historical data from similar experiments or pilot studies can provide reasonable approximations for $\Delta$, which encompasses the baseline means as well as the expected effect sizes.

Beyond $\Delta$, historical data can also inform $\gamma$, quantifying the ``effective spillover''. This is done by comparing outcomes under different treatment saturation levels. For instance, in \citep{baird2018optimal}, the authors examine the difference in responses for treated units under varying neighborhood exposure levels, i.e., $Y(1, p) - Y(1, 0)$, where $p$ represents the fraction of treated neighbors. Similarly, spillover effects on control units are assessed via $Y(0, p) - Y(0, 0)$. Similar hyperparameter estimation is also described in \citep{viviano2023causal}.

For the robustness parameter $\kappa$, historical data can again prove useful. Specifically, one can compute the sample covariance matrix from outcomes from prior experiments within, and across, treatment saturation levels and covariate groups. This involves aggregating products of residuals across various experimental conditions, and then computing the appropriate Schatten norm.

A natural starting point for estimating the homophily parameter $\eta$ is to use pre-experiment baseline measurements $b$. We can compute the average of the baselines $\hat{b}$ across units with similar covariates which are believed to be predictive of the outcomes, so that the number of unique elements in $\hat{b}$ is equal to the number of distinct covariate groups. Then, an initial estimate of $\eta$ is $\hat{b}^T L \hat{b}$. In fact, we could refine this further by the use a linear combination of these individual $\eta$s with coefficients learned through, for instance, a linear regression of outcomes from previous experiments on covariates. For example, regressing on gender, ethnicity, and age, one would obtain three coefficients $\beta_1, \beta_2, \beta_3$ with which we can estimate $\eta$ to be $\beta_1 \hat{\eta}_{\text{gender}} +\beta_2\hat{\eta}_{\text{ethnicity}} + \beta_3\hat{\eta}_{\text{age}}.$ This gives us an estimate of $\eta$ in the baselines. To estimate $\eta$ in $\phi$, one could do the same with a sample restricted to nodes that are treated and for which all of their neighbors are treated, as well as nodes that are controlled and for which all of their neighbors are controlled,  from a previous experiment. For our final $\eta$, we can simply take the maximum of these two values. To generalize this heuristic, we can frame the estimation procedure in regression terms. Taking $V$ to be the covariate matrix, and $x$ to be the treatment assignment vector, regressing historical outcomes
\begin{align*}
    y = V \beta + x \beta' + D^{-1}W x \beta'' + \varepsilon,
\end{align*}
we define $\eta= \hat{\beta}^T V^T L V \hat{\beta}$, $\kappa = \| S_\varepsilon \|_{q}$, where $S_\varepsilon$ is the sample covariance matrix of the $\varepsilon$ residuals. In particular, a simple and general approach is to specify candidate potential outcome models, such as those linear in some covariates, for example. Then, one could try different hyperparameter configurations and select the ones that minimize a prespecified loss.

The regression approach is now reminiscent of the covariate balancing formalization in \citep{harshaw2024balancing}. We also point the reader to the discussion in \citep{basse2018model} addressing the specification of a similar parameter in their framework. 

Ultimately, the strength of our framework lies in its adaptability. Much like the Gram-Schmidt Walk algorithm in \cite{harshaw2024balancing}, our approach does not prescribe a rigid estimation protocol. Instead, it provides a structured yet flexible methodology that practitioners can tailor to their specific experimental context, whether by refining parameter estimates through machine learning, incorporating alternative homophily metrics, or integrating auxiliary data sources. This adaptability ensures that domain experts can efficiently explore trade-offs and arrive at designs that balance interference, homophily, and robustness in a principled manner.

\section{Difference-in-Means estimator with Fixed Treatment Groups Sizes} \label{section_dim}

In this section, we consider our proposed framework applied to the difference-in-means estimator. Let $n_1 := \sum_{i=1}^n z_i$, and $n_0 := n-n_1$, i.e., the number of treated and control units respectively. Then,
\begin{align*}
    \hat{\tau}_{\text{DiM}} &= \frac{1}{n_1} \dotp{z, Y} - \frac{1}{n_0} \dotp{1 - z, Y} \\
    &=\frac{1}{n_1} \dotp{z,  \alpha + \Diag(\phi)z - s(z)} - \frac{1}{n_0} \dotp{1 - z,  \alpha + \Diag(\phi)z -  s(z)} \\
    &= \left\langle\frac{z}{n_1} - \frac{1 - z}{n_0}, \alpha - s(z) \right\rangle + \left\langle\frac{z}{n_1}, \phi \right \rangle. 
\end{align*}

Therefore, 
\begin{align*}
    \hat{\tau}_{\text{DiM}} - \tau = \left\langle\frac{z}{n_1} - \frac{1 - z}{n_0}, \alpha - s(z) \right\rangle + \left\langle\frac{z}{n_1} - \frac{\mathbf{1}}{n}, \phi \right \rangle. 
\end{align*}

\begin{theorem}[MSE bound] \label{thm:mse_bound_general_DiM}
Let $x \in \{-1, 1\}^n$ by defining $x := 2z-1$. Define $X := \mathbb{E}[xx^T]$. Let $\mathbf{1}^T X \mathbf{1} - \mathbf{1}^T \mu \mu^T \mathbf{1} = 0$, where $\mu = \frac{n_1 - n_0}{n} \mathbf{1}$, and let $\beta_1:=   n/(2n_1 n_0)$, and  $\beta_2:= 1/(2 n_1)$. Then, for $q \geq 1$,
\begin{align*}
\mathbf{E}[(\hat{\tau}_{\text{DiM}} - \tau)^2] \leq 7 \{ \operatorname{Tr}((\gamma \beta_1^2 L + \eta[\beta_1^2 + \beta_2^2]L^\dagger) X) + \kappa[\beta_1^2 + \beta_2^2]\|X\|_{q} \}.
\end{align*}
\end{theorem}

The proof is in \Cref{proof_mse_bound_general_DiM}.

\begin{corollary}[MSE bound (Symmetric case)] \label{thm:mse_bound_dim_symmetric}
For the symmetric case, i.e., when $n_1 = n_0 = n/2$, let $x \in \{-1, 1\}^n$ by defining $x := 2z-1$. Let $\mathbf{1}^T X \mathbf{1} = 0.$ Define $X := \mathbb{E}[xx^T]$. Then, for $q \geq 1$,
\begin{align*}
\mathbf{E}[(\hat{\tau}_{\text{DiM}} - \tau)^2] \leq \frac{5}{n^2} \{ \operatorname{Tr}((4\gamma L + 5\eta L^\dagger) X) + 5\kappa\|X\|_{q} \}.
\end{align*}
\end{corollary}

Let $\mu = \frac{n_1 - n_0}{n} \mathbf{1}$. Reparameterizing with $\Tilde{\gamma}$ and $\Tilde{\eta}$ so that
\begin{align*}
    \Tilde{\gamma} := \frac{\gamma \beta_1^2}{\kappa[\beta_1^2 + \beta_2^2]} \text{ and } \Tilde{\eta} := \frac{\eta}{\kappa},
\end{align*} we can write an SDP more explicitly in terms of the new trade-off parameters:
\begin{align} \label{sdp_relaxation_mse_p_norm_dim}
    \min_{X\in\mathbb{R}^{n\times n}} \quad & \Tilde{\eta}\text{Tr}(L^\dagger X)  +   \Tilde{\gamma} \text{Tr}(L X) + \|X\|_{q} \\
    \text{s.t.} \quad & X \succeq  0 \notag\\
    & \text{diag}(X) = 1 \notag \\
    & \mathbf{1}X\mathbf{1} = \mathbf{1}^T \mu \mu^T \mathbf{1} \notag
\end{align}

Let $X^*$ be the solution to the SDP above. Finally, we generate our treatment assignment vector $x^*$ by running \Cref{alg:gaussian_round} on $X^*$.

\begin{remark} \label{section_MIP} 
When $\kappa = 0$,  we can proceed with a matrix lifting and a SDP relaxation as previously demonstrated, or solve the problem more exactly with a mixed-integer program as described below. We note however that solving an SDP is polynomial-time with efficient solvers that can be sped up by exploiting problem structure. This is not true for the mixed-integer programs.
From the calculations in the proof of \Cref{thm:mse_bound_general_DiM}, when $\kappa =0$, and $\Tilde{\gamma} := \frac{\gamma \beta_1^2}{[\beta_1^2 + \beta_2^2]} \text{ and } \Tilde{\eta} := \eta [\beta_1^2 + \beta_2^2]$, we have the following MIP problem:
\begin{align}
\label{integer_program}
    \min_{x \in \mathbb{R}^n}  \quad & \Tilde{\eta} x^TL^{\dagger}x  + \Tilde{\gamma} x^T L x \\
    \text{s.t.} \quad & x_i = \pm 1 \quad \forall i \\
    & x^T \mathbf{1} = \mathbf{1}^T \mu 
\end{align}
\end{remark}

\subsection{Gaussian Rounding}

One approach to proving theoretical guarantees of this approximation procedure is by leveraging and adapting results from \citep{raghavendra2012approximating} who provide approximation guarantees to the min-bisection problem. For convenience, we restate the min-bisection problem followed by the main result of \citep{raghavendra2012approximating}.

Recall that the min-bisection problem is:
\begin{align*}
    \minimize_{x \in \mathbb{R}^n} \quad & \frac{1}{2} \sum_{i < j} w_{ij} (1 - x_i \cdot x_j) \\
    \text{s.t.} \quad & x_i \in \{ -1, 1 \} \quad \forall i \\
    & \sum_{i} x_i = 0
\end{align*}

The SDP relaxation is given by,
\begin{align*}
    \minimize \quad & \mathbb{E} \big[\sum_{i < j} w_{ij} (v_i - v_j)^2 \big] \\
    \text{s.t.} \quad & \| v_i\|^2 = 1 \quad \forall i \in V \\
    & \sum_{i} v_i = 0
\end{align*} or
\begin{align*}
    \minimize_{X\in\mathbb{R}^{n\times n}} \quad & \operatorname{Tr}(L X) \\
    \text{s.t.} \quad & X_{ii} = 1 \quad \forall i \in V \\
    & \mathbf{1}^T X \mathbf{1} = 0 \\
    &  X \succeq 0
\end{align*}

For this min-bisection problem, \citet{raghavendra2012approximating} propose an algorithm and provide a corresponding approximation error bound. 

\begin{theorem} [{\citealp[Theorem 1.2]{raghavendra2012approximating}}]
    For every $\delta > 0$, there exists an algorithm, which given a graph with a bisection cutting $\epsilon$-fraction of the edges, finds a bisection cutting at most $\mathcal{O}(\sqrt{\epsilon}) + \delta$-fraction of edges with high probability.
\end{theorem}

Their algorithm comprises of two parts. First, they lift the problem through the Lasserre hierarchy and iteratively solve by conditioning on assignments of a subset of nodes, until a pre-specified $\alpha$-level of independence is achieved between the solutions across the nodes. The second part of their approach, given this $\alpha$-independence solution, is a rounding procedure reminiscent of \citep{goemans1995improved}. Finally, to meet the cardinality constraints, they flip the assignments of the nodes with the smallest degrees on the larger side of the cut.

In proving their guarantees from this rounding, they leverage the $\alpha$-independence conditioning. While their $\alpha$-conditioning in the first-part of their procedure  does not apply to our setting, one could adapt their ideas to proving guarantees in our setting under a $\alpha_\kappa$-independence level, dependent on the randomization tuning parameter $\kappa$ as per \Cref{prop:X_norm_covariance_min}. To adapt their final ``flipping'' procedure to achieve cardinality constraints to our settings, we propose targeting the nodes with the smallest $a_i$ with $a_i$ corresponding to the columns vectors of the matrix $A$, instead of the nodes of the smallest degrees. Then, one could prove analogous results for this difference-in-means setting.

\subsection{Adapted Gram Schmidt Walk algorithm}

As a heuristic, one could use the adapted Gram Schmidt Walk algorithm as in the Horvitz-Thompson section, followed by the assignment ``flipping'' procedure targeting the nodes, on the larger side of the cut, with the smallest $Q_{ii}$  corresponding to columns vectors in the matrix $Q$. We also refer to \citep[S8.4]{harshaw2024balancing}. In particular, the authors describe how appending a constant vector to the covariate matrix helps make the treatment group sizes, $n_1, n_0$ closer to equal. A variant proposed by \citet[S8.4]{harshaw2024balancing} is by adding a cardinality constraint within the Gram-Schmidt Walk algorithm. This variant, however, does not inherit the theoretical results from the Bernoulli rounding procedures in \citep{harshaw2024balancing} due to violation of orthogonality of updates between pivot phases in the algorithm.

\section{Proofs to trade-off results} \label{proof:trade-offs}

\subsection{Proof to \Cref{prop:X_norm_covariance_min}} \label{proof_prop:X_norm_covariance_min}
We restate \Cref{prop:X_norm_covariance_min} and prove it below.

\begingroup
\renewcommand{\theproposition}{\ref{prop:X_norm_covariance_min}}
\begin{proposition}
Let $r \in \{-1,1\}^n$ denote the treatment assignment under a Bernoulli randomization design. Then, for fixed $p= \mathbb{P}(r_i = 1)$ for all $i \in [n]$, and $q>1$
    \begin{align*}
        \arg \min_{\substack{X \succeq 0; \\ X_{ii}=1}} \|X\|_{q} = \frac{1}{4p(1-p)} \operatorname{Cov}(r).
    \end{align*}
\end{proposition}
\addtocounter{proposition}{-1}
\endgroup

\begin{proof}[Proof]
    Let $X^* := I.$ Then, $X^*$ has $n$ non-zero eigenvalues which are all equal to 1. 
    We first show that $X^*$ is the unique minimizer of $\|X\|_{q}^{q}$ over $X\in \mathcal{C}:= \{X:X \succeq 0, X_{ii}=1$\}, and then we show that $X^* = \frac{1}{4p(1-p)} \operatorname{Cov}(r).$ 
    It is easy to see that $X^* \in \mathcal{C}$. For $q > 1$, and any $X\in\mathcal{C}$, with eigenvalues $\lambda_i$, 
    \begin{align*}
       \|X\|_{q}^{q} = (n) \times  \frac{1}{n} \sum_{i=1}^n \lambda_i^{q}  \geq (n) \times  (\frac{1}{n} \sum_{i=1}^n \lambda_i)^{q} = (n) \times  1 = \|X^*\|_{q}^{q},
    \end{align*}
    as $ f: x \to x^{q}$ is convex on $x\geq 0$, and $\sum_i \lambda_i = \sum_i X_{ii} = n$. That $X^*$ is the unique minimizer follows from the strict convexity of $\| \cdot \|_{q}^{q}$ on $\mathcal{C}$ for $q > 1$. 
    Indeed, if we express $\|X\|_{q}^{q} = g(\lambda_1,\dots,\lambda_{n}) = \sum_{i=1}^{n} \lambda_i^{q}$,
    then 
    \[
    \nabla^2 g = q(q-1) 
    \begin{bmatrix}
    \lambda_1^{q-2} & 0 & \dots & 0\\
    0 & \lambda_2^{q-2} & \dots & 0\\
    \vdots & \vdots & \ddots & \vdots \\
    0 & 0 & \dots & \lambda_{n}^{q-2}
    \end{bmatrix},
    \]
    which is positive definite at $X^*$ with its $n$ eigenvalues $\lambda_i = 1$ for $i = 1, 2, ..., n$.

    It is now sufficient to show that this matrix $X^*$ is equivalent to the $ \frac{1}{4p(1-p)}\operatorname{Cov}(r)$ under Bernoulli randomization with uniform treatment assignment probability. We begin by observing that $\operatorname{Cov}(r) = \mathbb{E}[rr^T] - \mathbb{E}[r]\mathbb{E}[r]^T$. Therefore, since $r \in \{-1,1\}$ with $\mathbb{P}(r_i = 1) = p$ for all $i$, we have that $\operatorname{Cov}(r)_{ii} = 4p(1-p)$. Recall that unit-level Bernoulli randomization is a randomization scheme where each unit is independently assigned to treatment or control. Therefore, $\operatorname{Cov}(r)_{ij} = 0$ for all $j \neq i$. 
\end{proof}

We state below the analogous result for complete randomization.

\begin{proposition} \label{prop:X_norm_covariance_min_complete}
Let $r \in \{-1,1\}^n$ denote the treatment assignment under a complete randomization design. Then, for fixed $n_1 = n_0$, i.e., $r^T \mathbf{1} = 0$, and $q>1$
    \begin{align*}
        \arg \min_{\substack{X\mathbf{1} =  \mathbf{0};\\ X \succeq 0; \\ X_{ii}=1}} \|X\|_{q} = \operatorname{Cov}(r).
    \end{align*}
\end{proposition}

\begin{proof}[Proof]
    Let $X^* := (1 + \frac{1}{n-1}) I - \frac{1}{n-1} \mathbf{1}\mathbf{1}^T = \frac{n}{n-1} I - \frac{n}{n-1} (\mathbf{1}/\sqrt{n}) (\mathbf{1}/\sqrt{n})^T.$ Then, $X^*$ has $n-1$ non-zero eigenvalues which are all equal to $\frac{n}{n-1}$, and one zero eigenvalue. Indeed, the first term has $n$ eigenvalues that are equal to $\frac{n}{n-1}$, while the second term has one eigenvalue that is equal to $\frac{n}{n-1}$, and $n-1$ eigenvalues that are equal to 0, and as they share the same eigenvectors, their difference has the desired property.
    We first show that $X^*$ is the unique minimizer of $\|X\|_{q}^q$ over $X\in \mathcal{C}:= \{X:X\mathbf{1} = \mathbf{0},X \succeq 0, X_{ii}=1$\}, and then we show that $X^* = \operatorname{Cov}(r).$ 
    It is easy to see that $X^* \in \mathcal{C}$. For $q > 1$, and any $X\in\mathcal{C}$, with eigenvalues $\lambda_i$, 
    \begin{align*}
        \|X\|_{q}^{q} = (n-1) \times  \frac{1}{n-1} \sum_{i=1}^n \lambda_i^{q}  \geq (n-1) \times  (\frac{1}{n-1} \sum_{i=1}^n \lambda_i)^{q} = (n-1) \times  (\frac{n}{n-1})^{q} = \|X^*\|_{q}^{q},
    \end{align*}
    as $ f: x \to x^{q}$ is convex on $x\geq 0$, and $\sum_i \lambda_i = \sum_i X_{ii} = n$. That $X^*$ is the unique minimizer follows from the strict convexity of $\| \cdot \|_{q}^{q}$ on $\mathcal{C}$ for $q > 1$. 
    Indeed, if we express $\|X\|_{q}^{q} = g(\lambda_1,\dots,\lambda_{n-1}) = \sum_{i=1}^{n-1} \lambda_i^{q}$,
    then 
    \[
    \nabla^2 g = q(q-1) 
    \begin{bmatrix}
    \lambda_1^{q-2} & 0 & \dots & 0\\
    0 & \lambda_2^{q-2} & \dots & 0\\
    \vdots & \vdots & \ddots & \vdots \\
    0 & 0 & \dots & \lambda_{n-1}^{q-2}
    \end{bmatrix},
    \]
    which is positive definite at $X^*$ with its first $n-1$ eigenvalues $\lambda_i = \frac{n}{n-1}$ for $i = 1, 2, ..., n-1$.

    It is now sufficient to show that this matrix $X^*$ is equivalent to the $\cov(r)$ under complete randomization, with equal number of treated and control units, i.e., $r^T \mathbf{1} = 0$. We begin by observing that $\cov(r) = \mathbb{E}[rr^T]$, and that $\cov(r) \mathbf{1} = \mathbb{E}[rr^T \mathbf{1}] = 0$. Therefore, since $r \in \{-1,1\}$ with equal probability, we have that $\cov(r)_{ii} = 1$, and that the row sums of $\cov(r) = 0$. This must imply that $\sum_{j \neq i} \cov(r)_{ij} = -1$, for all $i \in [n]$. Recall that unit-level complete randomization is a uniform randomization process. Therefore, $\cov(r)_{ij} = \cov(r)_{ik}$ for all $j \neq k \neq i$. Therefore, this together with $\sum_{j \neq i} \cov(r)_{ij} = -1$ imply that $\cov(r)_{ij} = -\frac{1}{n-1}$ for all $j \neq i$, for every $i$.
\end{proof}

\subsection{Proof to \Cref{thm:mse_bound_general_HT}} \label{proof_mse_bound_general_HT}

We first state and prove lemmas we use to prove \Cref{thm:mse_bound_general_HT}.

\begin{lemma} \label{lemma:homophily_bd}
    Suppose that $h$ is $\eta$-homophilous, $\dotp{h, \mathbf{1}} = 0$, and $X:= \mathbb{E}[xx^T]$, then
    \[
    \mathbb{E}[\dotp{x, h}^2] \leq \eta\operatorname{Tr}(L^{\dagger}X).
    \]
\end{lemma}

\begin{proof}
    Observe, as $h$ is orthogonal to the kernel of $L$, and $L$ is PSD,
    \begin{align*}
       \sup_{\dotp{h, Lh} \leq \eta} \dotp{x, h} = \sup_{\dotp{h, Lh} \leq \eta} \dotp{x, L^{\dagger/2} L^{1/2}h} =  \sup_{\dotp{h, Lh} \leq \eta} \dotp{L^{\dagger/2}x, L^{1/2}h}.
    \end{align*}

    Therefore, by Cauchy-Schwarz, 
    \begin{align*}
       \sup_{\dotp{h, Lh} \leq \eta}  \mathbb{E}[\dotp{x, h}^2] = \eta\operatorname{Tr}(L^{\dagger}X).
    \end{align*}
\end{proof}

\begin{lemma} \label{lemma:interference_bd}
    Suppose that $s$ is $\gamma$-interferent, $\dotp{s(z), \mathbf{1}/n} = 0$ for all $z$, and $X:= \mathbb{E}[xx^T]$, then
    \[
    \mathbb{E}[\dotp{x, s(z)}^2] \leq \gamma\operatorname{Tr}(LX).
    \]
\end{lemma}

\begin{proof}
    Observe, as $s(z)$ is orthogonal to the kernel of $L$, and $L$ is PSD,
    \begin{align*}
       \sup_{\dotp{s(z), L^\dagger s(z)} \leq \gamma} \dotp{x, s(z)} = \sup_{\dotp{s(z), L^{\dagger}s(z)} \leq \gamma} \dotp{x,L^{1/2} L^{\dagger/2} s(z)} =  \sup_{\dotp{s(z), L^{\dagger}s(z)} \leq \gamma} \dotp{ L^{1/2}x, L^{\dagger/2}s(z)}.
    \end{align*}

    Therefore, by Cauchy-Schwarz, 
    \begin{align*}
       \sup_{\dotp{s(z), L^{\dagger} s(z)} \leq \gamma}  \mathbb{E}[\dotp{x, s(z)}^2] = \gamma \operatorname{Tr}(L X).
    \end{align*}
\end{proof}

\begin{lemma} \label{lemma:robustness_bd}
    Suppose that $\varepsilon$ is independent of $x$ and $\Sigma := \mathbb{E}[\varepsilon \varepsilon^T]$ is such that $\|\Sigma\|_{q^*} \leq \kappa$, $1/q^* + 1/q = 1$. Let $X:= \mathbb{E}[xx^T]$. Then,
    \[
    \mathbb{E}[\dotp{x, \varepsilon}^2] \leq \kappa \|X\|_{q}.
    \]
\end{lemma}

\begin{proof}
We begin by writing 
\begin{align*}
     \mathbb{E}[\dotp{x, \varepsilon}^2] = \operatorname{Tr}(\mathbb{E}[\varepsilon\varepsilon^T]\mathbb{E}[xx^T]) = \operatorname{Tr}(\Sigma X).
\end{align*}
Then, the proof follows from \Cref{prop:general_schatten_p_bound}.
\end{proof}

In fact, the results in \Cref{lemma:homophily_bd} can be tightened. We write this formally in the following lemma.
\begin{lemma}
 \label{lemma:tighter_homophily_bd}
    Suppose that $h$ is $\eta$-homophilous, $\dotp{h, \mathbf{1}} = 0$, and $X:= \mathbb{E}[xx^T]$, then
    \[
    \mathbb{E}[\dotp{x, h}^2] \leq \eta \, \lambda_{\max}(L^{\dagger/2}XL^{\dagger/2}).
    \]   
\end{lemma}

\begin{proof}
    Observe, as $h$ is orthogonal to the kernel of $L$, and $L$ is PSD,
    \begin{align*}
       \sup_{\dotp{h, Lh} \leq \eta} \mathbb{E}[\dotp{x, h}^2] =  \sup_{\dotp{h, Lh} \leq \eta} h^T X h &= \sup_{\| L^{1/2} h\| \leq \eta^{1/2}} h^T L^{1/2} L^{\dagger/2 } X L^{\dagger/2} L^{1/2} h \\
       &= \sup_{\| w\| \leq 1}  \eta w^T L^{\dagger/2 } X L^{\dagger/2} w \\
       &=  \eta \| L^{\dagger/2 } X L^{\dagger/2} \|_{\operatorname{op}}.
    \end{align*}
\end{proof}

We restate \Cref{thm:mse_bound_general_HT} below.

\begingroup
\renewcommand{\thetheorem}{\ref{thm:mse_bound_general_HT}}
\begin{theorem}[(General) MSE bound]
Let $p:= \mathbb{P}(z_i = 1)$ for all $i \in [n]$, $\beta_1 := 1/(2np(1-p))$, $\beta_2 := 1/(2np)$, $a = \langle \alpha, \mathbf{1}/n\rangle$, $b = \langle \phi, \mathbf{1}/n\rangle$, and $\langle s(z), \mathbf{1}/n\rangle \leq \sqrt{\delta}$ for all $z$. Let also $x \in \{-1, 1\}^n$ by defining $x := 2z-1$. Define $X:=\mathbb{E}[xx^T] - \mu \mu^T$, $\mu := \mathbb{E}[x]$. Then, 
     \begin{align*}
          \mathbb{E}[(\hat{\tau} -\tau)^2] \leq 7 \{ \operatorname{Tr}((\gamma \beta_1^2 L + \eta[\beta_1^2 + \beta_2^2] L^\dagger) X) + \kappa[\beta_1^2 + \beta_2^2] \|X\|_{q} + ([a \beta_1 + b\beta_2]^2 + \beta_1^2 \delta) \operatorname{Tr}(\mathbf{1}\mathbf{1}^TX)\}.
     \end{align*}
\end{theorem}
\addtocounter{theorem}{-1}
\endgroup

\begin{proof}
    The relevant quantities of interest are
    \begin{align}
    &\hat{\tau} = \frac{1}{n}\dotp{ Y, z/p - (1-z)/(1-p)}  = \frac{1}{n}\dotp{\alpha + s(z), z/p - (1-z)/(1-p) } + \frac{1}{n}\dotp{ \phi, z/p },\\
    &\tau = \dotp{ \phi, \mathbf{1}/n }, \\
    &\hat{\tau} - \tau = \dotp{\alpha + s(z), z/(np) - (1-z)/(n(1-p))} + \dotp{ \phi,z/(np) - \mathbf{1}/n }.
    \end{align}
    
    Notice that we can re-write
    \begin{align*}
    &z/(np) - (1-z)/(n(1-p)) \\
    &= \left[z/(np) - (1-z)/(n(1-p)) - \frac{1}{2}(1/(np) - 1/(n(1-p))) \mathbf{1} \right] +  \frac{1}{2}(1/(np) - 1/(n(1-p))) \mathbf{1}\\
    &=: \beta_1 x + c_1 \mathbf{1}
    \end{align*}
    where $c_1 := \frac{1}{2}(1/(np) - 1/(n(1-p)))$. Similarly,
    \begin{align*}
        z/(np) - \mathbf{1}/n = [z/(np) - \mathbf{1}/n - (1/(2np)-1/n) \mathbf{1}] + (1/(2np)-1/n)\mathbf{1} =: \beta_2 x + c_2 \mathbf{1}
    \end{align*}
    where $c_2 := (1/(2np)-1/n)$.
     Before proceeding, we first center $\alpha, \phi$, decomposing them as $\alpha = \bar{\alpha} + \alpha',\, \phi = \bar{\phi} + \phi'$, $\bar{\alpha},\, \bar{\phi}$ denoting the projection onto the constant vector for each term respectively. Similarly, we decompose $s(z) = \Bar{s}(z) + \Tilde{s}(z)$. Then, we can write,
     \begin{align*}
        \mathbb{E}[(\hat{\tau} - \tau )^2]  &=  \mathbb{E}[(\dotp{\alpha + s(z), z/(np) - (1-z)/(n(1-p))} + \dotp{ \phi,z/(np) - \mathbf{1}/n })^2]\\
        &= 7 \mathbb{E}[(\dotp{\bar{\alpha}, z/(np) - (1-z)/(n(1-p))} + \dotp{ \bar{\phi},z/(np) - \mathbf{1}/n })^2 \\
        &\quad + \dotp{\bar{s}(z), z/(np) - (1-z)/(n(1-p))}^2\\
        &\quad + \langle h_\alpha, \beta_1 (x - \mu)\rangle^2 + \langle \Tilde{s}(z), \beta_1 (x - \mu)\rangle^2 + \langle \varepsilon_\alpha, \beta_1 (x - \mu)\rangle^2 \\
        &\quad + \langle h_\phi, \beta_2 (x - \mu) \rangle^2 + \langle \varepsilon_\phi, \beta_2 (x - \mu)\rangle^2],
     \end{align*} where the last two lines on the right-hand-side come from the mean-zero components of $\alpha$, $s(z)$, and $\phi$. Indeed, as these are orthogonal to $\mathbf{1}$ we can drop the $c \mathbf{1}$ terms from the expression, and add the $\mu$ terms. Thus, we can write
     \begin{align*}
         &\langle \alpha' + \Tilde{s}(z), z/(np) - (1-z)/(n(1-p))\rangle + \langle \phi',z/(np) - \mathbf{1}/n\rangle \\
         &= \langle h_\alpha, \beta_1 (x - \mu)\rangle + \langle \Tilde{s}(z), \beta_1 (x - \mu)\rangle + \langle \varepsilon_\alpha, \beta_1 (x - \mu)\rangle + \langle h_\phi, \beta_2 (x - \mu) \rangle + \langle \varepsilon_\phi, \beta_2 (x - \mu)\rangle.
     \end{align*}

     We can upper-bound the contribution of the terms in the first two lines, as we notice that by the definition of our marginal probabilities,
     \[\mathbb{E}[
      \langle \bar{\alpha}, z/(np) - (1-z)/(n(1-p))\rangle] = 0 = \mathbb{E}[\langle \bar{\phi},z/(np) - \mathbf{1}/n\rangle],
     \]
    hence,
     \begin{align*}
         &\mathbb{E}[(\langle \bar{\alpha}, z/(np) - (1-z)/(n(1-p))  \rangle + \langle \bar{\phi}, z/(np) - \mathbf{1}/n \rangle)^2] \\
         &= \operatorname{Var}(\langle \bar{\alpha}, z/(np) - (1-z)/(n(1-p))\rangle + \langle \bar{\phi}, z/(np) - \mathbf{1}/n \rangle )\\
         &= \operatorname{Var}(\langle \bar{\alpha}, \beta_1 (x-\mu) \rangle + \langle \bar{\phi}, \beta_2 (x-\mu) \rangle)\\
         &= \operatorname{Var}([\beta_1a + \beta_2 b]\langle \mathbf{1},  (x-\mu)  \rangle) + \operatorname{Var}(\langle \bar{s}(z),  (x-\mu)  \rangle)\\
         &= (\beta_1 a + \beta_2 b)^2 \operatorname{Tr}(\mathbf{1}\mathbf{1}^TX),
     \end{align*}
     for $X:=\mathbb{E}[xx^T] - \mu \mu^T$, $\mu := \mathbb{E}[x]$, $a = \langle \alpha, \mathbf{1}/n\rangle$, $b = \langle \phi, \mathbf{1}/n\rangle$. 

    Similarly, 
         \begin{align*}
         \mathbb{E}[\langle \bar{s}(z), z/(np) - (1-z)/(n(1-p)) \rangle^2] &\leq \delta \mathbb{E}[\langle \mathbf{1}, z/(np) - (1-z)/(n(1-p))  \rangle^2]\\
         &= \delta \operatorname{Var}(\langle \mathbf{1}, z/(np) - (1-z)/(n(1-p))\rangle) \\
         &= \delta \operatorname{Var}(\langle \mathbf{1}, \beta_1 (x-\mu) \rangle)\\
         &\leq \beta_1^2 \delta \operatorname{Tr}(\mathbf{1}\mathbf{1}^TX).
     \end{align*}
     Thus, by \Cref{lemma:homophily_bd,lemma:interference_bd,lemma:robustness_bd}, we get
     \begin{align*}
         \mathbb{E}[(\hat{\tau} - \tau )^2] \leq \operatorname{Tr}((\gamma \beta_1^2 7 L + \eta[\beta_1^2 + \beta_2^2]7L^\dagger) X) + \kappa[\beta_1^2 + \beta_2^2]7 \|X\|_{q} + ([a\beta_1 + b\beta_2]^2 + \beta_1^2 \delta) 7 \operatorname{Tr}(\mathbf{1}\mathbf{1}^TX).
     \end{align*}    
\end{proof}

\subsection{Tighter bounds} \label{subsection_tighter_bounds} 

In \Cref{thm:mse_bound_general_HT}, we relied on the inequality from \Cref{lemma:homophily_bd}. If we, instead, used the tighter inequality from \Cref{lemma:tighter_homophily_bd}, we obtain the following worst-case bound.

\begin{theorem}[(Tighter) MSE bound] \label{thm:tight_mse_bound_general_HT}
Let $p:= \mathbb{P}(z_i = 1)$ for all $i \in [n]$, $\beta_1 := 1/(2np(1-p))$, $\beta_2 := 1/(2np)$, $a = \langle \alpha, \mathbf{1}/n\rangle$, $b = \langle \phi, \mathbf{1}/n\rangle$, and $\langle s(z), \mathbf{1}/n\rangle \leq \sqrt{\delta}$ for all $z$. Let also $x \in \{-1, 1\}^n$ by defining $x := 2z-1$. Define  $X:=\mathbb{E}[xx^T] - \mu \mu^T$, $\mu := \mathbb{E}[x]$. Then, for $q \geq 1$,
     \begin{align*}
          \mathbb{E}[(\hat{\tau} -\tau)^2] \leq 7 \{ &\gamma \beta_1^2 \operatorname{Tr}(L X) +  \eta[\beta_1^2 + \beta_2^2] \lambda_{\max}(L^{\dagger/2} X L^{\dagger/2}) + \kappa[\beta_1^2 + \beta_2^2] \|X\|_{q} \\
          &+ ([a \beta_1 + b\beta_2]^2 + \beta_1^2 \delta) \operatorname{Tr}(\mathbf{1}\mathbf{1}^TX)\}.
     \end{align*}
\end{theorem}

However, writing this as an SDP introduces an additional PSD constraint, and the resulting optimization problem is much more computationally intensive.

\subsection{Proof to \Cref{thm:mse_bound_general_DiM}} \label{proof_mse_bound_general_DiM}

We restate \Cref{thm:mse_bound_general_DiM} below.

\begingroup
\renewcommand{\thetheorem}{\ref{thm:mse_bound_general_DiM}}
\begin{theorem}[MSE bound]
Let $x \in \{-1, 1\}^n$ by defining $x := 2z-1$. Define $X := \mathbb{E}[xx^T]$. Let $\mathbf{1}^T X \mathbf{1} - \mathbf{1}^T \mu \mu^T \mathbf{1} = 0$, where $\mu = \frac{n_1 - n_0}{n} \mathbf{1}$, and let $\beta_1:=   n/(2n_1 n_0)$, and  $\beta_2:= 1/(2 n_1)$. Then, for $q \geq 1$,
\begin{align*}
\mathbf{E}[(\hat{\tau}_{\text{DiM}} - \tau)^2] \leq 7 \{ \operatorname{Tr}((\gamma \beta_1^2 L + \eta[\beta_1^2 + \beta_2^2]L^\dagger) X) + \kappa[\beta_1^2 + \beta_2^2]\|X\|_{q} \}.
\end{align*}
\end{theorem}
\addtocounter{theorem}{-1}
\endgroup

\begin{proof}
    The relevant quantities of interest are
    \begin{align}
    &\hat{\tau}_{\text{DiM}} = \dotp{ Y, z/n_1 - (1-z)/n_0}  = \dotp{\alpha + s(z), z/n_1 - (1-z)/n_0 } + \dotp{ \phi, z/n_1 },\\
    &\tau = \dotp{ \phi, \mathbf{1}/n }, \\
    &\hat{\tau}_{\text{DiM}} - \tau = \dotp{\alpha + s(z), z/n_1 - (1-z)/n_0} + \dotp{ \phi,z/n_1 - \mathbf{1}/n }.
    \end{align}
    
    Notice that we can re-write
    \begin{align*}
    z/n_1 - (1-z)/n_0 &= \left[z/n_1 - (1-z)/n_0 - \frac{1}{2}(1/n_1 - 1/n_0) \mathbf{1} \right] +  \frac{1}{2}(1/n_1 - 1/n_0) \mathbf{1}\\
    &=: \beta_1 x + c_1 \mathbf{1}
    \end{align*}
    where $c_1 := \frac{1}{2}(1/n_1 - 1/n_0)$. Similarly,
    \begin{align*}
        z/n_1 - \mathbf{1}/n = [z/n_1 - \mathbf{1}/n - (1/(2n_1)-1/n) \mathbf{1}] + (1/(2n_1)-1/n)\mathbf{1} =: \beta_2 x + c_2 \mathbf{1}
    \end{align*}
    where $c_2 := (1/(2n_1)-1/n)$.
     Before proceeding, we first center $\alpha, \phi$, decomposing them as $\alpha = \bar{\alpha} + \alpha',\, \phi = \bar{\phi} + \phi'$, $\bar{\alpha},\, \bar{\phi}$ denoting the projection onto the constant vector for each term respectively. Similarly, we decompose $s(z) = \bar{s}(z) + \Tilde{s}(z)$. Then, we can write
     \begin{align*}
         &\mathbb{E}[(\tau - \hat{\tau}_{\text{DiM}})^2] \\
         &= 7\mathbb{E}[(\dotp{\bar{\alpha}, z/n_1 - (1-z)/n_0} + \dotp{\bar{\phi},z/n_1 - \mathbf{1}/n })^2 + \dotp{\bar{s}(z), z/n_1 - (1-z)/n_0}^2 \\
         &\quad + \langle h_\alpha, \beta_1 x\rangle^2 + \langle \Tilde{s}(z), \beta_1 x\rangle^2 + \langle \varepsilon_\alpha, \beta_1 x\rangle^2 + \langle h_\phi, \beta_2 x\rangle^2 + \langle \varepsilon_\phi, \beta_2 x\rangle^2],
     \end{align*}
     where the terms in the second line of the right-hand-side come from the mean-zero components of $\alpha$, $s(z)$, and $\phi$. Indeed, as these are orthogonal to $\mathbf{1}$ we can drop the $c \mathbf{1}$ terms from the expression. Thus, we can write
     \begin{align*}
         \langle \alpha' + \Tilde{s}(z), z/n_1 - (1-z)/n_0 \rangle + \langle \phi',z/n_1 - \mathbf{1}/n\rangle &= \langle h_\alpha, \beta_1 x\rangle + \langle \Tilde{s}(z), \beta_1 x\rangle + \langle \varepsilon_\alpha, \beta_1 x\rangle\\
         &\quad\quad + \langle h_\phi, \beta_2 x\rangle + \langle \varepsilon_\phi, \beta_2 x\rangle.
     \end{align*}
     We can upper-bound the contribution of the terms in the first line, as we notice that by the definition of our marginal probabilities,
     \[\mathbb{E}[
      \langle \bar{\alpha}, z/n_1 - (1-z)/n_0)\rangle] = 0 = \mathbb{E}[\langle \bar{\phi},z/n_1 - \mathbf{1}/n\rangle],
     \]
     hence,
     \begin{align*}
         \mathbb{E}[(\langle \bar{\alpha}, z/n_1 - (1-z)/n_0 \rangle + \langle \bar{\phi},z/n_1 - \mathbf{1}/n\rangle)^2] &= \operatorname{Var}(\langle \bar{\alpha}, z/n_1 - (1-z)/n_0)\rangle + \langle \bar{\phi},z/n_1 - \mathbf{1}/n \rangle)\\
         &= \operatorname{Var}(\langle \bar{\alpha}, \beta_1 x\rangle + \langle \bar{\phi},z/n_1 - \mathbf{1}/n \rangle)\\
         &=(\beta_1 a + \beta_2 b)^2 \operatorname{Tr}(\mathbf{1}\mathbf{1}^T [X - \mu \mu^T])
     \end{align*}
     for $X:=\mathbb{E}[xx^T]$, $\mu := \mathbb{E}[x]$, $a = \langle \alpha, \mathbf{1}/n\rangle$, $b = \langle \phi, \mathbf{1}/n\rangle$. 

     By our constraints $\mathbf{1}^T X \mathbf{1} - \mathbf{1}^T \mu \mu^T \mathbf{1} = 0$, 
     \[\mathbb{E}[(\langle \bar{\alpha}, z/n_1 - (1-z)/n_0 \rangle + \langle \bar{\phi},z/n_1 - \mathbf{1}/n\rangle)^2] = 0.\]

     Similarly, 
     \begin{align*}
         \mathbb{E}[\langle \bar{s}(z), z/n_1 - (1-z)/n_0 \rangle^2] &\leq \delta \mathbb{E}[\langle \mathbf{1}, z/n_1 - (1-z)/n_0 \rangle^2] \\
         &= \delta \operatorname{Var}(\langle \mathbf{1}, z/n_1 - (1-z)/n_0\rangle) \\
         &= \delta \operatorname{Var}(\langle \mathbf{1}, \beta_1 x\rangle) \\
         &=\beta_1^2 \delta \operatorname{Tr}(\mathbf{1}\mathbf{1}^T [X - \mu \mu^T]),
     \end{align*} which is constrained to be zero by our constraints $\mathbf{1}^T X \mathbf{1} - \mathbf{1}^T \mu \mu^T \mathbf{1} = 0.$
     
     Thus, by \Cref{lemma:homophily_bd,lemma:interference_bd,lemma:robustness_bd}, we get
     \begin{align*}
         \mathbb{E}[(\tau - \hat{\tau}_{\text{DiM}})^2] \leq \operatorname{Tr}((\gamma \beta_1^2 7 L + \eta[\beta_1^2 + \beta_2^2]7L^\dagger) X) + \kappa[\beta_1^2 + \beta_2^2]7 \|X\|_{q}.
     \end{align*}    
\end{proof}

\section{Proofs to Theoretical Guarantees}
\label{proof:thm_approximation_error}

Our proof of \Cref{thm:approximation_error} utilizes the seminal ideas underlying the approximation bound for MAXCUT, as introduced in \citet{goemans1995improved}. For convenience, we reiterate the MAXCUT problem in the following subsection.

\subsection{The MAXCUT problem} \label{maxcut_context}
\begin{align*}
    \maximize_{x \in \mathbb{R}^n} \quad & \frac{1}{2} \sum_{i < j} w_{ij} (1 - x_i \cdot x_j) \\
    \text{s.t.:} \quad & x_i \in \{ -1, 1 \} \quad \forall i \in V \\
\end{align*}

The SDP relaxation is given by
\begin{align*}
    \maximize_{X \in \mathbb{R}^{n \times n}} \quad & \operatorname{Tr}(L X) \\
    \text{s.t.:} \quad & X_{ii} = 1 \quad \forall i \in V \\
    &  X \succeq 0
\end{align*}

Let $\zeta = \operatorname{sign}(\xi)$, $\xi \sim \calN(0, X^*)$, where $X^*$ is the solution to the SDP relaxation of MAXCUT above. \citet{goemans1995improved} proved that 
\begin{align*}
    \mathbb{E} \zeta^T L \zeta \geq 0.878 \max_{x \in \{-1,1 \}^n} x^T L x,
\end{align*} by leveraging Grothendieck's identity $$\frac{\pi}{2}\mathbb{E}[\zeta_i \zeta_j] = \arcsin(X_{i,j}^*).$$

We now prove our results generalizing Grothendieck's identity when $p \neq 1/2$, in \Cref{non_symmetric_rounding_cov}. When $p = 1/2$, the special case of Grothendieck's identity is recovered.

\begin{lemma} \label{non_symmetric_rounding_cov}
    Let $p:= \mathbb{P}(x_i  = 1)$, and $X$ be the SDP solution to \ref{sdp_relaxation_mse_p_norm}. Let also $v_i$, $i=1,2, \ldots, n$ be the column vectors of $X^{1/2}$. Denote the vector returned by \Cref{alg:gaussian_round} by $\zeta$. Then,
    \[\cov(\zeta_i, \zeta_j) = \frac{8p^2}{\pi} \arcsin{\dotp{v_i, v_j}},\] for $i \neq j$.
\end{lemma}

\begin{proof}
Write $p_\xi^i := \mathbb{P}(\zeta_i = 1 | \dotp{v_i, \xi} > 0)$. Since $\mathbb{P}(\dotp{v_i, \xi} > 0) = 0.5$, and since $\mathbb{P}(\zeta_i = 1 | \dotp{v_i, \xi} \leq 0) = 0$, we know that 
\begin{align*}
    p := \mathbb{P}(\zeta_i = 1) &= \mathbb{P}(\zeta_i = 1 | \dotp{v_i, \xi} > 0) \mathbb{P}(\dotp{v_i, \xi} > 0) + \mathbb{P}(\zeta_i = 1 | \dotp{v_i, \xi} \leq 0) \mathbb{P}(\dotp{v_i, \xi} \leq 0) \\
    &= p_\xi^i/2.
\end{align*}
Therefore, we have that $p_\xi^i = 2p$ for all $i$. Define $\theta$ to be the angle corresponding to the region of vectors $\{v_k, v_l\}$ such that $\dotp{v_k, \xi} >0$, and $\dotp{v_l, \xi} >0$,  i.e., $\theta:= \{\max_{u,v} \arccos{\dotp{u, v}: \dotp{u, \xi} >0, \dotp{v, \xi} >0}, \xi \in \operatorname{Unif}(S^{n-1})\}$, where $S^{n-1}$ is the $(n-1)$-dimensional sphere. Then, $\theta= \pi - \arccos{\dotp{v_i, v_j}},$ and
\begin{align*}
   \mathbb{P}(\zeta_i \neq \zeta_j) &= \frac{4p(\pi-\theta) + 4p(1-2p)\theta}{2 \pi} \\
   &= \frac{2p}{\pi}\left(\pi - \theta + (1-2p) \theta) \right) \\
   &=  \frac{2p}{\pi}\left[\arccos{\dotp{v_i, v_j}} + (1-2p) (\pi - \arccos{\dotp{v_i, v_j}}) \right].
\end{align*}
Thus,
\begin{align*}
    \mathbb{E}[\zeta_i \zeta_j] &= 1 - 2 \mathbb{P}(\zeta_i \neq \zeta_j) \\
    &= 1 - \frac{4p}{\pi}\left[\arccos{\dotp{v_i, v_j}} + (1-2p) (\pi - \arccos{\dotp{v_i, v_j}}) \right] \\
    &= \frac{\pi(1- 4p (1-2p)) -8p^2 \arccos{\dotp{v_i, v_j}}}{\pi}.
\end{align*}

\begin{align*}
  \cov(\zeta_i, \zeta_j) &=   \frac{\pi(1- 4p (1-2p)) -8p^2 \arccos{\dotp{v_i, v_j}}}{\pi} - (2p -1)^2 \\
  &= \frac{8p^2}{\pi} \arcsin{\dotp{v_i, v_j}}.
\end{align*}
\end{proof}

\begin{figure}
    \centering
    \includegraphics[width=0.5\linewidth]{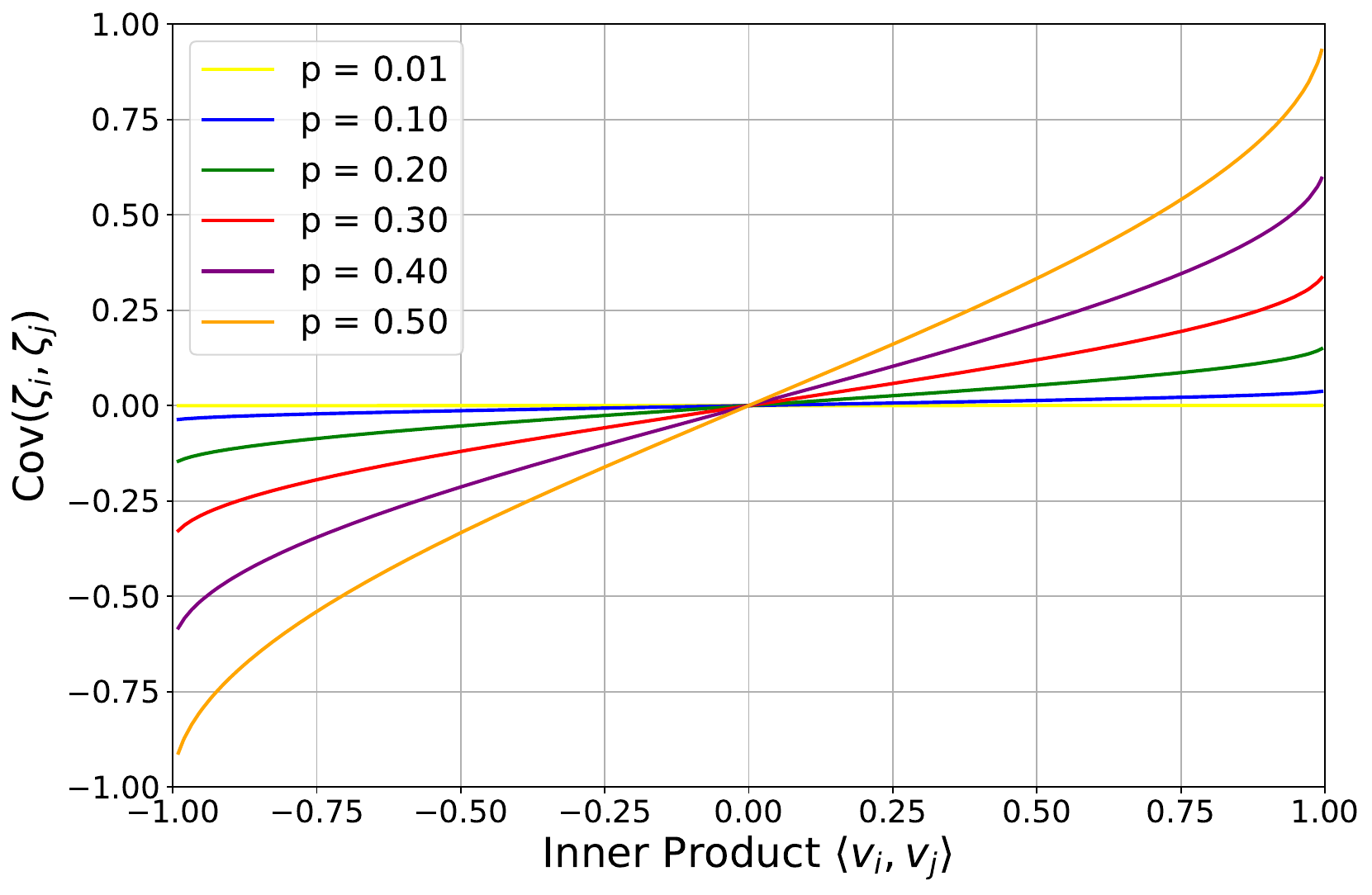}
    \caption{The relationship between $\cov(\zeta_i,\zeta_j)$ and vector inner-products $v_i, v_j$ in the SDP relaxation for various treatment assignment probabilities $p$.}
    \label{fig:geometric_pf_curves}
\end{figure} 
\Cref{fig:geometric_pf_curves} displays this relationship for various treatment assignment probabilities $p$.

We first prove \Cref{thm:approximation_error}.

\begin{theorem}[Approximation bound] \label{thm:approximation_error}
Let $\Xi = \operatorname{Cov}(\zeta)$ be the covariance matrix of the associated Rademacher random variables. Let $X$ be the SDP solution. We assume that $X$ is non-degenerate, i.e., $X \succeq \delta I$, $\delta > 0$. For $c = \left( \frac{8p^2}{\pi}  + \frac{8p^2 \psi_\kappa}{\pi \delta} + \frac{4p(1-p) - 8p^2/\pi}{\delta}  \right)$ for $\psi_\kappa= n \alpha_\kappa^{3},$ where $\alpha_\kappa$ is the maximum off-diagonal entry of $X$,
\begin{align*}
    \Xi \preceq c_\kappa X.
\end{align*}
\end{theorem}

\begin{proof}
Recall that for Rademacher random variables $\zeta_i \in \{-1,1\}$ with $\mathbb{P}(\zeta_i = 1),$ $\var(\zeta_i) = 4p(1-p)$. Therefore, we know that $\Xi_{ii} = 4p(1-p)$, and what is left is to consider the off-diagonals. From our derivation above for $(i,j)$ pairs, we know that for the off-diagonals,
\begin{align*}
   \Xi = \mathbb{E}[\zeta \zeta^T] - \mu \mu^T = \frac{8p^2}{\pi} \arcsin{(X)}, 
\end{align*}
where $\arcsin{(X)}$ is written to be the entry-wise function on the matrix $X$, with Taylor expansion,
\begin{align*}
    \arcsin{(X)} = X + \sum_{k=1}^\infty \frac{\binom{2k}{k}}{4^k (2k + 1)} X^{\circ (2k+1)}.
\end{align*}
Let $\alpha_\kappa$ be the maximum entry in the off-diagonals $X - I$. Here we note the dependence on $\kappa$; by \Cref{prop:X_norm_covariance_min}, as $\kappa$ increases in the objective function relative to the other tradeoff parameters $\gamma$, $\eta$, $\alpha_\kappa$ decreases. Let also $f(X):= \sum_{k=1}^\infty \frac{\binom{2k}{k}}{4^k (2k + 1)} X^{\circ (2k+1)}$, where $X^{\circ \beta}$ denotes all entries of $X$ being raised to the power of $\beta$. 

Denote the off-diagonal matrix of $X$ by $\Tilde{R}$. Therefore,
\begin{align*}
     \Xi &= \frac{8 p^2}{\pi} (\arcsin{(X)}) + 4p(1-2p)I \\
     &=  4p^2((\frac{2}{\pi}\arcsin{(X)} -I) + I) + 4p(1-2p)I \\
     &= \frac{8 p^2}{\pi} (\Tilde{R} + \calO(\Tilde{R}^{\circ 3})) + 4p(1-p)I, \\
     &= \frac{8 p^2}{\pi} X + \frac{8 p^2}{\pi} \calO(\Tilde{R}^{\circ 3})  + (4p(1-p) - \frac{8p^2}{\pi})I.
\end{align*} 

Let $\psi_\kappa = n \alpha_\kappa^{3}$, then $\| \calO(\Tilde{R}^{\circ 3})\|_{\text{op}} \leq \psi_\kappa$, and $\Tilde{R}^{\circ 3} \preceq \frac{\psi_\kappa}{\delta} X$. In particular, for $\kappa$ large, if $\alpha_\kappa = \calO(n^{-1/3})$, then $\psi_\kappa = \calO(1).$ It is easy to then see that
\begin{align*}
   \left( \frac{8p^2}{\pi}  + \frac{8p^2 \psi_\kappa}{\pi \delta} + \frac{4p(1-p) - 8p^2/\pi}{\delta}  \right)X \succeq \Xi. 
\end{align*}
\end{proof}
We now prove \Cref{corr:approximation_bound} as a corollary. First, we restate it below.
\begingroup

\renewcommand{\thetheorem}{\ref{corr:approximation_bound}}

\addtocounter{theorem}{-1} 

\begin{theorem}
    Let $\Xi := \cov(\zeta)$ be the covariance matrix of the associated Rademacher random variables. Let $X^*$ be the optimal achievable covariance matrix of the design in \Cref{thm:mse_bound_general_HT}. 
    Let also $X_{\text{SDP}}$ denote the SDP solution matrix. We assume that $X_{\text{SDP}}$ is non-degenerate, i.e., $X_{\text{SDP}} \succeq \delta_\kappa I$, $\delta_\kappa > 0$. For $c_\kappa =\frac{1}{4p(1-p)} \left( \frac{8p^2}{\pi}  + \frac{8p^2 \psi_\kappa}{\pi \delta_\kappa} + \frac{4p(1-p) - 8p^2/\pi}{\delta_\kappa}  \right)$ for $\psi_\kappa= n \alpha_\kappa^{3},$ where $\alpha_\kappa$ is the maximum off-diagonal entry of $X_{\text{SDP}}$, in absolute value, 
    \begin{align*}
        \operatorname{Tr}(L\Xi) + \operatorname{Tr}(L^\dagger\Xi ) +  \|\Xi \|_{q} + \operatorname{Tr}(\mathbf{1} \mathbf{1}^T \Xi) \leq c_\kappa \left( \operatorname{Tr}(LX^*) +  \operatorname{Tr}(L^\dagger X^*)+  \| X^*\|_{q}  + \operatorname{Tr}(\mathbf{1} \mathbf{1}^T X^*)\right).
    \end{align*}
\end{theorem}
\endgroup

\begin{proof}
We see that $L, L^{\dagger}, \mathbf{1} \mathbf{1}^T$ are PSD. It is enough to notice that for PSD matrices $A, B, C$, with $B \succeq C$, $\operatorname{Tr}(A(B-C)) \geq 0$. Finally, notice that for $X = 4p(1-p) X_{\operatorname{SDP}}$, where $X_{\operatorname{SDP}}$ is the SDP solution, and for $X^*$ is the optimal achievable covariance matrix of a design, $\operatorname{Tr}(LX) +  \operatorname{Tr}(L^\dagger X)+  \| X\|_{q}  + \operatorname{Tr}(\mathbf{1} \mathbf{1}^T X) \leq \operatorname{Tr}(LX^*) +  \operatorname{Tr}(L^\dagger X^*)+  \| X^*\|_{q}  + \operatorname{Tr}(\mathbf{1} \mathbf{1}^T X^*).$   
\end{proof}

\begin{proposition}[{\citealp[Theorem 6.4]{harshaw2024balancing}}] \label{thm_gsw}
The Gram-Schmidt Walk design with parameter $\lambda \in [0,1]$ provides balance-robustness guarantee
$\| \operatorname{Cov}(Ax) \|_\infty \leq \frac{\max_{i \in [n]} \| x_i\| _2^2}{1-\lambda}$ and $\| \operatorname{Cov}(x) \|_\infty \leq \frac{1}{\lambda}$.
\end{proposition}

We now state and prove the general version of \Cref{thm:gsw}, allowing for $p \neq 1/2$.
\begin{theorem} \label{thm:gsw_nonsym}
Let $A:=(\Tilde{\eta}L^{\dagger} +\Tilde{\gamma} L + \Delta \mathbf{1} \mathbf{1}^T)^{1/2}$, where $\Tilde{\eta} = \eta[\beta_1^2 + \beta_2^2]$, $\Tilde{\gamma} = \gamma \beta_1^2$, $\Tilde{\kappa} =\kappa[\beta_1^2 + \beta_2^2]$, and $\Delta = (a \beta_1 + b \beta_2)^2$.
The Gram-Schmidt Walk algorithm, with randomization parameter $\lambda$, returns an assignment $x$ such that $\|\operatorname{Cov}(Ax)\|_{{p'}} \leq \omega_A^2/(1-\lambda)\|I \|_{{p'}}$, and $\|\operatorname{Cov}(x)\|_{q}\leq \|I\|_{q}/\lambda=n^{1/q}/\lambda$, for $\omega_A := \max_{i\in n} \|A_i\|$, $A_i$ the column $i$ of the matrix $A$. Thus, for $q \geq 1$,
\[
\sup_{\chi(\eta, \gamma, \kappa)} \mathbb{E}[(\tau - \hat{\tau})^2] \leq 7 \left[\mathbb{E}[\|A(x-\mu)\|^2] + \Tilde{\kappa}\|X\|_{q} \right] \leq 7 \left[\frac{\omega_A^2 n}{1-\lambda} + \Tilde{\kappa}\frac{n^{1/q}}{\lambda}\right],
\] where $\mu_i = \mathbb{E}[x_i]$, for all $i \in [n].$
This is minimized for $\lambda = \frac{\sqrt{\Tilde{\kappa} n^{1/q}}}{ \sqrt{\omega_A^2n} + \sqrt{\Tilde{\kappa}n^{1/q}}}$. 
\end{theorem}

\begin{proof} \label{pf_thm_gsw}
From \Cref{thm:mse_bound_general_HT}, we have 
\begin{align*}
    \sup_{\chi(\eta, \gamma, \kappa)} \mathbb{E}[(\tau - \hat{\tau})^2] \leq 7 \left[ \operatorname{Tr}(X AA^T) + \Tilde{\kappa} \| X\|_q \right]= 7 \left[\mathbb{E}[\|A(x-\mu)\|^2 + \Tilde{\kappa} \| X\|_q \right]
\end{align*}

From the proof of \citet[Theorem 6.4]{harshaw2024balancing}, $\operatorname{Cov}(x) \preceq \lambda^{-1}I$, and (2) $\operatorname{Cov}(Ax) \preceq \omega_A^2 (1-\lambda)^{-1}I$. Thus, from (1) we have $\| X \|_q \leq \lambda^{-1}\|I \|_q = \lambda^{-1} n^{1/q}$. From (2), $\operatorname{Tr}(XAA^T) \leq \omega_A^2 (1-\lambda)^{-1} \operatorname{I} = \omega_A^2 (1-\lambda)^{-1} n$. Therefore, 
\begin{align*}
    7 \left[ \operatorname{Tr}(X AA^T) + \Tilde{\kappa} \| X\|_q \right] \leq 7 \left[  \frac{\omega_A^2 n}{(1-\lambda)}+ \Tilde{\kappa} \frac{n^{1/q}}{\lambda} \right].
\end{align*}

Next, to minimize the right-hand side above, we take the derivative with respect to $\lambda$ and set it equal to zero. Let $m = \omega_A^2 n (1-\lambda)^{-1}+ \Tilde{\kappa} n^{1/q}\lambda^{-1}$. Then, 
\begin{align*}
    \frac{dm}{d\lambda} &= \frac{\omega_A^2 n}{(1-\lambda)^2} - \frac{\tilde{\kappa} n^{1/q}}{\lambda^2} = 0 \\
    \lambda &= \frac{\sqrt{\tilde{\kappa}n^{1/q}}}{\sqrt{\tilde{\kappa}n^{1/q}} \pm \sqrt{\omega_A^2 n}}.
\end{align*}

Since $\lambda \in [0,1]$, $\lambda = \frac{\sqrt{\tilde{\kappa}n^{1/q}}}{\sqrt{\tilde{\kappa}n^{1/q}} + \sqrt{\omega_A^2 n}}.$ Additionally, since $\omega_A^2 n(1-\lambda)^{-1}+ \Tilde{\kappa} n^{1/q}\lambda^{-1}$ is convex in $\lambda$, $\lambda = \frac{\sqrt{\tilde{\kappa}n^{1/q}}}{\sqrt{\tilde{\kappa}n^{1/q}} + \sqrt{\omega_A^2 n}}$ is its minimizer, yielding the claim.
\end{proof}

The result above also holds when $A^T:= [\Tilde{\eta}^{1/2}L^{\dagger/2} \quad \Tilde{\gamma}^{1/2} L^{1/2}\quad \Delta^{1/2} {\mathbf{1} \mathbf{1}^T}^{1/2}].$

\section{Simulations}

\begin{figure}[h!]
    \centering
    \includegraphics[width=\linewidth]{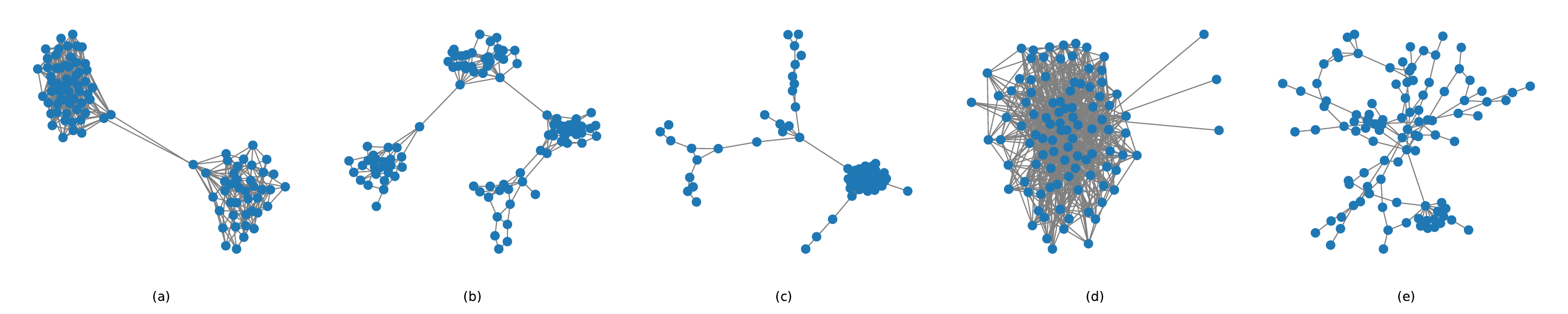}
    \caption{From left to right: Graphs generated from (a): SBM (within cluster connection probability $20/n$, across connection probability $\frac{0.1}{n}$), with two clusters of equal membership assignment probability; (b): SBM (within cluster connection probability $20/n$, across connection probability $\frac{0.1}{n}$) with four clusters of equal membership assignment probability; (c): SBM (within cluster connection probability $20/n$, across connection probability $\frac{0.1}{n}$) with four clusters of membership assignment probabilities 0.7, 0.1, 0.1, 0.1; (d): Barabasi-Albert (preferential-attachment) with initial $n/10$ (=10) nodes connected according to the Erdös-Rényi model with connection probability $10/n$; Graph 5: Erdös-Rényi with connection probability $2/n$.}
    \label{fig:all_graphs}
\end{figure}

\begin{table}[ht]
\centering
\begin{tabular}{@{}lp{11cm}@{}}
\toprule
\textbf{Design} & \textbf{Description} \\
\midrule
SDP & SDP solution $X^*$ \\
SDP + GaussRound & SDP followed by Gaussian rounding \\
SDP + QuantRound & SDP followed by quantile rounding (see \Cref{sec-quantile_rounding})\\
GSW & Adapted Gram-Schmidt Walk design \citep{harshaw2024balancing} \\
GSW + covariates & Adapted Gram-Schmidt Walk design \citep{harshaw2024balancing} using covariate-encoding directly instead of $L^\dagger$ \\
Louvain & Cluster-randomization with default Python implementation of the Louvain clustering algorithm \\
$\epsilon$-net & Cluster-randomization under $(\epsilon{=}1)$-net clustering \citep{ugander2013graph} \\
Spectral & Cluster-randomization under spectral clustering \\
Causal & Cluster-randomization under Causal clustering \citep{viviano2023causal} with a grid-search for optimal hyperparameters \\
Bernoulli & Uniform Bernoulli randomization \\
\bottomrule
\end{tabular}
\caption{Key of designs compared in this paper.}
\label{tab:design_methods}
\end{table}

\input{content/real_data_simulations}

\subsection{Worst-case MSE bounds comparisons on simulated networks}
Throughout this subsection and the next, we compare the performance of our designs against various designs from the literature. We describe the different designs we compare in \Cref{tab:design_methods}. We note that the different cluster-randomized designs were also considered in \citep{viviano2023causal}.

\begin{figure}
    \centering
    \begin{subfigure}[b]{0.3\textwidth}
        \centering
        \includegraphics[width=\linewidth]{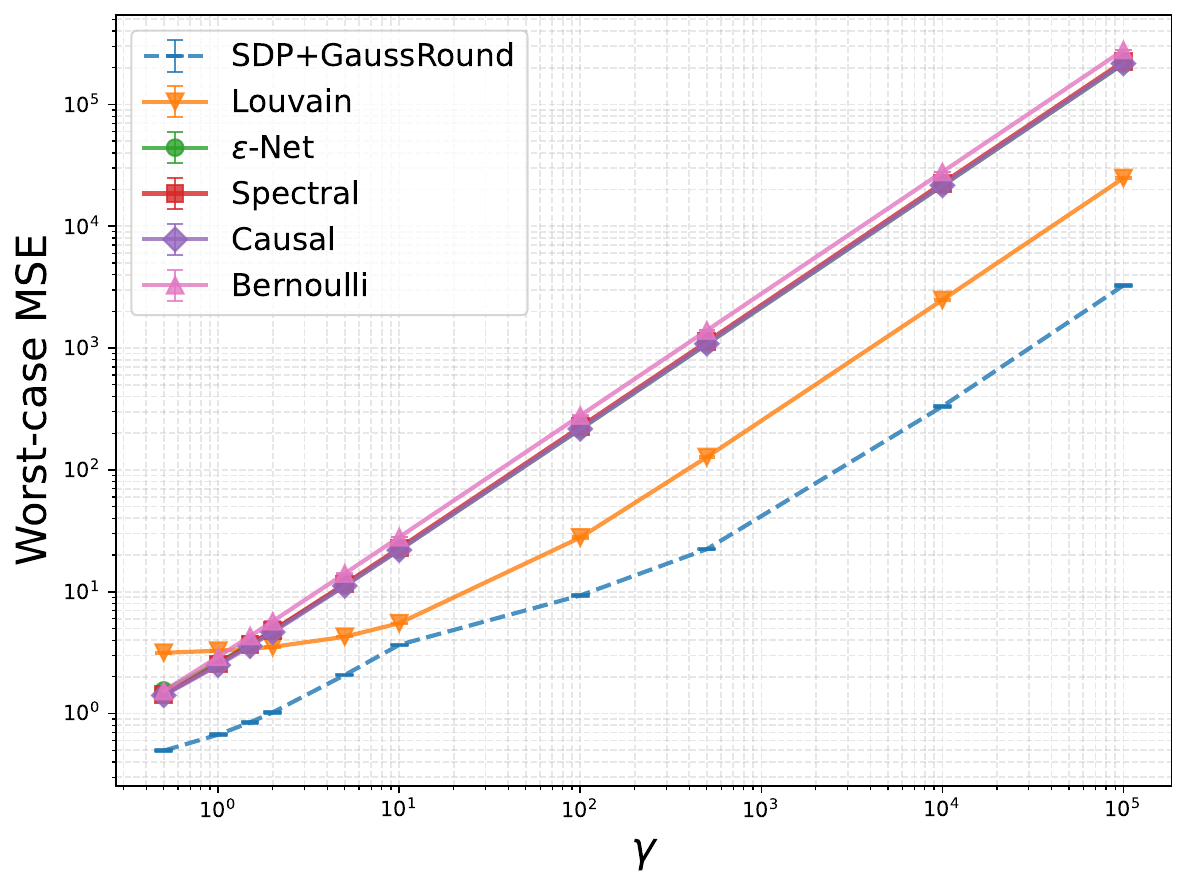}
    \end{subfigure}
    \hfill
    \begin{subfigure}[b]{0.3\textwidth}
        \centering
        \includegraphics[width=\linewidth]{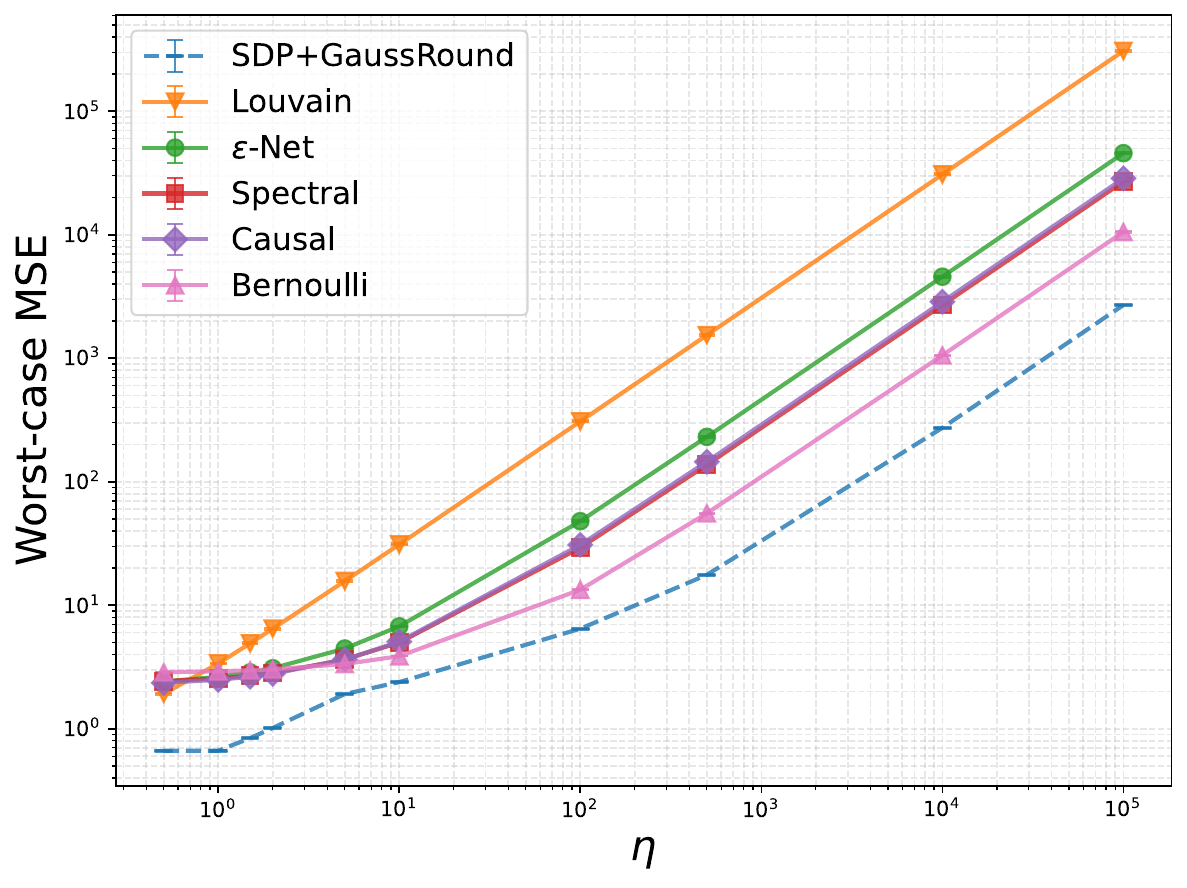}
    \end{subfigure}
    \hfill
    \begin{subfigure}[b]{0.3\textwidth}
        \centering
        \includegraphics[width=\linewidth]{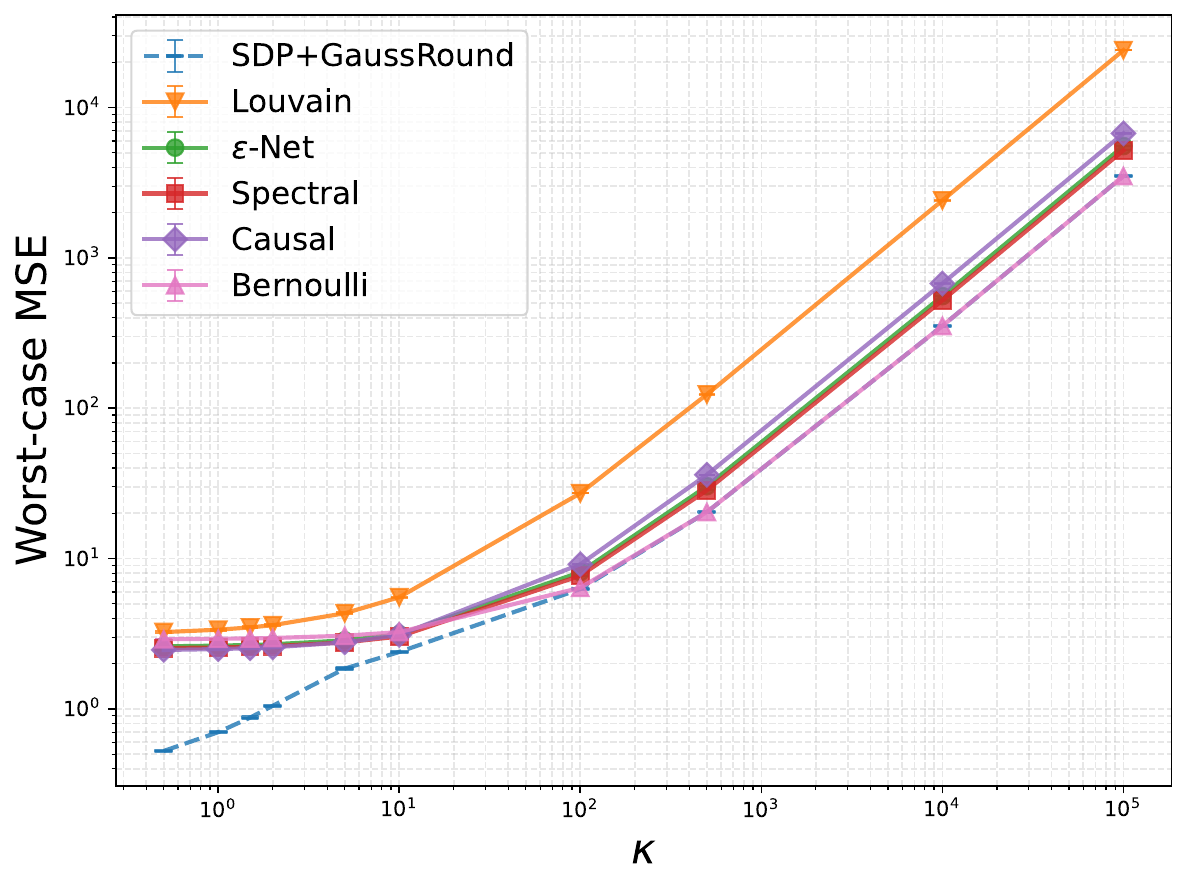}
    \end{subfigure}
    \caption{Comparison of worst-case MSE bounds between our designs, cluster randomization designs (with various clusterings) and Bernoulli randomization averaged over graphs generated from an SBM (within cluster connection probability $20/n$, across connection probability $\frac{0.1}{n}$), with two clusters of equal membership assignment probability (see Graph 1 in \Cref{fig:all_graphs}). The other two trade-off parameters are set to one. Error bars represent the standard error.}
    \label{fig:worst_case_graph_one}
\end{figure}

\begin{figure}
    \centering
    \begin{subfigure}[b]{0.3\textwidth}
        \centering
        \includegraphics[width=\linewidth]{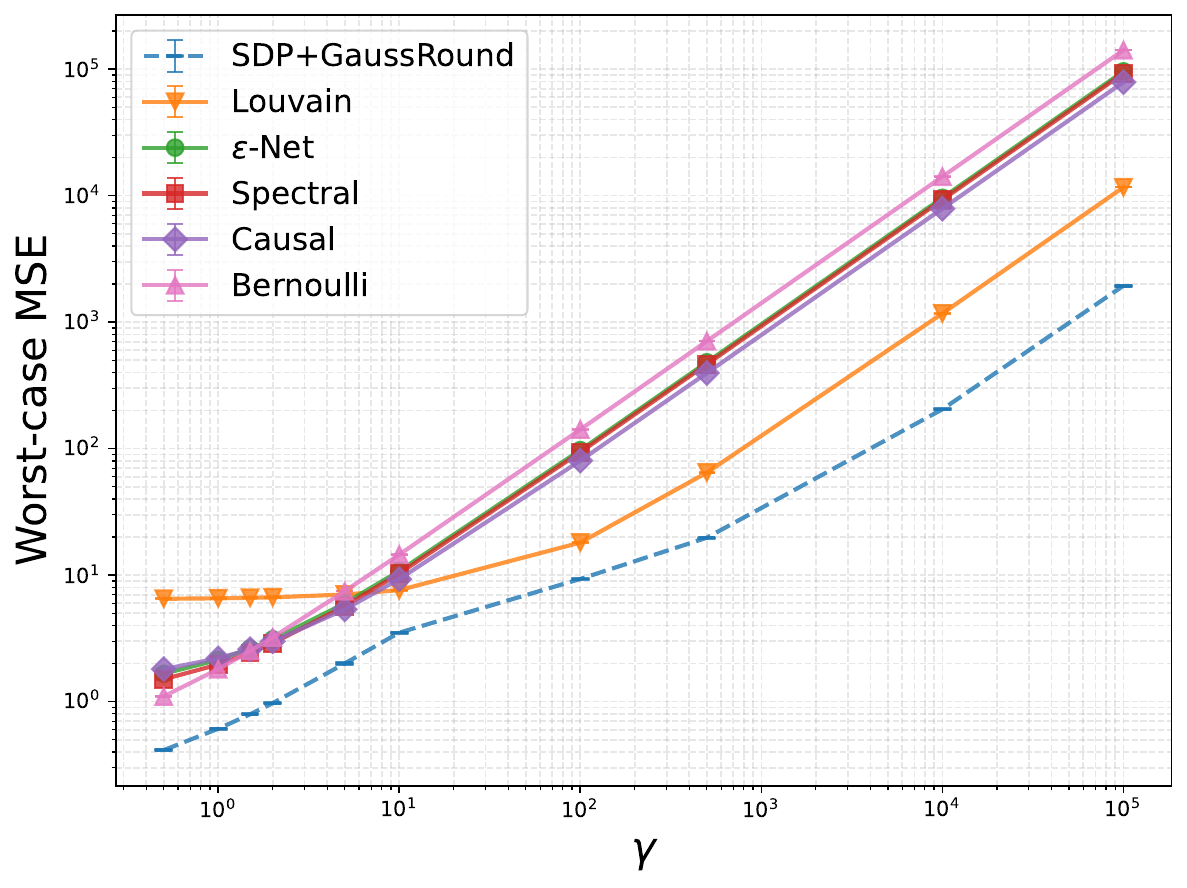}
    \end{subfigure}
    \hfill
    \begin{subfigure}[b]{0.3\textwidth}
        \centering
        \includegraphics[width=\linewidth]{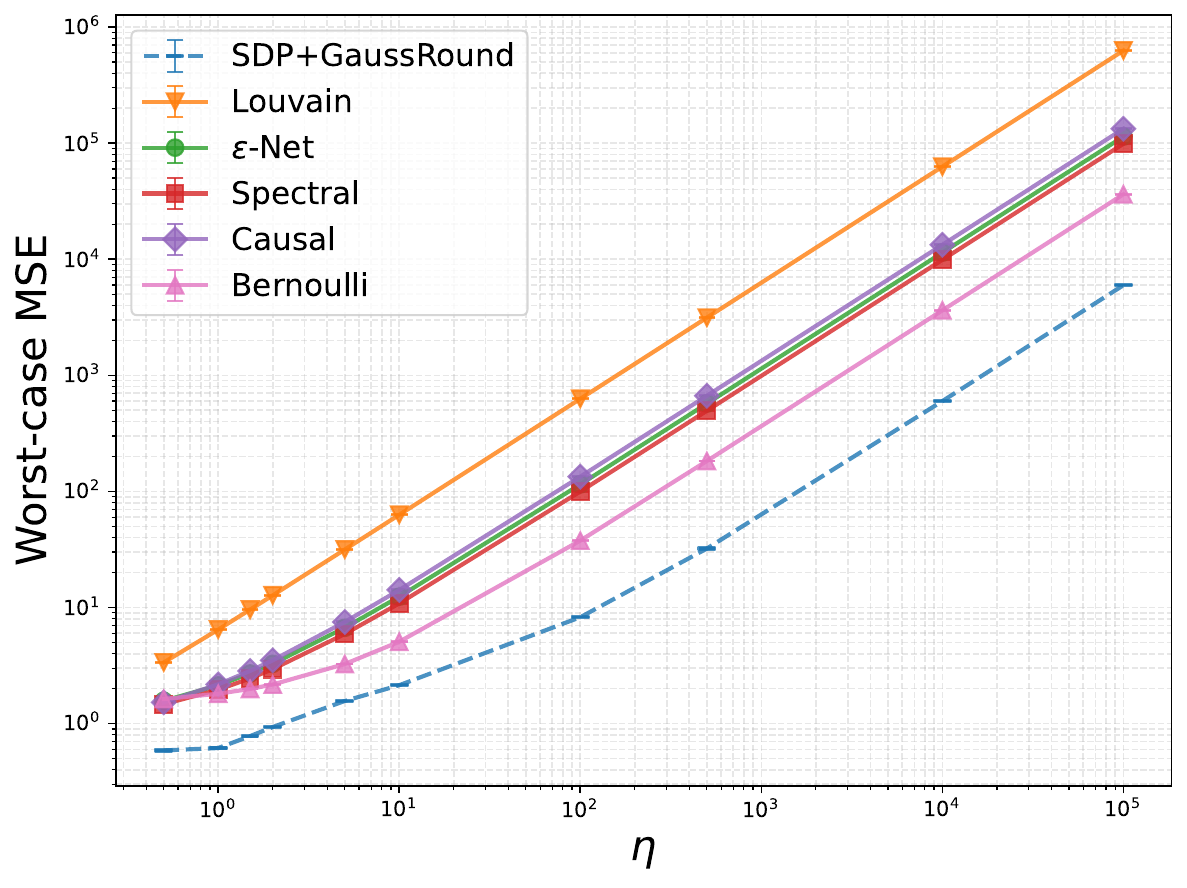}
    \end{subfigure}
    \hfill
    \begin{subfigure}[b]{0.3\textwidth}
        \centering
        \includegraphics[width=\linewidth]{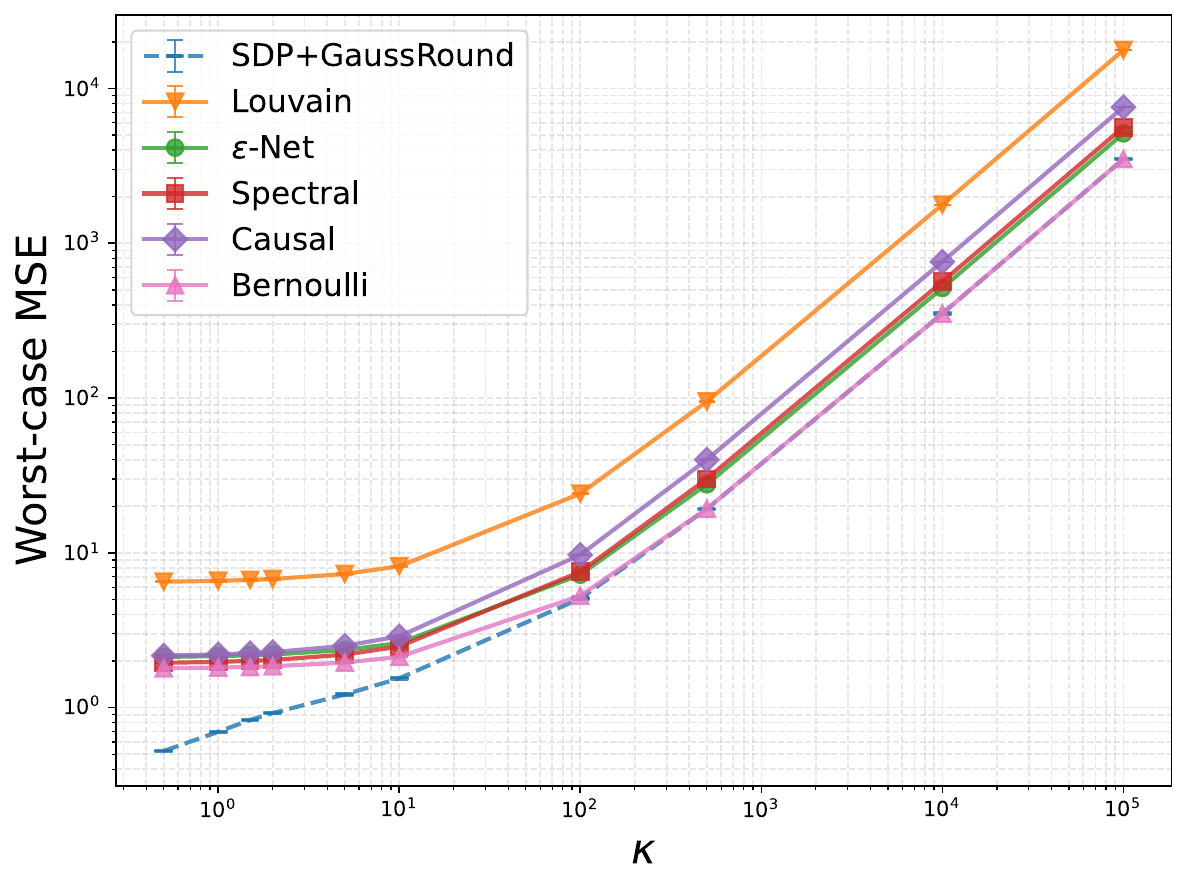}
    \end{subfigure}
    \caption{Comparison of worst-case MSE bounds between our designs, cluster randomization designs (with various clusterings) and Bernoulli randomization averaged over graphs generated from an SBM (within cluster connection probability $20/n$, across connection probability $\frac{0.1}{n}$), with two clusters of equal membership assignment probability (see Graph 2 in \cref{fig:all_graphs}). The other two trade-off parameters are set to one. Error bars represent the standard error.}
    \label{fig:worst_case_graph_two}
\end{figure}

\begin{figure}
    \centering
    \begin{subfigure}[b]{0.3\textwidth}
        \centering
        \includegraphics[width=\linewidth]{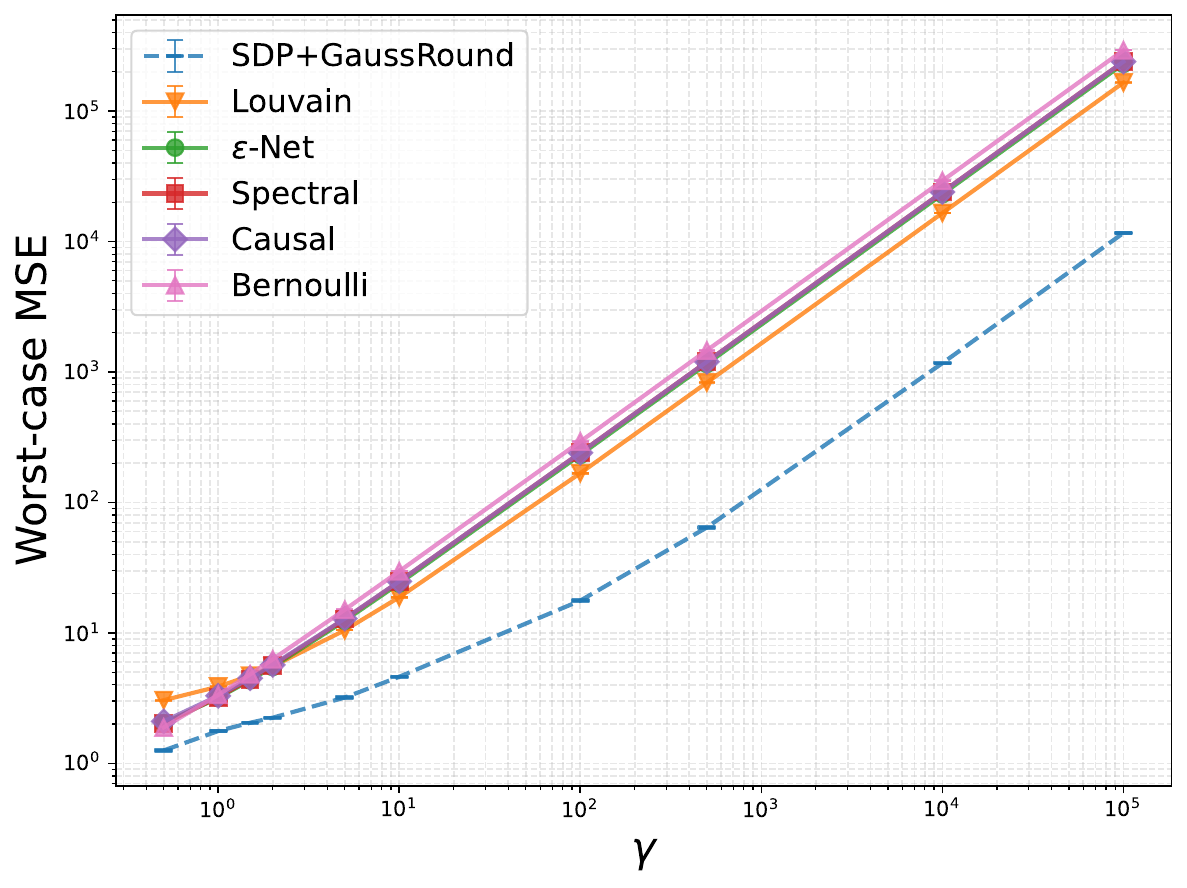}
    \end{subfigure}
    \hfill
    \begin{subfigure}[b]{0.3\textwidth}
        \centering
        \includegraphics[width=\linewidth]{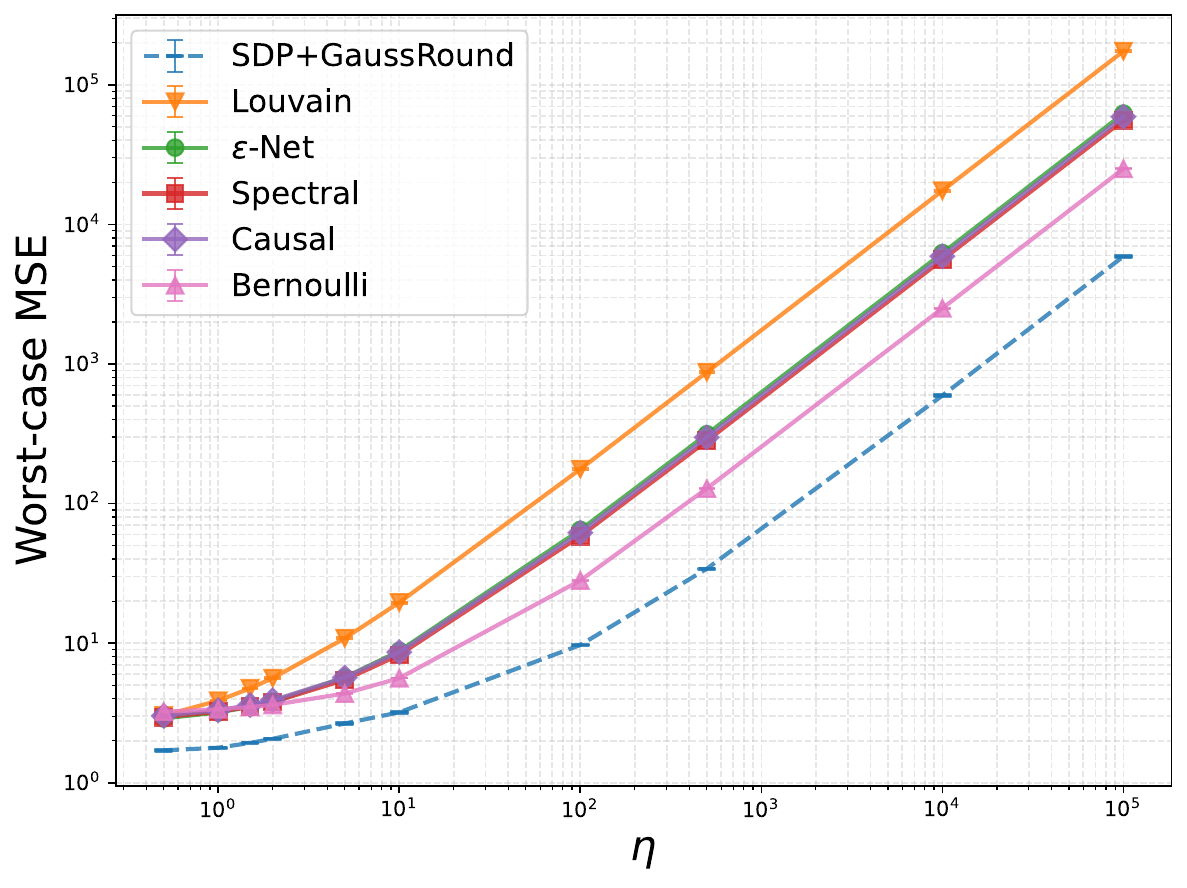}
    \end{subfigure}
    \hfill
    \begin{subfigure}[b]{0.3\textwidth}
        \centering
        \includegraphics[width=\linewidth]{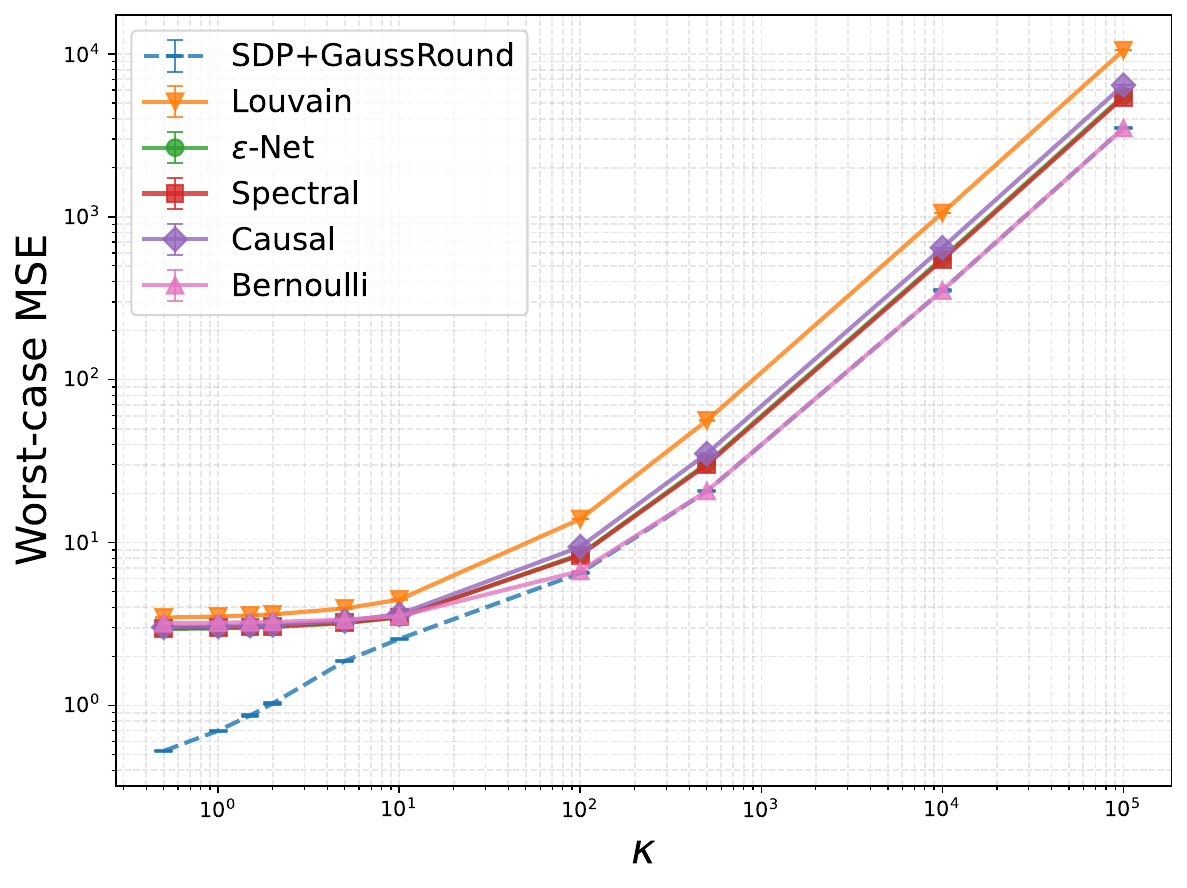}
    \end{subfigure}
    \caption{Comparison of worst-case MSE bounds between our designs, cluster randomization designs (with various clusterings) and Bernoulli randomization averaged over graphs generated from an SBM (within cluster connection probability $20/n$, across connection probability $\frac{0.1}{n}$) with four clusters of membership assignment probability (0.70, 0.1, 0.1, 0.1) (see Graph 3 in \cref{fig:all_graphs}). The other two trade-off parameters are set to one. Error bars represent the standard error.}
\label{fig:worst_case_graph_three}
\end{figure}

\begin{figure}
    \centering
    \begin{subfigure}[b]{0.3\textwidth}
        \centering
        \includegraphics[width=\linewidth]{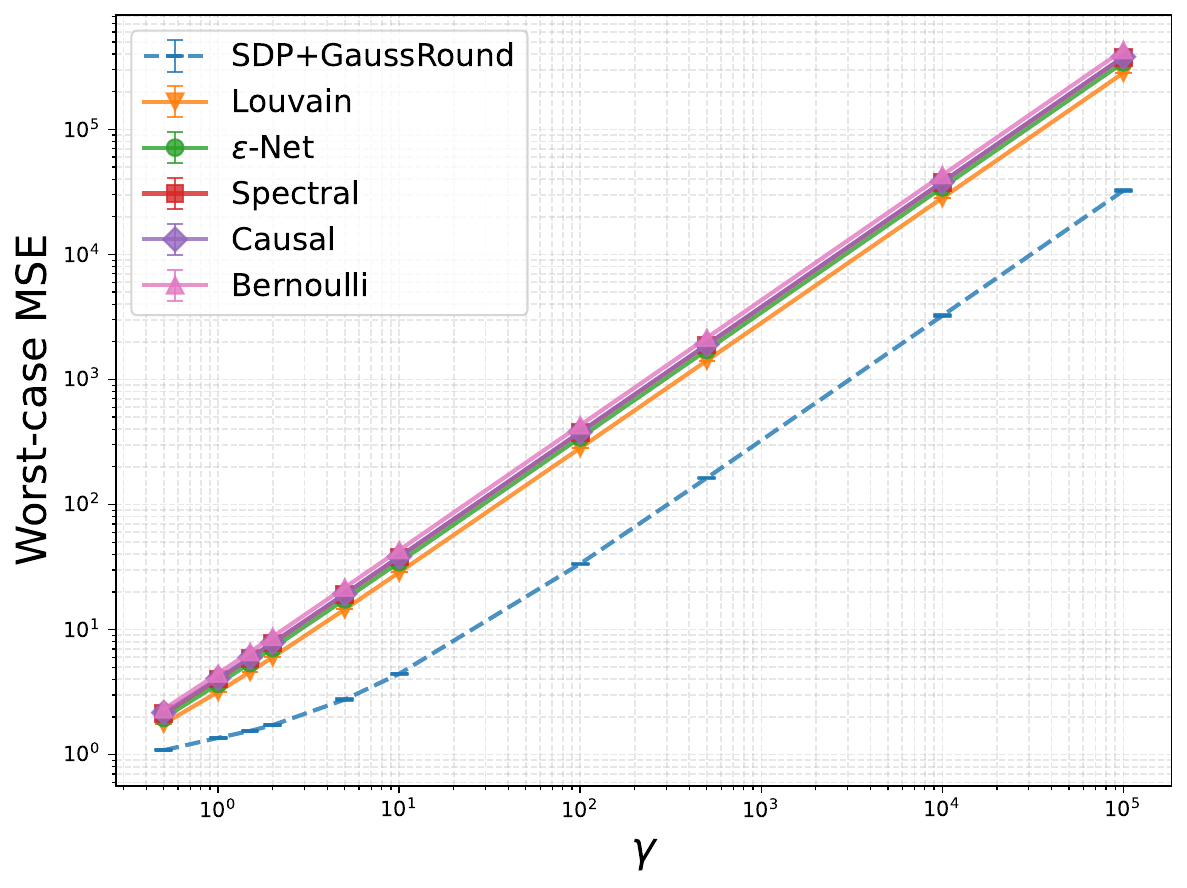}
    \end{subfigure}
    \hfill
    \begin{subfigure}[b]{0.3\textwidth}
        \centering
        \includegraphics[width=\linewidth]{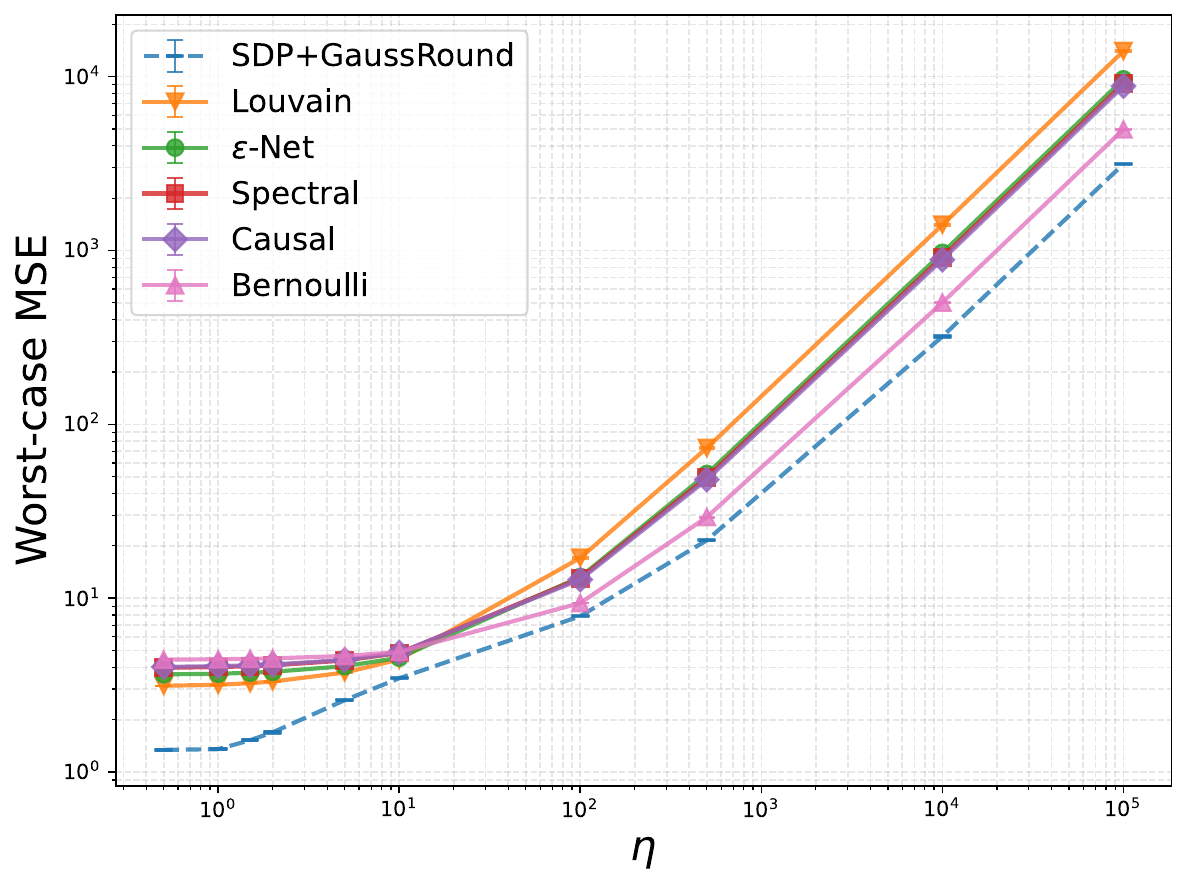}
    \end{subfigure}
    \hfill
    \begin{subfigure}[b]{0.3\textwidth}
        \centering
        \includegraphics[width=\linewidth]{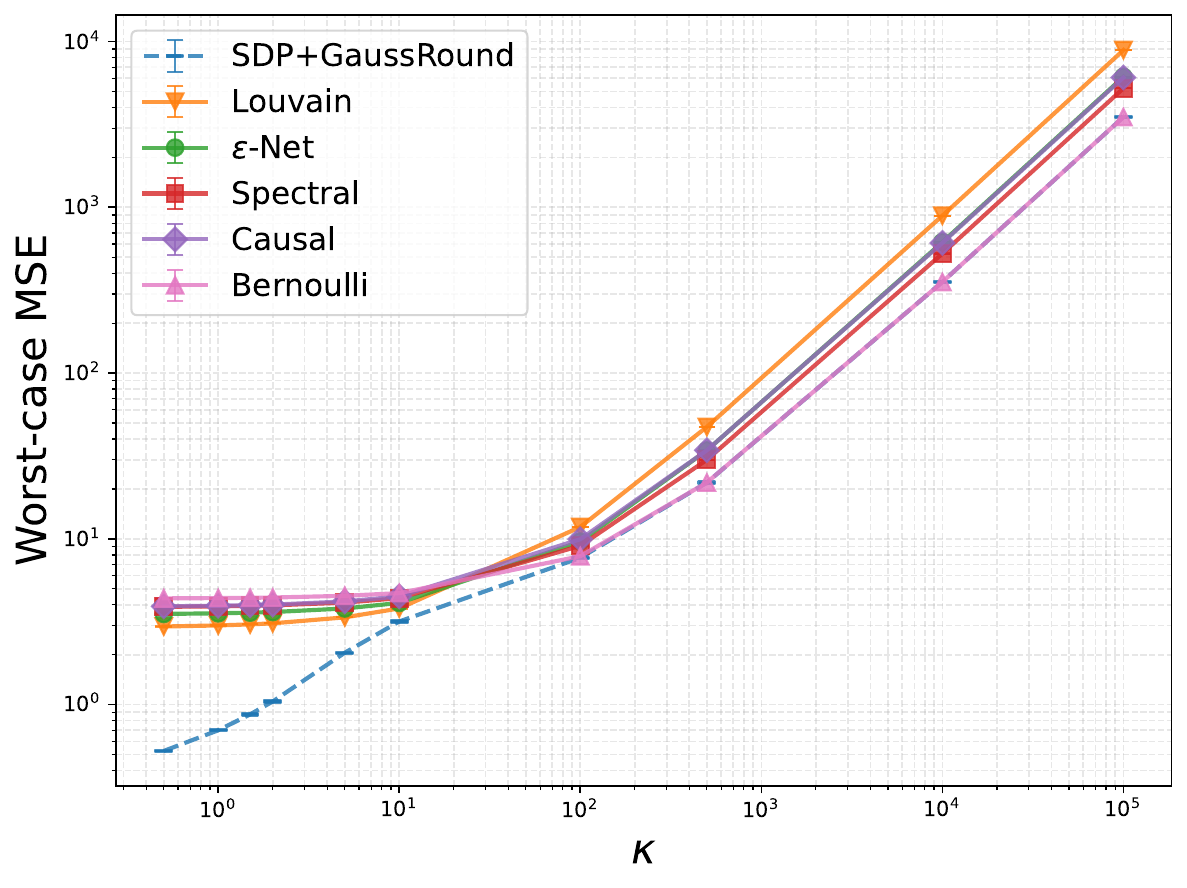}
    \end{subfigure}
    \caption{Comparison of worst-case MSE bounds between our designs, cluster randomization designs (with various clusterings) and Bernoulli randomization averaged over graphs generated from a Barabasi-Albert (preferential-attachment) model with initial $n/10$ (=10) nodes connected according to the Erdös-Rényi model with connection probability $10/n$ (see Graph 4 in \cref{fig:all_graphs}). The other two trade-off parameters are set to one. Error bars represent the standard error.}
    \label{fig:worst_case_graph_four}
\end{figure}

\begin{figure}
    \centering
    \begin{subfigure}[b]{0.3\textwidth}
        \centering
        \includegraphics[width=\linewidth]{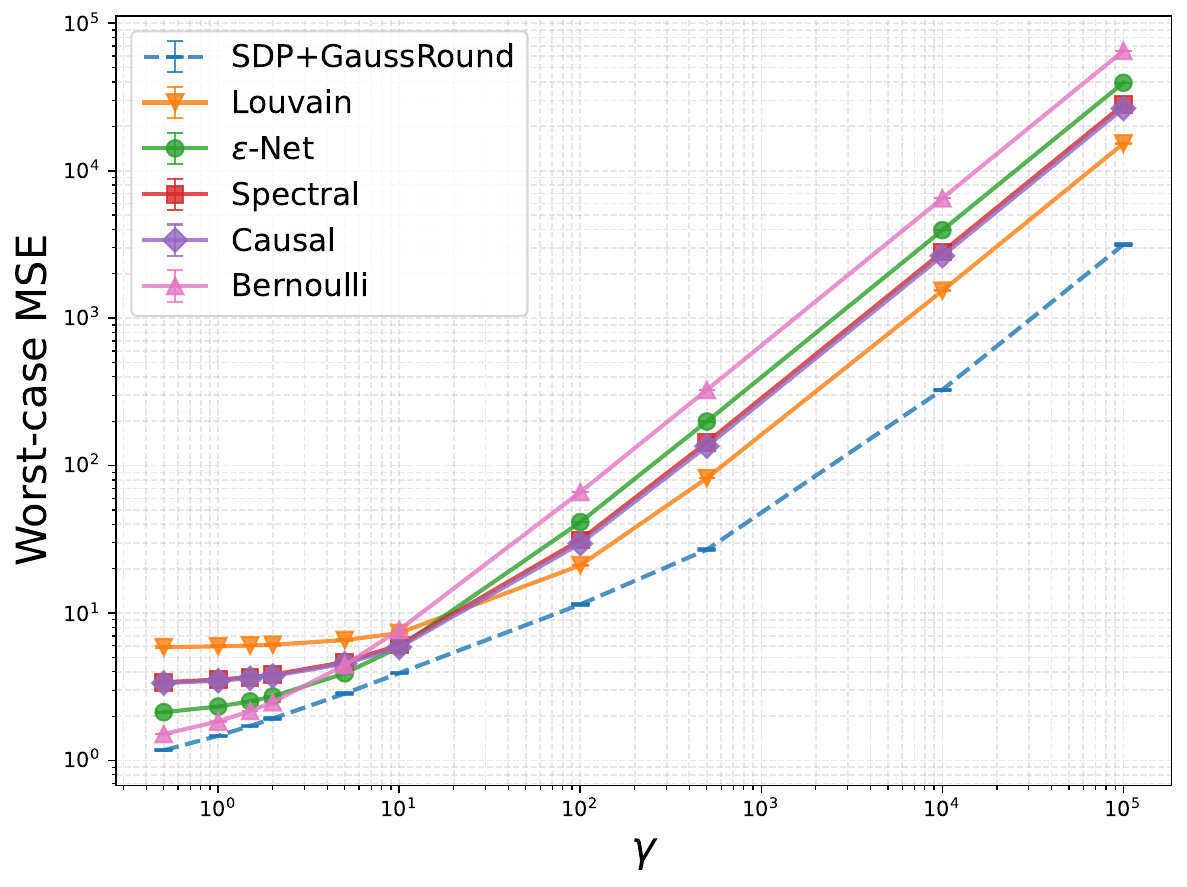}
    \end{subfigure}
    \hfill
    \begin{subfigure}[b]{0.3\textwidth}
        \centering
        \includegraphics[width=\linewidth]{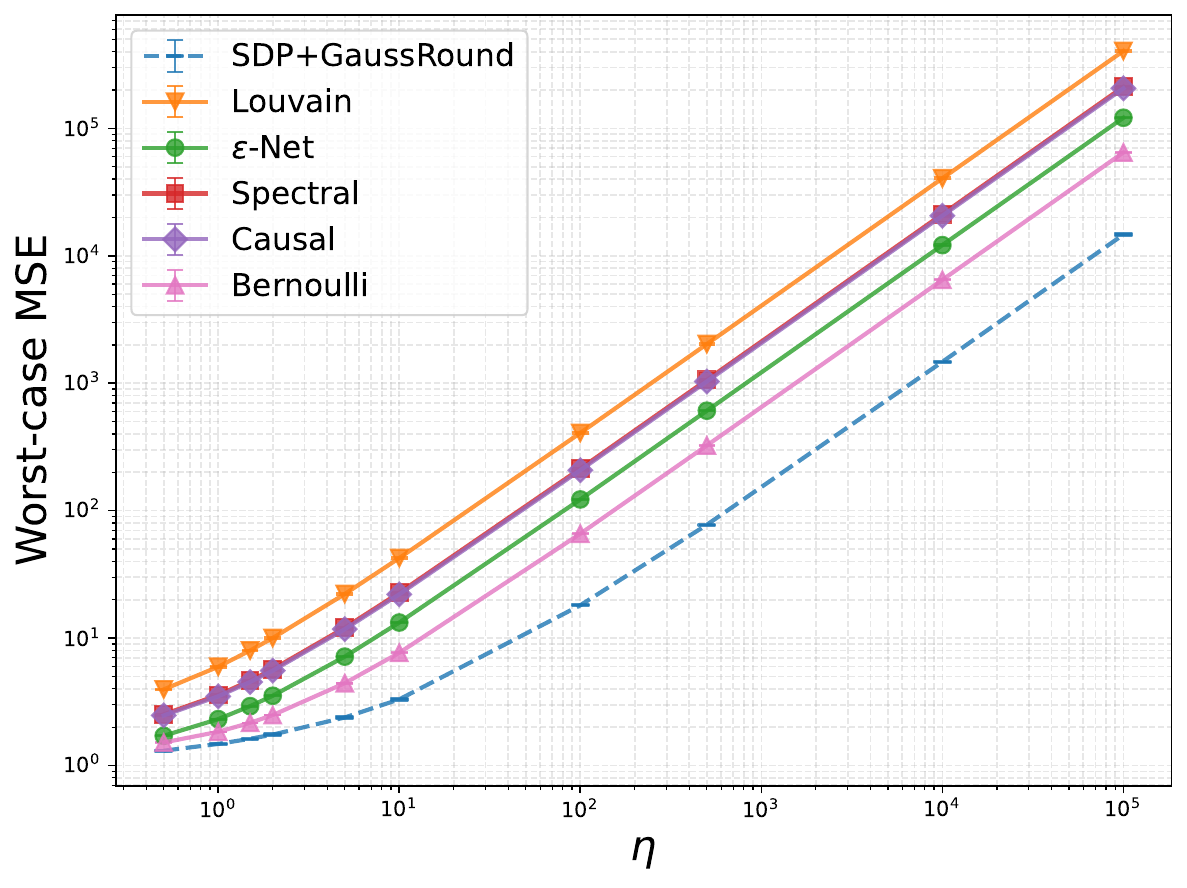}
    \end{subfigure}
    \hfill
    \begin{subfigure}[b]{0.3\textwidth}
        \centering
        \includegraphics[width=\linewidth]{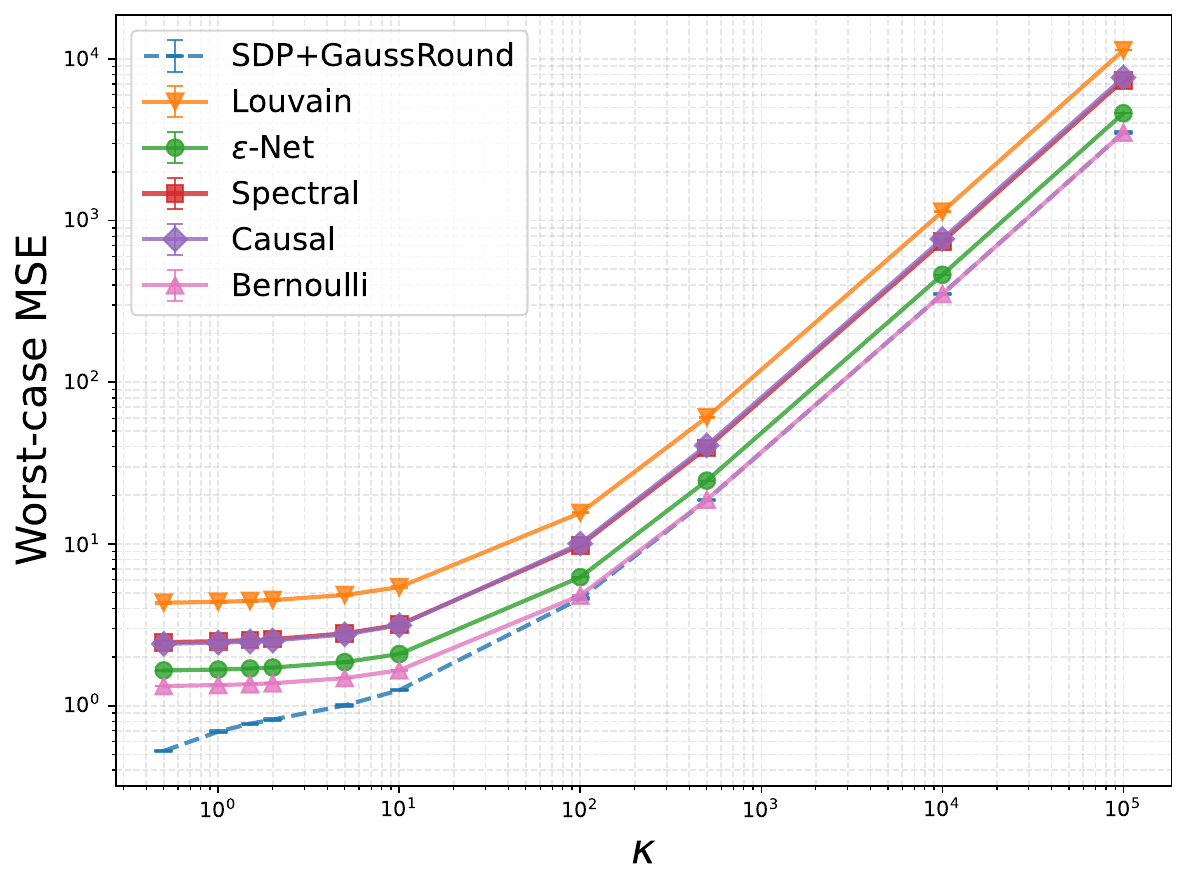}
    \end{subfigure}
    \caption{Comparison of worst-case MSE bounds between our designs, cluster randomization designs (with various clusterings) and Bernoulli randomization averaged over graphs generated from an Erdös-Rényi model with connection probability $2/n$ (see Graph 5 in \cref{fig:all_graphs}). The other two trade-off parameters are set to one. Error bars represent the standard error.}
    \label{fig:worst_case_graph_five}
\end{figure}

In \Cref{fig:worst_case_graph_one,fig:worst_case_graph_two,fig:worst_case_graph_three,fig:worst_case_graph_four,fig:worst_case_graph_five}, we compare the worst case bounds by plugging into the bound in \Cref{thm:mse_bound_general_HT} the covariance matrices from various designs. We compare the worst-case MSE bounds for five different graph generation models, as described in \Cref{fig:all_graphs}, averaged across 1000 graph instances of each graph model. 
For procedures where we cannot directly compute the design covariance, we generate the sample covariance matrices of the designs over 1000 instances of the design vectors for each graph instance. 
We note that the gap between the optimal solution from solving the SDP and the other designs are due to unattainability of the optimal SDP solution given that the SDP formulation is a relaxation of the problem. In these settings, we set the fixed tradeoff parameters to be 1.

\begin{figure}[ht]
    \centering
    \includegraphics[width=0.5\linewidth]{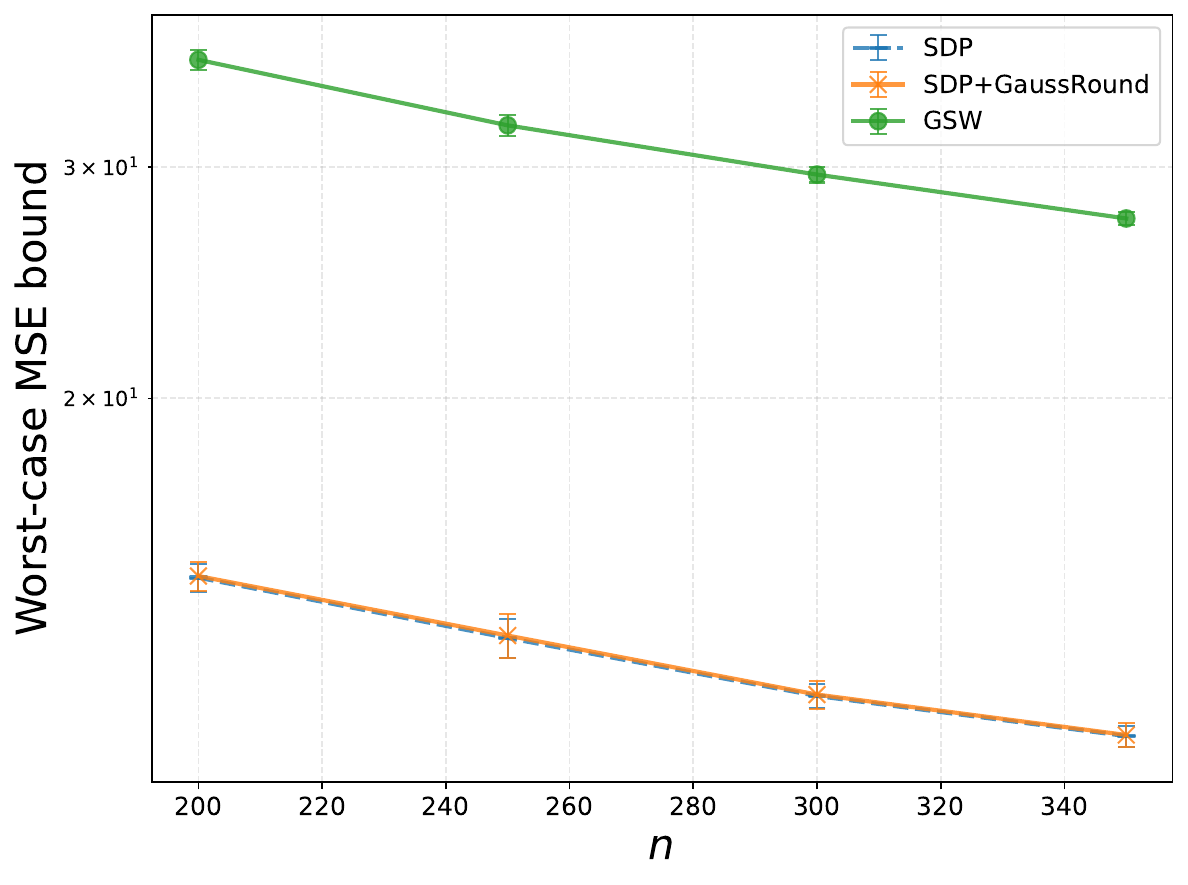}
    \caption{Comparisons with adapted Gram-Schmidt algorithm averaged over graphs generated from an SBM (within cluster connection probability $20/n$, across connection probability $\frac{0.1}{n}$), with two clusters of equal membership assignment probability (see Graph 1 in \cref{fig:all_graphs}), for different $n$. Error bars represent the standard error.}
    \label{fig:gsw_comparisons_graph1}
\end{figure}

\begin{figure}[ht]
    \centering
    \includegraphics[width=0.5\linewidth]{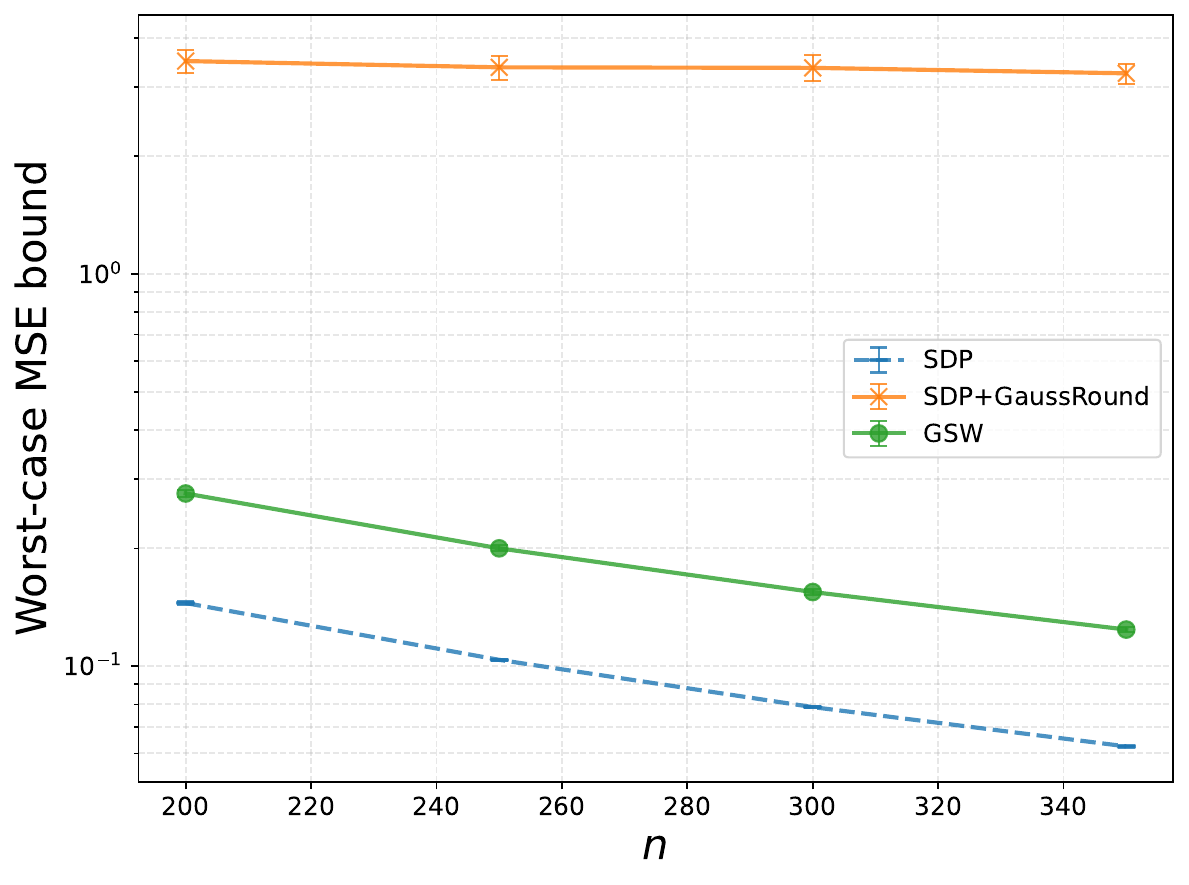}
    \caption{Comparisons with adapted Gram-Schmidt algorithm averaged over graphs generated from an SBM (within cluster connection probability $20/n$, across connection probability $\frac{0.1}{n}$), with two clusters of equal membership assignment probability (see Graph 2 in \cref{fig:all_graphs}), for different $n$. Error bars represent the standard error.}
    \label{fig:gsw_comparisons_graph2}
\end{figure}

\begin{figure}[ht]
    \centering
    \includegraphics[width=0.5\linewidth]{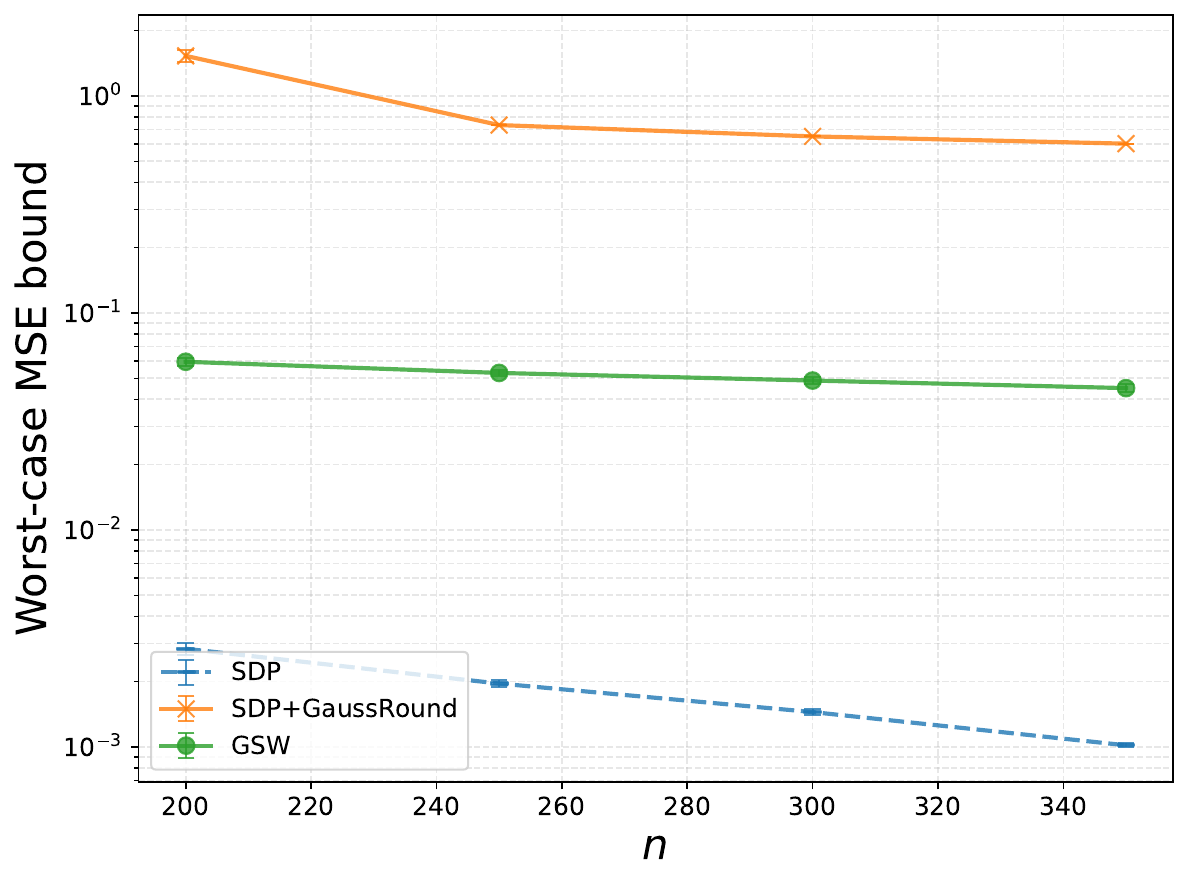}
    \caption{Comparisons with adapted Gram-Schmidt algorithm averaged over graphs generated from an SBM (within cluster connection probability $20/n$, across connection probability $\frac{0.1}{n}$) with four clusters of membership assignment probability (0.70, 0.1, 0.1, 0.1) (see Graph 3 in \cref{fig:all_graphs}), for different $n$. Error bars represent the standard error.}
    \label{fig:gsw_comparisons_graph3}
\end{figure}

\begin{figure}[ht]
    \centering
    \includegraphics[width=0.5\linewidth]{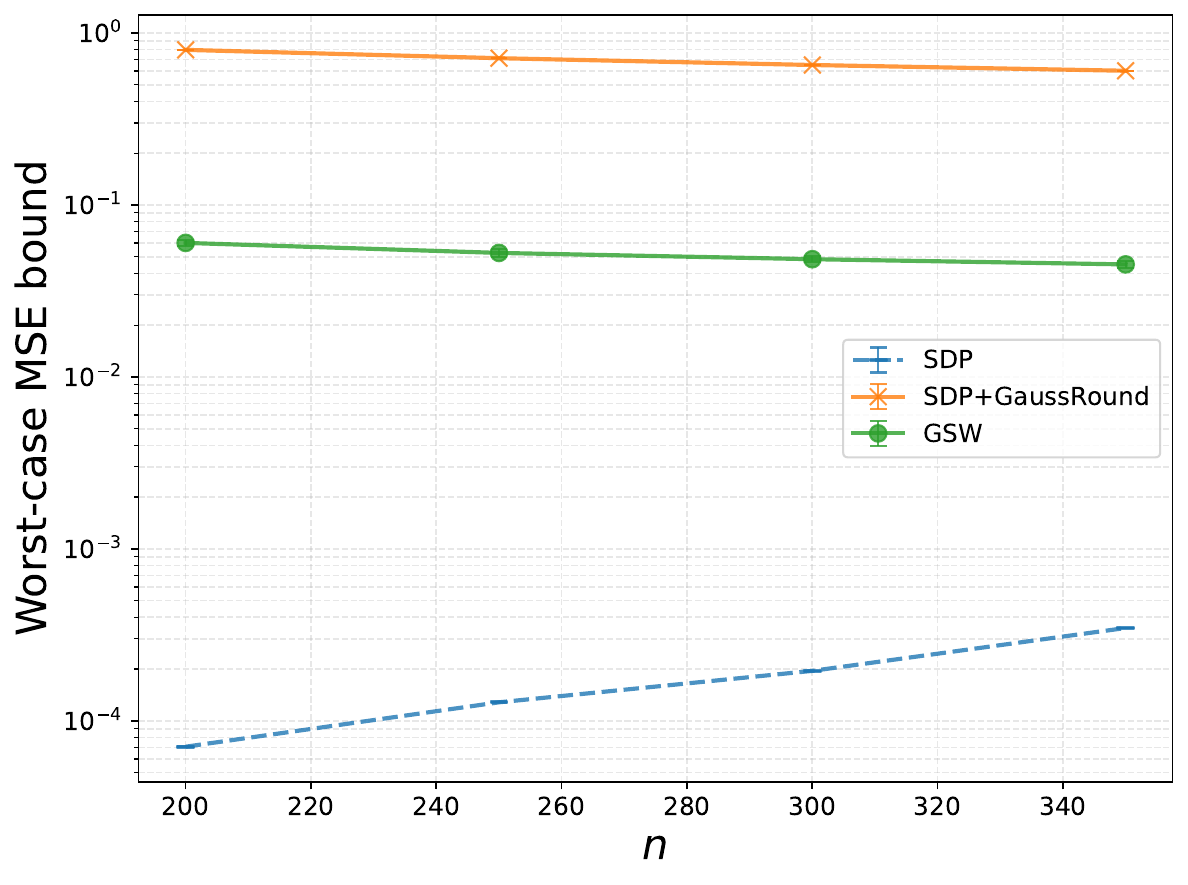}
    \caption{Comparisons with adapted Gram-Schmidt algorithm averaged over graphs generated from a Barabasi-Albert (preferential-attachment) model with initial $n/10$ (=10) nodes connected according to the Erdös-Rényi model with connection probability $10/n$ (see Graph 4 in \cref{fig:all_graphs}), for different $n$ . Error bars represent the standard error.}
    \label{fig:gsw_comparisons_graph4}
\end{figure}

\begin{figure}[ht]
    \centering
    \includegraphics[width=0.5\linewidth]{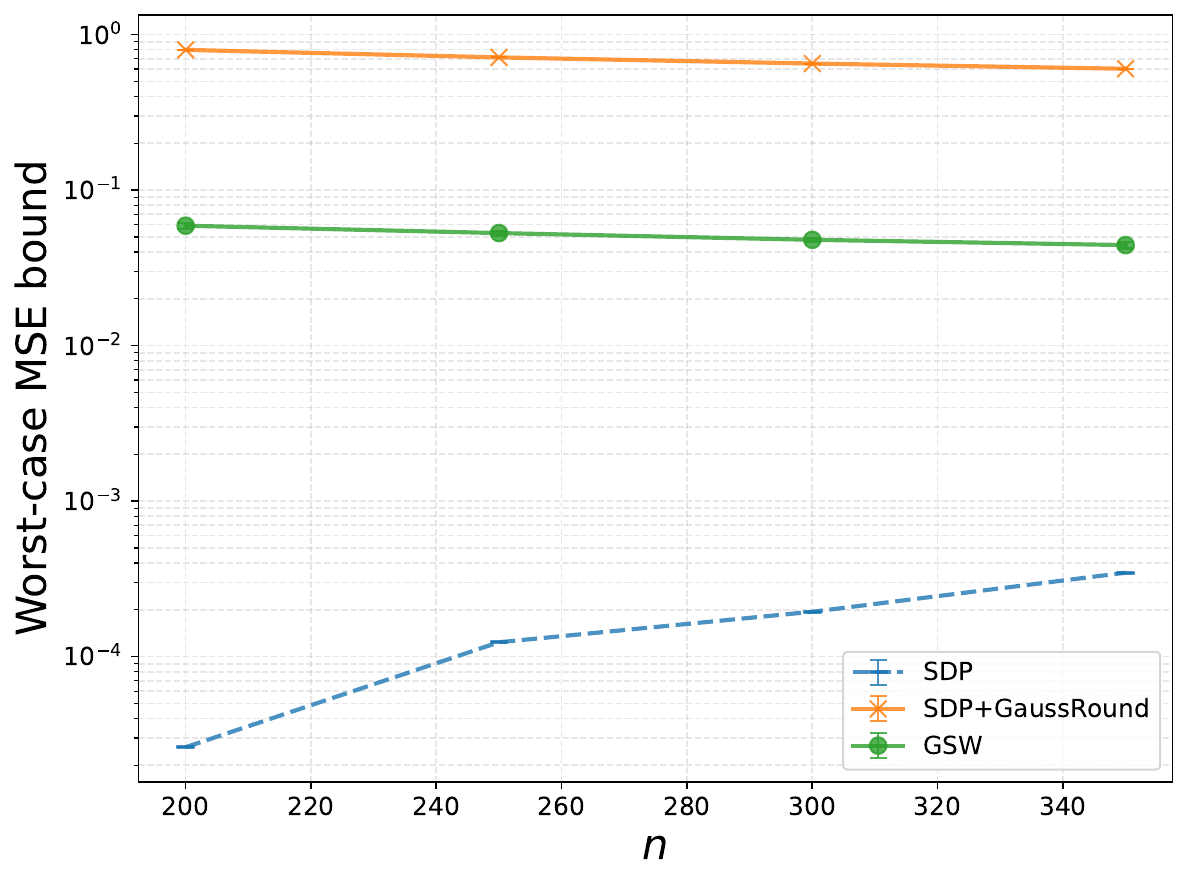}
    \caption{Comparisons with adapted Gram-Schmidt algorithm averaged over graphs generated from an Erdös-Rényi model with connection probability $2/n$ (see Graph 5 in \cref{fig:all_graphs}), for different $n$. Error bars represent the standard error.}
    \label{fig:gsw_comparisons_graph5}
\end{figure}

\Cref{fig:gsw_comparisons_graph1,fig:gsw_comparisons_graph2,fig:gsw_comparisons_graph3,fig:gsw_comparisons_graph4,fig:gsw_comparisons_graph5} compare worst-case MSE bounds using our SDP approach followed by Gaussian rounding, and our adapted Gram-Schmidt Walk approach, across various $n$. Here, we set $\kappa = 100, \eta = 250, \gamma = 1000, q = 2$, and $\Delta = 0.01 \gamma$. We observe that the adapted Gram-Schmidt Walk designs have comparable worst-case error bounds to the SDP solutions, for $n$ ranging between 200 and 400. The SDP solutions followed by Gaussian rounding perform worse than the adapted Gram-Schmidt Walk designs for settings of types Graphs 2 -- 5 in \Cref{fig:all_graphs}, while in settings of type Graph 1 in \Cref{fig:all_graphs}, the Gaussian rounding procedure almost perfectly approximates the SDP solution.    

For completeness, we also compare the MSEs under simulated worst-case potential outcomes on various graphs. We present the results of these experiments in the next section.

\subsection{MSE comparisons under synthetic worst-case potential outcomes} \label{mse_comparisons_existing_designs}
\begin{figure}
    \centering
    \begin{subfigure}[b]{0.3\textwidth}
        \centering
        \includegraphics[width=\linewidth]{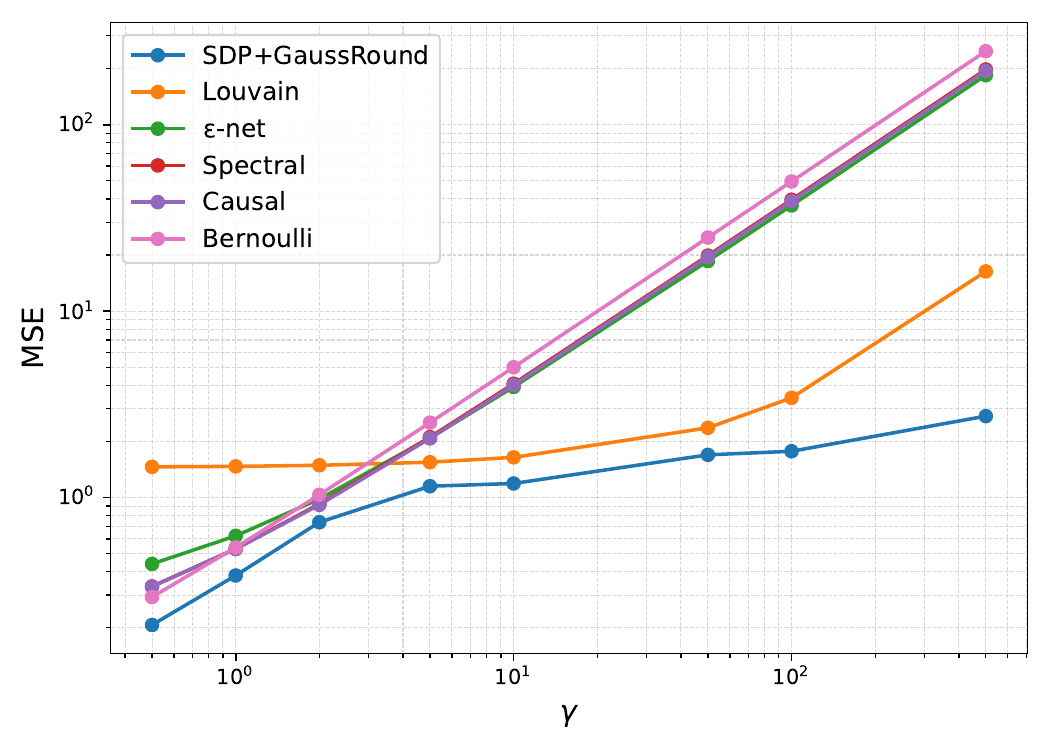}
    \end{subfigure}
    \hfill
    \begin{subfigure}[b]{0.3\textwidth}
        \centering
        \includegraphics[width=\linewidth]{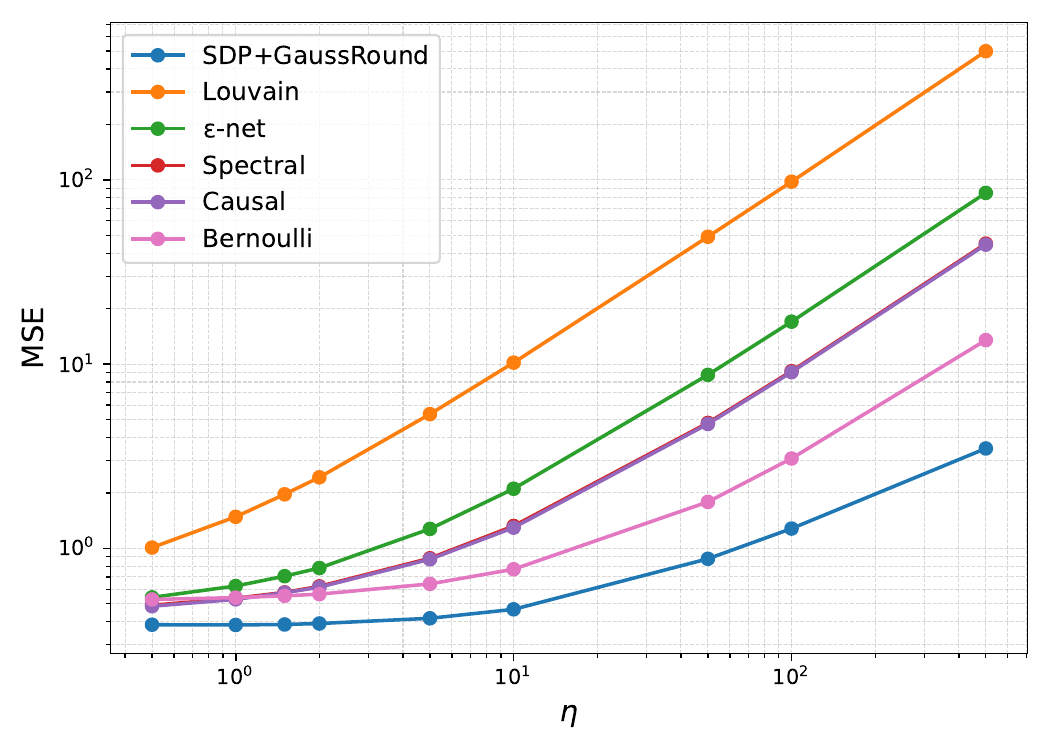}
    \end{subfigure}
    \hfill
    \begin{subfigure}[b]{0.3\textwidth}
        \centering
        \includegraphics[width=\linewidth]{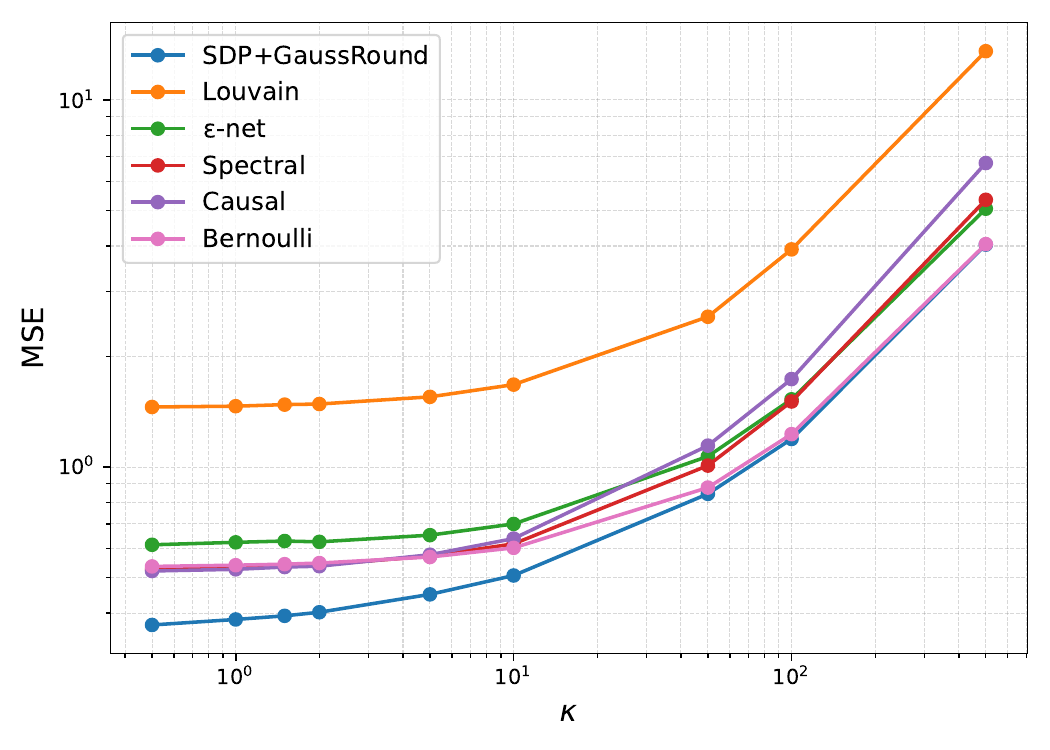}
    \end{subfigure}
    \caption{Comparison of MSEs between our designs, cluster randomization designs (with various clusterings) and Bernoulli randomization averaged over graphs generated from an SBM (within cluster connection probability $20/n$, across connection probability $\frac{0.1}{n}$), with two clusters of equal membership assignment probability (see Graph 1 in \Cref{fig:all_graphs}). The other two trade-off parameters are set to one. Error bars represent the standard error.}
    \label{fig:sdp_ave_case_graph_one}
\end{figure}

\begin{figure}
    \centering
    \begin{subfigure}[b]{0.3\textwidth}
        \centering
        \includegraphics[width=\linewidth]{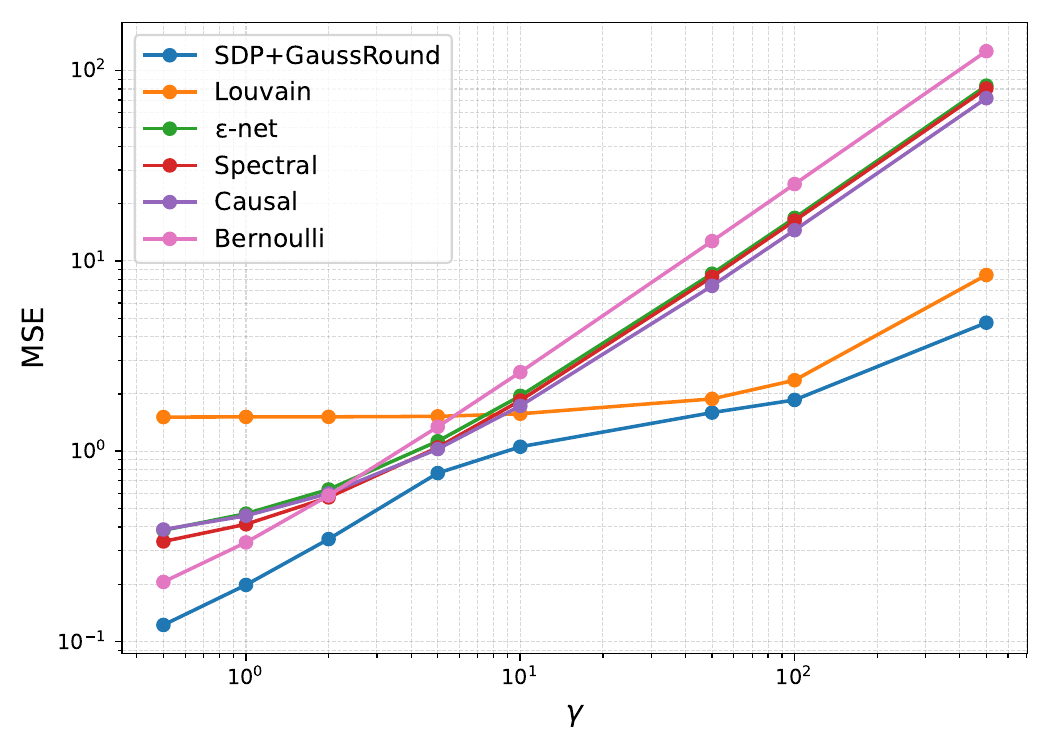}
    \end{subfigure}
    \hfill
    \begin{subfigure}[b]{0.3\textwidth}
        \centering
        \includegraphics[width=\linewidth]{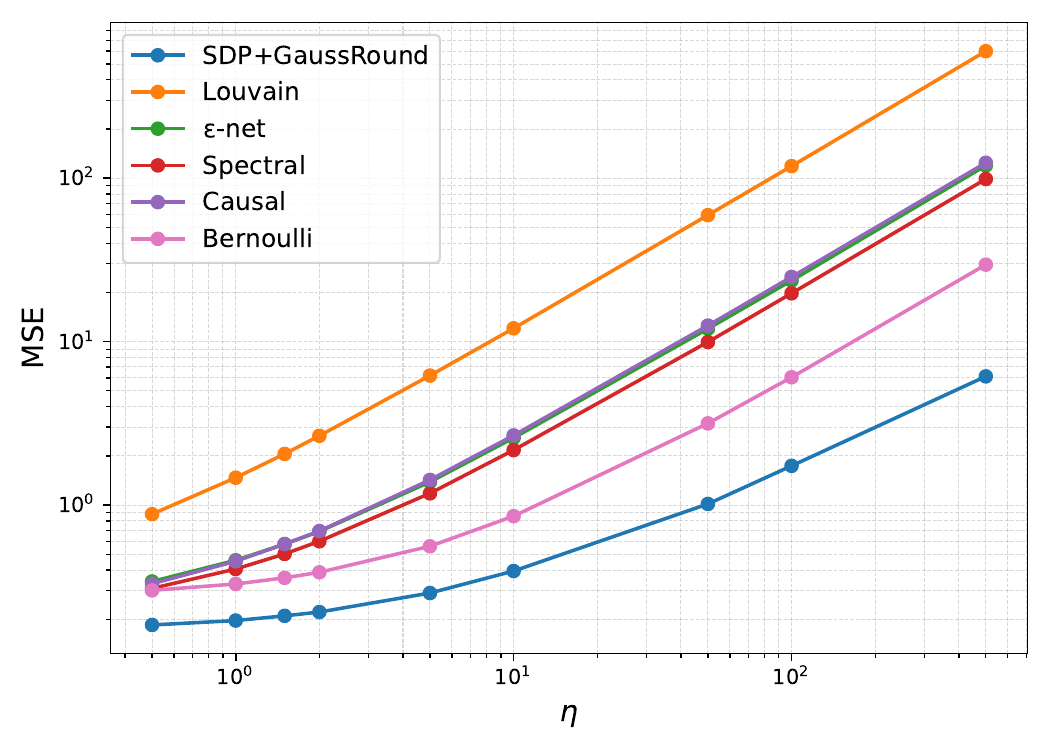}
    \end{subfigure}
    \hfill
    \begin{subfigure}[b]{0.3\textwidth}
        \centering
        \includegraphics[width=\linewidth]{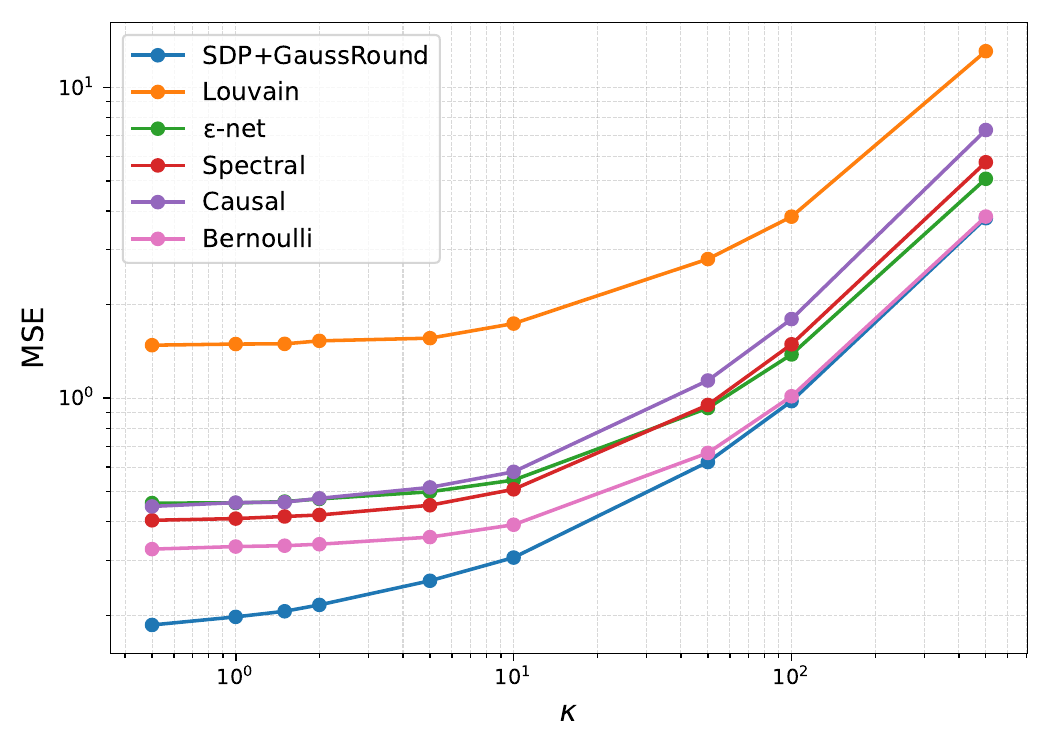}
    \end{subfigure}
    \caption{Comparison of MSEs between our designs, cluster randomization designs (with various clusterings) and Bernoulli randomization averaged over graphs generated from an SBM (within cluster connection probability $20/n$, across connection probability $\frac{0.1}{n}$), with two clusters of equal membership assignment probability (see Graph 2 in \Cref{fig:all_graphs}). The other two trade-off parameters are set to one. Error bars represent the standard error.}
    \label{fig:sdp_ave_case_graph_two}
\end{figure}

\begin{figure}
    \centering
    \begin{subfigure}[b]{0.3\textwidth}
        \centering
        \includegraphics[width=\linewidth]{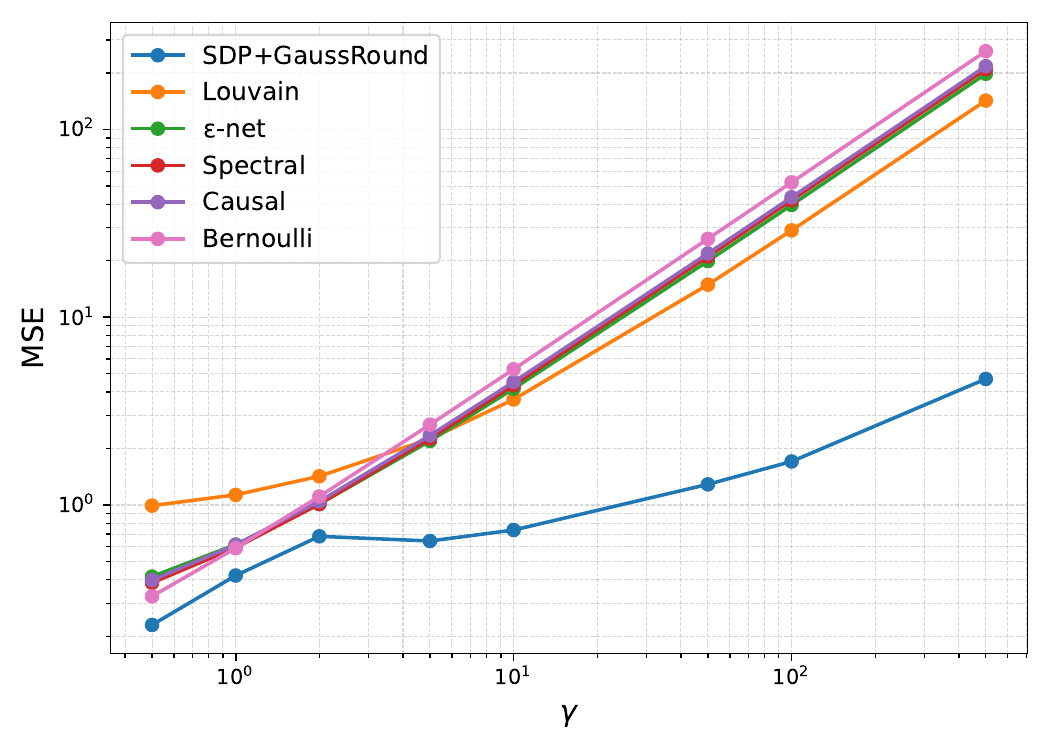}
    \end{subfigure}
    \hfill
    \begin{subfigure}[b]{0.3\textwidth}
        \centering
        \includegraphics[width=\linewidth]{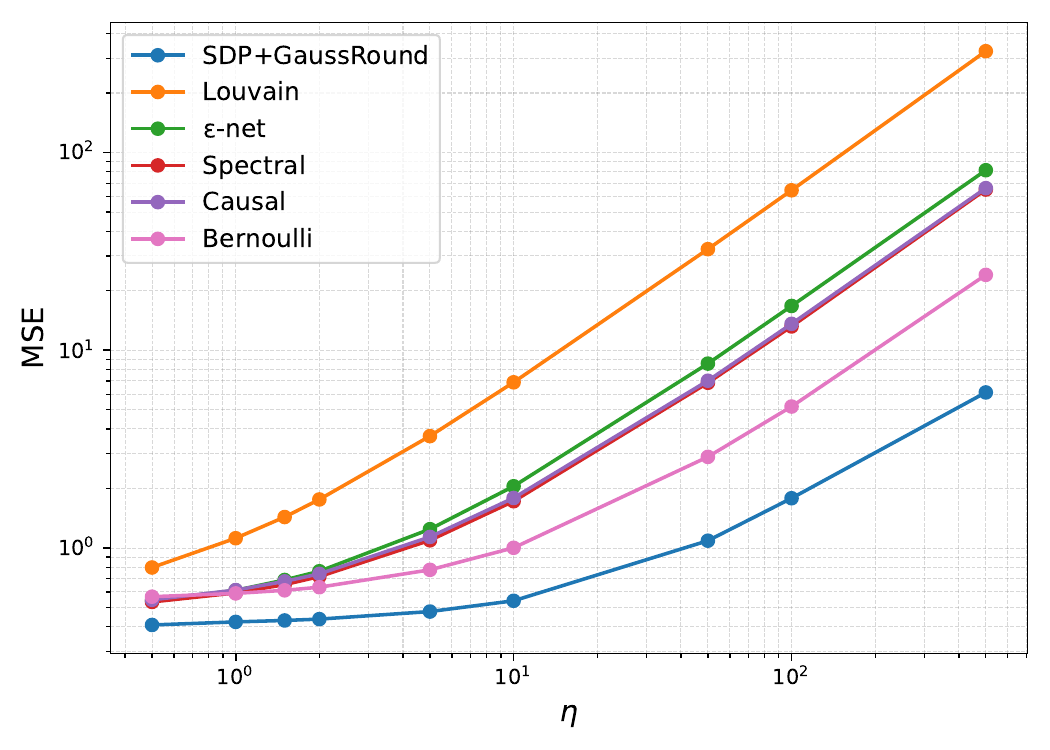}
    \end{subfigure}
    \hfill
    \begin{subfigure}[b]{0.3\textwidth}
        \centering
        \includegraphics[width=\linewidth]{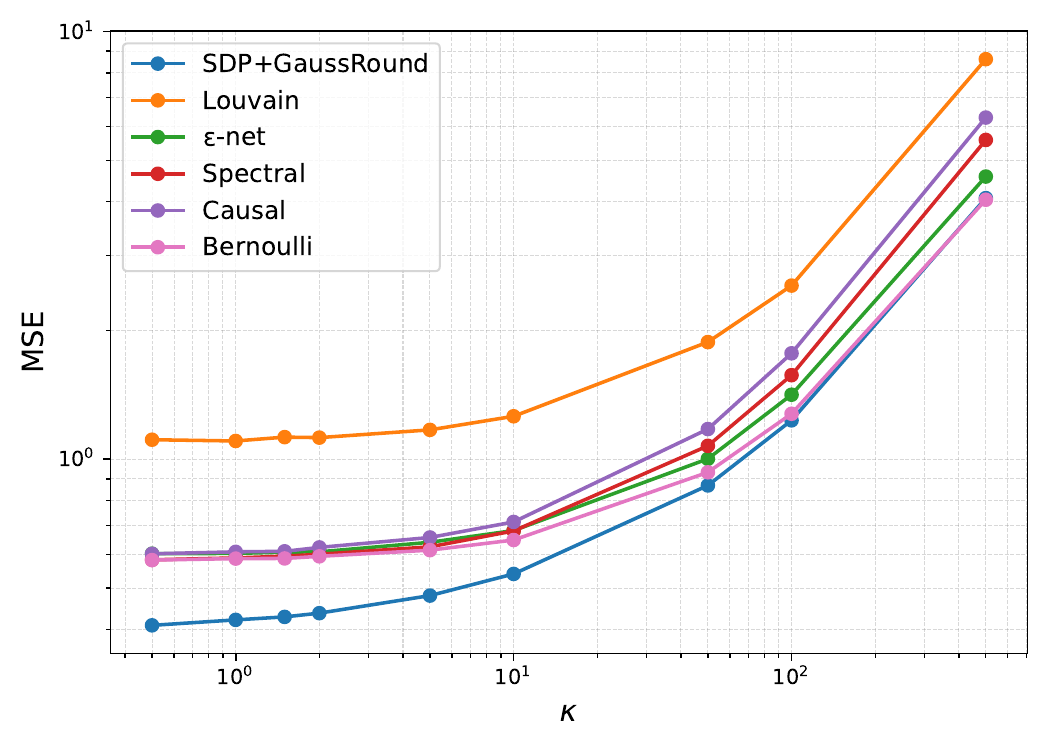}
    \end{subfigure}
    \caption{Comparison of MSEs between our designs, cluster randomization designs (with various clusterings) and Bernoulli randomization averaged over graphs generated from an SBM (within cluster connection probability $20/n$, across connection probability $\frac{0.1}{n}$) with four clusters of membership assignment probability (0.7, 0.1, 0.1, 0.1) (see Graph 3 in \Cref{fig:all_graphs}). The other two trade-off parameters are set to one. Error bars represent the standard error.}
    \label{fig:sdp_ave_case_graph_three}
\end{figure}

\begin{figure}
    \centering
    \begin{subfigure}[b]{0.3\textwidth}
        \centering
        \includegraphics[width=\linewidth]{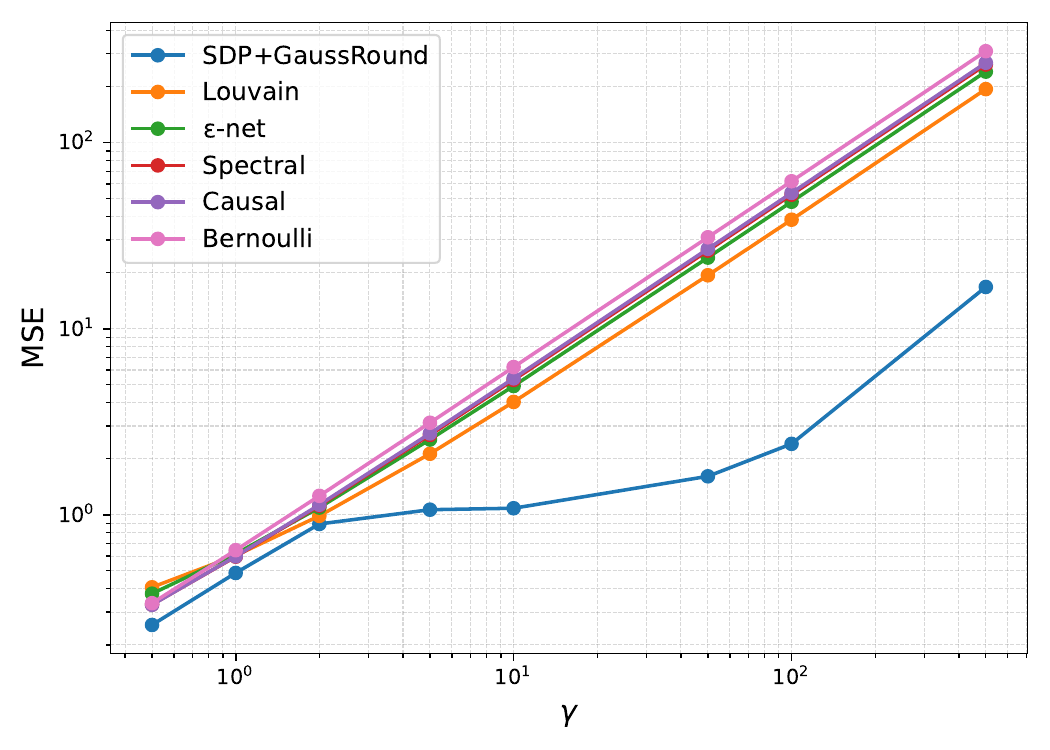}
    \end{subfigure}
    \hfill
    \begin{subfigure}[b]{0.3\textwidth}
        \centering
        \includegraphics[width=\linewidth]{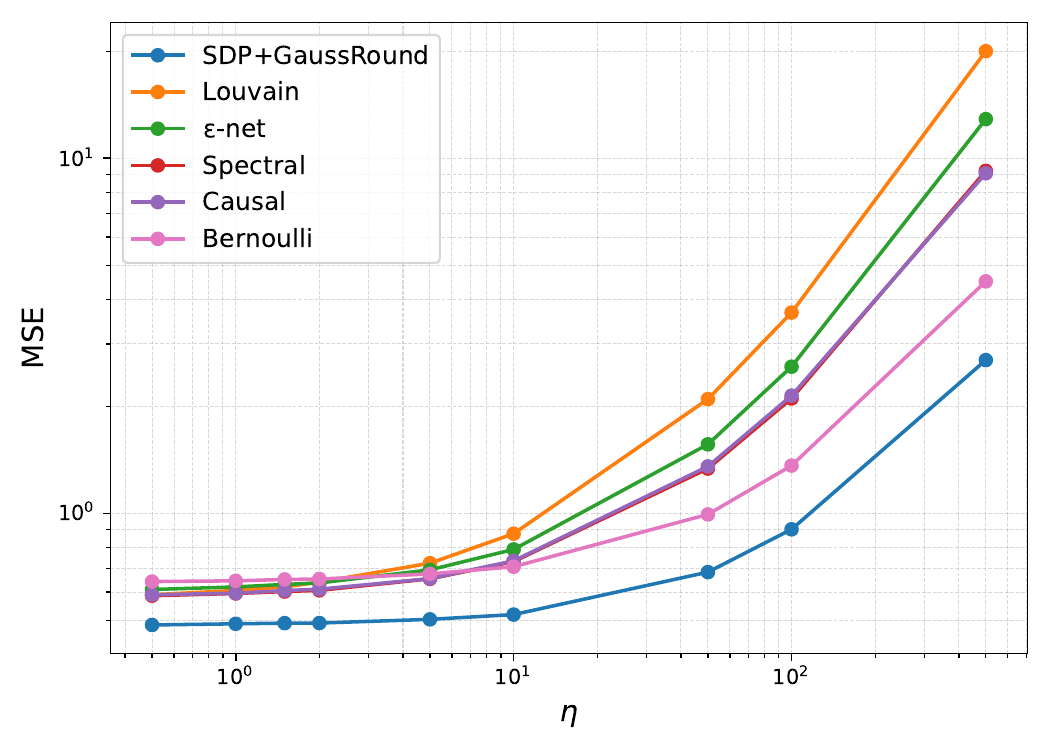}
    \end{subfigure}
    \hfill
    \begin{subfigure}[b]{0.3\textwidth}
        \centering
        \includegraphics[width=\linewidth]{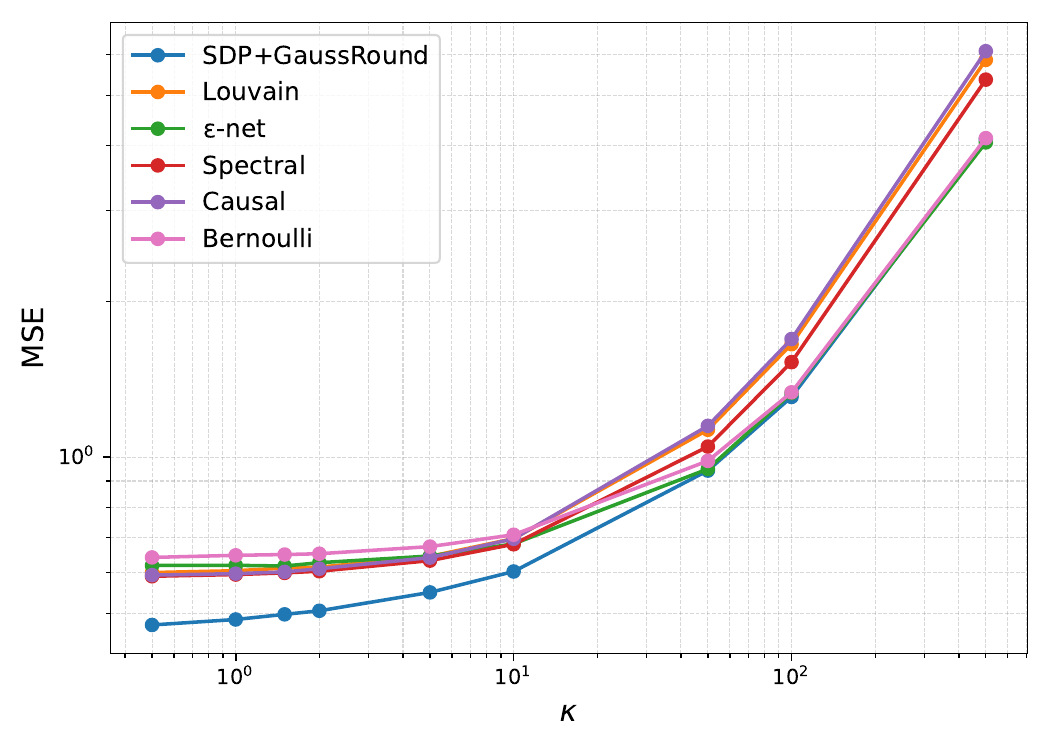}
    \end{subfigure}
    \caption{Comparison of MSEs between our designs, cluster randomization designs (with various clusterings) and Bernoulli randomization averaged over graphs generated from a Barabasi-Albert (preferential-attachment) model with initial $n/10$ (=10) nodes connected according to the Erdös-Rényi model with connection probability $10/n$ (see Graph 4 in \Cref{fig:all_graphs}). The other two trade-off parameters are set to one. Error bars represent the standard error.}
    \label{fig:sdp_ave_case_graph_four}
\end{figure}

\begin{figure}
    \centering
    \begin{subfigure}[b]{0.3\textwidth}
        \centering
        \includegraphics[width=\linewidth]{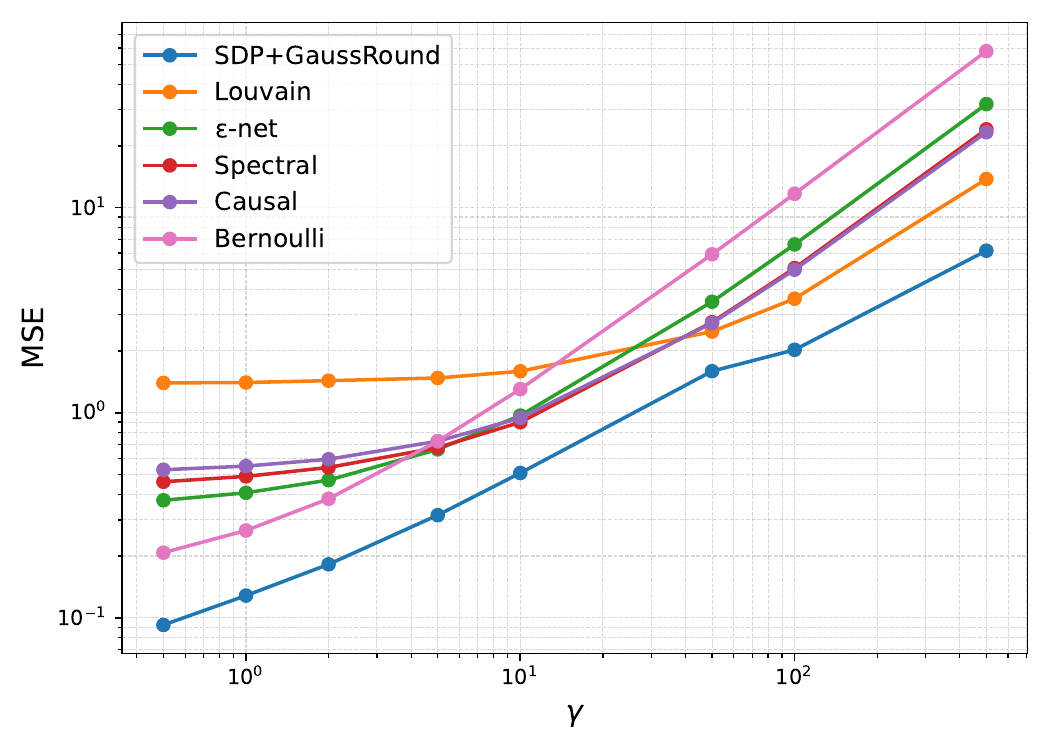}
    \end{subfigure}
    \hfill
    \begin{subfigure}[b]{0.3\textwidth}
        \centering
        \includegraphics[width=\linewidth]{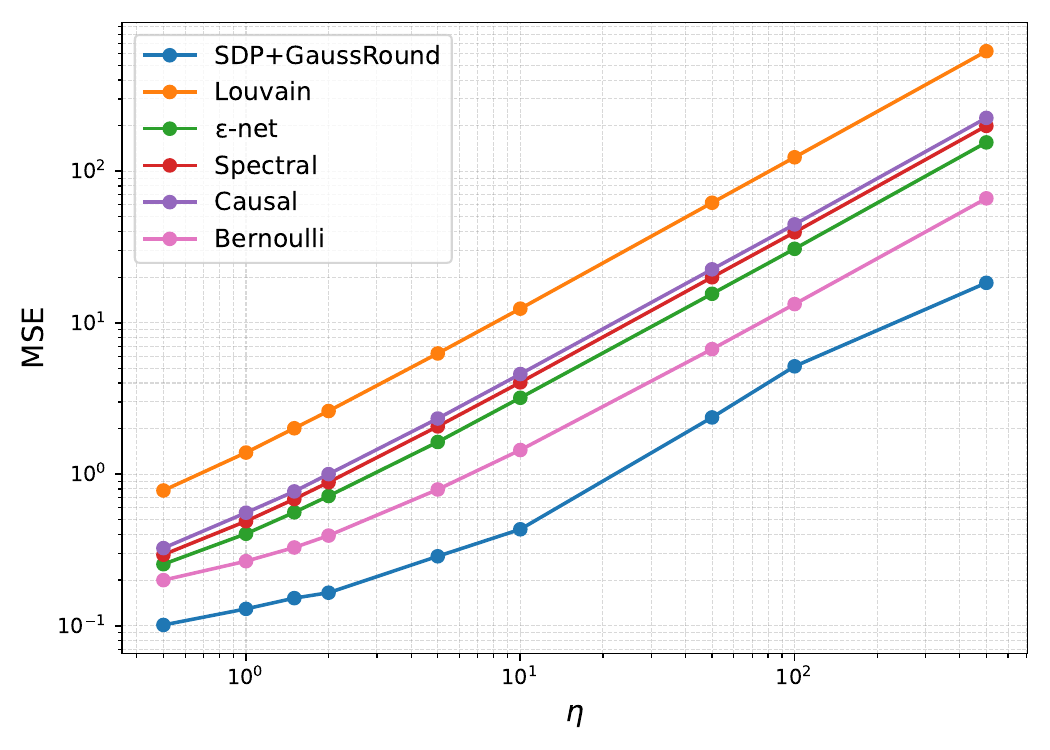}
    \end{subfigure}
    \hfill
    \begin{subfigure}[b]{0.3\textwidth}
        \centering
        \includegraphics[width=\linewidth]{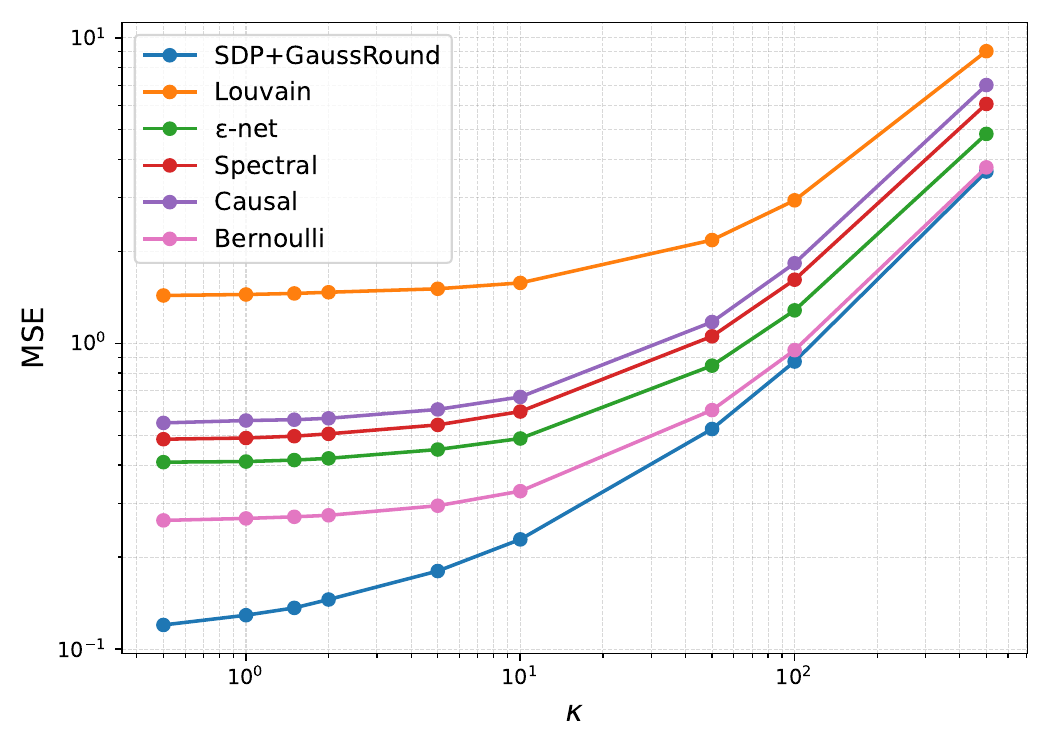}
    \end{subfigure}
    \caption{Comparison of MSEs between our designs, cluster randomization designs (with various clusterings) and Bernoulli randomization averaged over graphs generated from an Erdös-Rényi model with connection probability $2/n$ (see Graph 5 in \Cref{fig:all_graphs}). The other two trade-off parameters are set to one. Error bars represent the standard error.}
    \label{fig:sdp_ave_case_graph_five}
\end{figure}

In \Cref{fig:sdp_ave_case_graph_one,fig:sdp_ave_case_graph_two,fig:sdp_ave_case_graph_three,fig:sdp_ave_case_graph_four,fig:sdp_ave_case_graph_five}, we display the MSE using simulated potential outcomes across various designs, namely our design, cluster-randomization under the default Python implementation of the Louvain clustering, $(\varepsilon=1)$-net clustering \citep{ugander2013graph}, Spectral clustering, Causal Clustering \citep{viviano2023causal}, and Bernoulli Randomization. We compare the MSEs for five different graph generation models, averaged across 1000 graph instances of each graph model, as described in \Cref{fig:all_graphs}. In these settings, we set the fixed tradeoff parameters to be 1, and $a, b = 0.01 \gamma$. We display the results in \Cref{fig:sdp_ave_case_graph_one,fig:sdp_ave_case_graph_two,fig:sdp_ave_case_graph_three,fig:sdp_ave_case_graph_four,fig:sdp_ave_case_graph_five}.

The worst-case potential outcomes were generated according to the following model. For given $L, L^\dagger, \eta, \gamma, \kappa$, and randomly generated baselines $\alpha_0, \phi_0$ (and, thus, $\Delta$), we compute the covariance matrices $X$ corresponding to the different designs described above. Then, adversarially, for the worst-case, we compute the potential outcome components. We describe this in detail in the following:
To generate the homophily components $h_f$, $f \in \{\alpha, \phi\}$,
\begin{align*}
    v &= \text{ max eigenvector}(L^{\dagger/2} X L^{\dagger/2}), \\
    h_f &=  \sqrt{\eta}L^{\dagger/2} v, \\
    % h_f &=  \frac{\sqrt{\eta}v}{\| v^TL^{1/2}\|_2}, \\
    h_f &=  h_f - \dotp{h_f, \frac{\mathbf{1}}{n}} \mathbf{1}
\end{align*}

In the above, we leverage the proof to the bounds in \Cref{lemma:tighter_homophily_bd} to generate the worst-case homophily components. Indeed, let $w = L^{1/2} h_f/ \sqrt{\eta}$, then, the $h_f$ that achieves $\sup_{\|L^{1/2} h_f\|_2^2 \leq \eta} {h_f}^T L^{1/2} L^{\dagger/2} X L^{\dagger/2} L^{1/2} {h_f}$ is the same as the $\sqrt{\eta}L^{\dagger/2}w$ for the $w$ that achieves $\sup_{\|w\|_2^2 \leq 1} w^T L^{\dagger/2} X L^{\dagger/2} w$, which is simply the maximum eigenvector of $L^{\dagger/2} X L^{\dagger/2}$.

To generate the heterogeneous variation components $\epsilon_f$, $f \in \{\alpha, \phi \}$ $q^*=2$, we take
\begin{align*}
    \Sigma &= \kappa X/\|X\|_F, \\
    z &\sim \calN(0, I_n), \\
    \epsilon_f &= \Sigma^{1/2} z.
\end{align*}
If $q^* = 1$ instead, we replace $\Sigma$ above with $\Sigma = \kappa vv^T$, where $v$ is the leading eigenvector of $X$. Indeed, it is not difficult to see that these are the $\epsilon_f$ that achieve the bounds in \Cref{lemma:robustness_bd,prop:general_schatten_p_bound}. 

Finally, let $\Tilde{\alpha} = h_\alpha + \epsilon_\alpha$, and $\Tilde{\phi} = h_\phi + \epsilon_\phi$. Recall that here $a = \dotp{ \alpha_0, \frac{\mathbf{1}}{n}}$, and $b = \dotp{ \phi_0, \frac{\mathbf{1}}{n}}.$ We take $\alpha = a \mathbf{1} + (\Tilde{\alpha} - \dotp{\Tilde{\alpha}, \frac{\mathbf{1}}{n}} \mathbf{1})$ and $\phi = b \mathbf{1} + (\Tilde{\phi} - \dotp{\Tilde{\phi}, \frac{\mathbf{1}}{n}} \mathbf{1}).$

To generate the interference component for each treatment assignment vector $x = 2z - 1$ generated from $X$ from the corresponding covariance matrix, we take 
\begin{align*}
s(z) &=  \frac{\sqrt{\gamma}L x}{\| x^T L^{1/2} \|_2}, \\
s(z) &=  s(z) - \dotp{s(z), \frac{\mathbf{1}}{n}} \mathbf{1}.
\end{align*}

Indeed, through the proof the Cauchy-Schwarz inequality, it is not difficult to see that the $s(z)$ that achieves $\sup_{\|L^{\dagger/2} s(z) \|_2^2 \leq \gamma} \mathbb{E}[ \langle x, s(z)\rangle^2]$, must satisfy $L^{\dagger/2} s(z) = \gamma^{1/2} L^{1/2} x / \|L^{1/2} x\|_2$.

\subsection{Priors and Worst-Case}

We take this opportunity to illustrate how our framework allows for the incorporation of priors in the bounds to improve the performance of the design otherwise tailored to the worst-case setting. Indeed, if the practitioner strongly believed that the homophily components of the potential outcomes followed a certain type of distribution, then we can refine the trade-off parameter bounds. In particular, if the practitioner believed that the homophily components followed a uniform distribution on a sphere, much like in the data generation procedure in the previous paragraph, we can replace the $\eta$ parameter by $\eta/n$. Indeed, if $h \sim {L^\dagger}^{1/2} \text{Unif}(S_{n-1}^\eta)$ where $S_n^\eta$ is the $(n-1)$-sphere of radius $\sqrt{\eta}$, i.e., the scaled normalized gaussians in the previous paragraph, then,
\begin{align*}
    \mathbb{E}[(h^T x)^2] &=   \operatorname{Tr}(X \mathbb{E}[{L^\dagger}^{1/2}vv^T{L^\dagger}^{1/2}]) \\
    &=  \operatorname{Tr}( X {L^\dagger}^{1/2}\mathbb{E}[vv^T]{L^\dagger}^{1/2})\\
    &= \frac{\eta}{n}  \operatorname{Tr}( X L^{\dagger}).
\end{align*}

Similarly, if the practitioner believed that the interference $s(z)$ followed a uniform distribution on the sphere, we can replace $\gamma$ by $\gamma/n$. Indeed, if $s(z) \sim L^{1/2} \text{Unif}(S_{n-1}^\gamma)$, where $S_{n-1}^\gamma$ is the $(n-1)-$sphere of radius $\sqrt{\gamma}$, then,
\begin{align*}
    \mathbb{E}[(s(z)^T x)^2] &=   \operatorname{Tr}(X \mathbb{E}[{L}^{1/2}vv^T{L}^{1/2}]) \\
    &=  \operatorname{Tr}( X {L}^{1/2}\mathbb{E}[vv^T]{L}^{1/2})\\
    &= \frac{\gamma}{n}  \operatorname{Tr}( X L).
\end{align*}

\subsection{Quantile rounding} \label{sec-quantile_rounding}
One can also consider what we call ``quantile rounding'', which is equivalent to Gaussian rounding when $p=1/2$. Generally, quantile rounding for $p \leq 1/2$ is simply the procedure of selecting the threshold $t_q(i)$ for the rounding of $\xi \sim \calN(0, X^*), \, \xi \in \mathbb{R}^n$ to $ \zeta \in \{-1,1\}^n$ based on the quantiles of $\xi_i$. That is, we take $t_q(i) = \sqrt{X^*_{ii}} \cdot \Phi^{-1}(1-p)$, and let $\zeta_i = 1$ if $\xi_i \geq t_q(i)$, and $\zeta_i = -1$ otherwise.

\subsection{Comparisons between SDP and adapted Gram-Schmidt Walk designs}
In \Cref{fig:gsw_comparisons_graph1,fig:gsw_comparisons_graph2,fig:gsw_comparisons_graph3,fig:gsw_comparisons_graph4,fig:gsw_comparisons_graph5}, we compare the worst-case bounds of the designs produced by the SDP and adapted Gram-Schmidt Walk algorithms for $n \in \{200, 250, 300, 350\}$. As expected, the SDP solutions yield uniformly smaller worst-case bounds, since the SDP directly optimizes the worst-case objective. However, over this range of $n$, we do not observe a clear trend in the gap between the SDP- and Gram–Schmidt-Walk-based bounds that would reflect the predicted looseness of the adapted Gram–Schmidt Walk bound at large $n$. This is likely because we only consider a limited range of $n$, as solving the SDP becomes computationally prohibitive for larger $n$. Finally, for all the simulated networks besides the SBM graphs with two equal-membership clusters, the adapted Gram–Schmidt Walk algorithm yields smaller bounds than Gaussian rounding applied to the SDP solutions.

\end{document}

%% file: content/real_data_simulations.tex
\subsection{Real Data} \label{section_read_data}

\Cref{fig:house_visits_village} illustrates a single village network in rural Karnataka, South India \citep{banerjee2013diffusion,DVN/U3BIHX_2013}. For context, \citet{banerjee2013diffusion} collected network data from 75 villages before the introduction of financial services in these villages by a microfinance institute. 
% The representatives of this institute begin, in each village, by inviting particular ``leaders'' of the village to an information meeting. They then ask these ``leaders'' to spread this information.  
In particular, in 2006, six months before the microfinance institute began their work, \citet{banerjee2013diffusion} conducted surveys on subsamples of villagers gathering demographic information. The survey included a portion on social network data along 12 dimensions. In \Cref{fig:house_visits_village}, we display Village No.6 which we had randomly selected. In this network, the nodes represent village members, and an edge between any pair of nodes is present if the corresponding villagers visit each other's homes. In this figure, the nodes in the network are colored according to the castes of the corresponding villagers, illustrating homophily in the network, as members of the same caste are more connected.

\begin{figure}
    \centering
    \begin{subfigure}{0.45\textwidth}
        \centering
        \includegraphics[width=\linewidth]{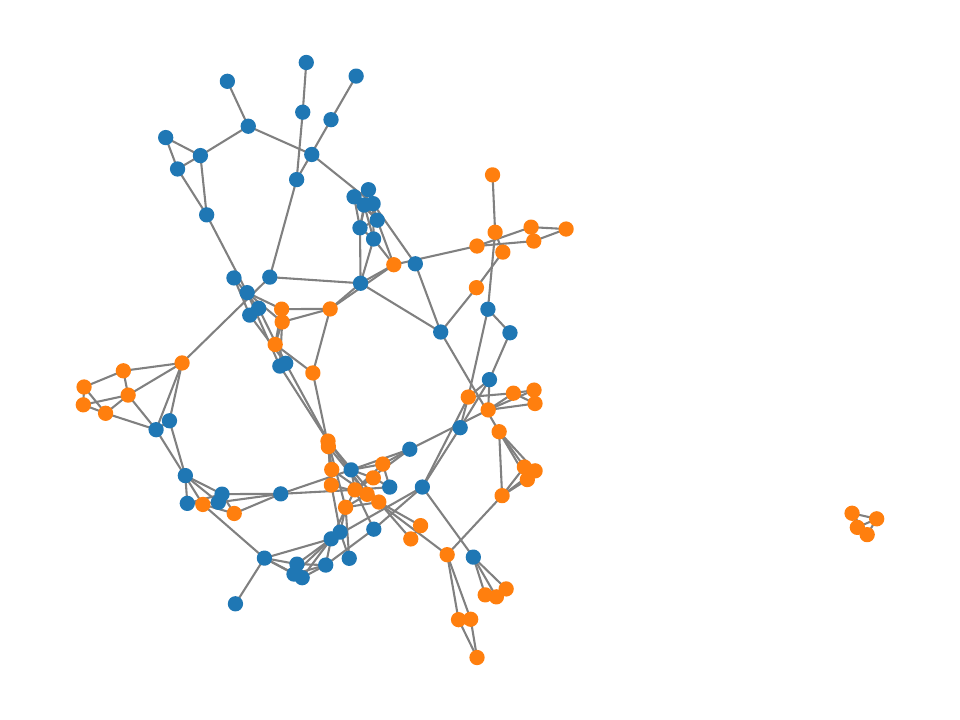}
    \end{subfigure}
    \hfill
    \begin{subfigure}{0.45\textwidth}
        \centering
        \includegraphics[width=\linewidth]{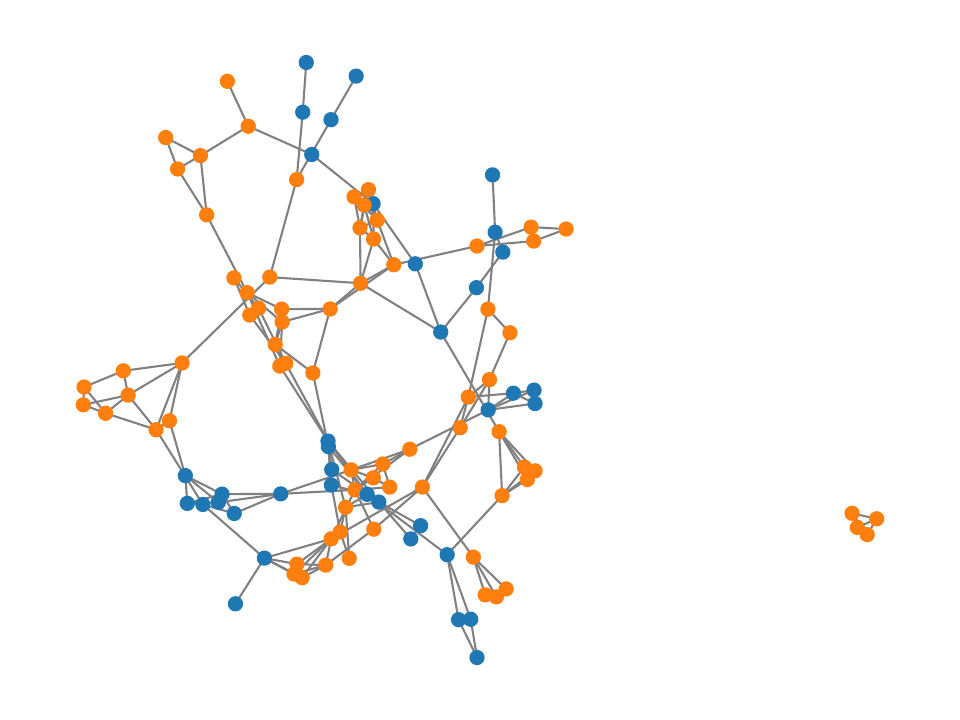}
    \end{subfigure}
    \vspace{0.2em}
    \begin{subfigure}{0.45\textwidth}
        \centering
        \includegraphics[width=\linewidth]{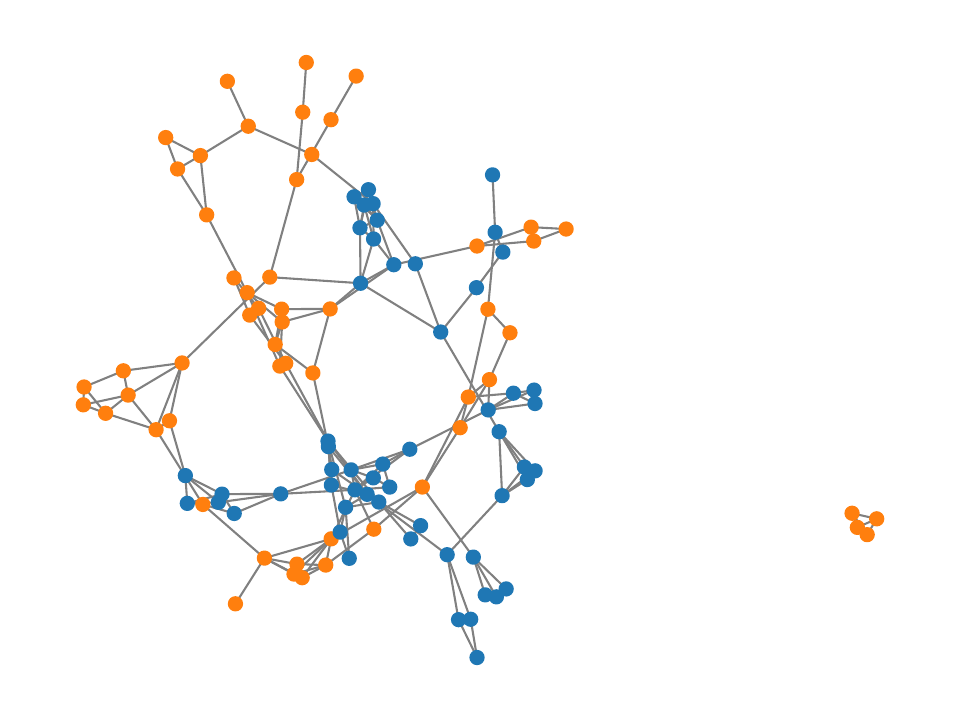}
    \end{subfigure}
    \hfill
    \begin{subfigure}{0.45\textwidth}
        \centering
        \includegraphics[width=\linewidth]{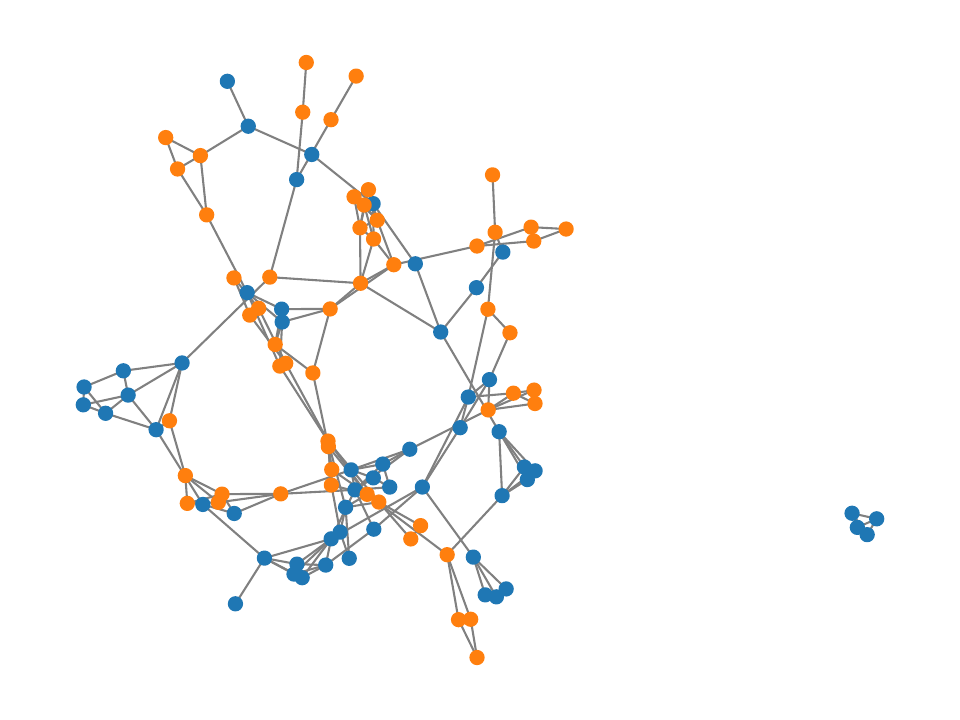}
    \end{subfigure}
    \caption{Assignments under various cluster randomization designs. Top (left to right): $\epsilon=1$-net, causal cluster randomizations. Bottom (left to right): Louvain, and spectral cluster randomizations.}
    \label{fig:comp_colorings}
\end{figure}

In \Cref{fig:comp_colorings}, we illustrate treatment assignments under various cluster-randomized designs, depicting almost complete overlap with the underlying homophily structure. In the worst case where the distinct homophilous groups have heterogeneous treatment effects, such cluster-randomized designs can lead to imprecise estimates of the GATE. Generally, this imprecision worsens as the number of distinct homophilous groups increases for a fixed population size.

% , which would lead to imprecise estimates of the GATE, especially as the number of distinct homophilous groups increases with heterogeneous treatment effects.

% \begin{figure}[H]
%     \centering
%     \includegraphics[width=0.5\linewidth]{figures/go_caste_6.pdf}
%     \caption{Network of home visits between members of a village (no.6) with different node colors representing different castes of village members.}
%     \label{fig:house_visits_village}
% \end{figure}

% \begin{figure}
%     \centering
%     \includegraphics[width=0.5\linewidth]{figures/real_opt_go_caste_6.pdf}
%     \caption{Caption}
%     \label{fig:enter-label}
% \end{figure}

\begin{figure}[!ht]
    \centering
    \begin{subfigure}{0.45\textwidth}
        \centering
        \includegraphics[width=\linewidth]{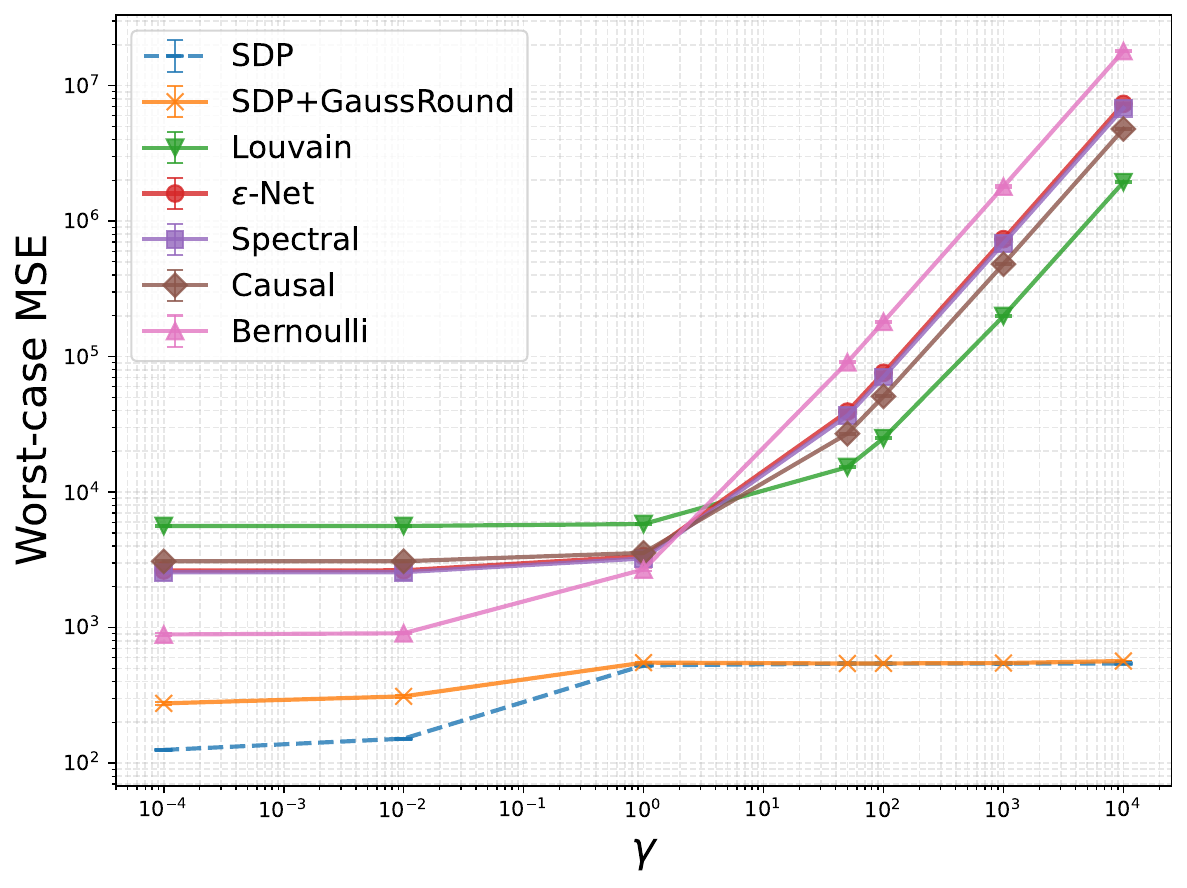}
    \end{subfigure}
    \begin{subfigure}{0.45\textwidth}
        \centering
        \includegraphics[width=\linewidth]{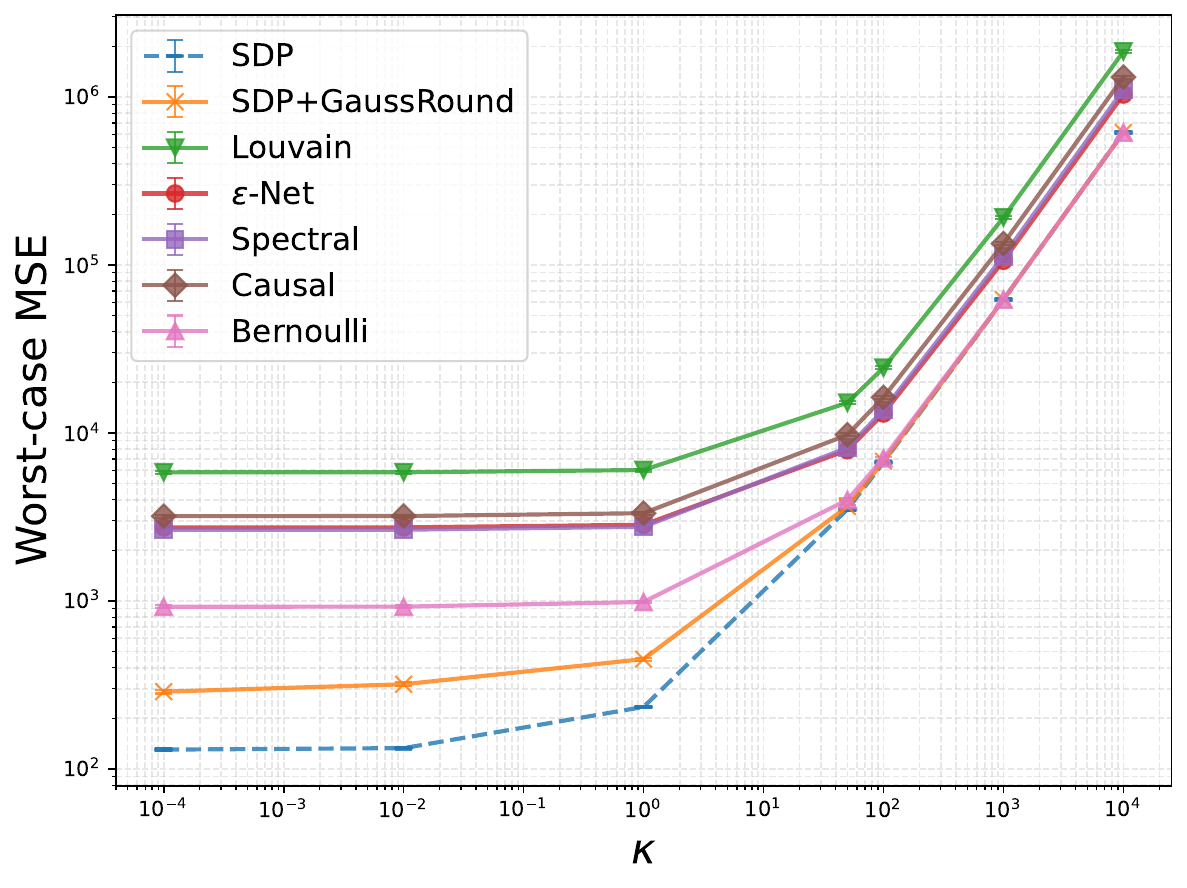}
    \end{subfigure}
    \caption{Comparisons of the worst-case MSE bounds across various designs, for different levels of $\gamma$ and $\kappa$, with the average (over 1000 trials) $\eta\approx 196$. Error bars represent the standard error.}
    \label{fig:comp_all_rd}
\end{figure}

\begin{figure}[!ht]
    \centering
    \begin{subfigure}{0.45\textwidth}
        \centering
        \includegraphics[width=\linewidth]{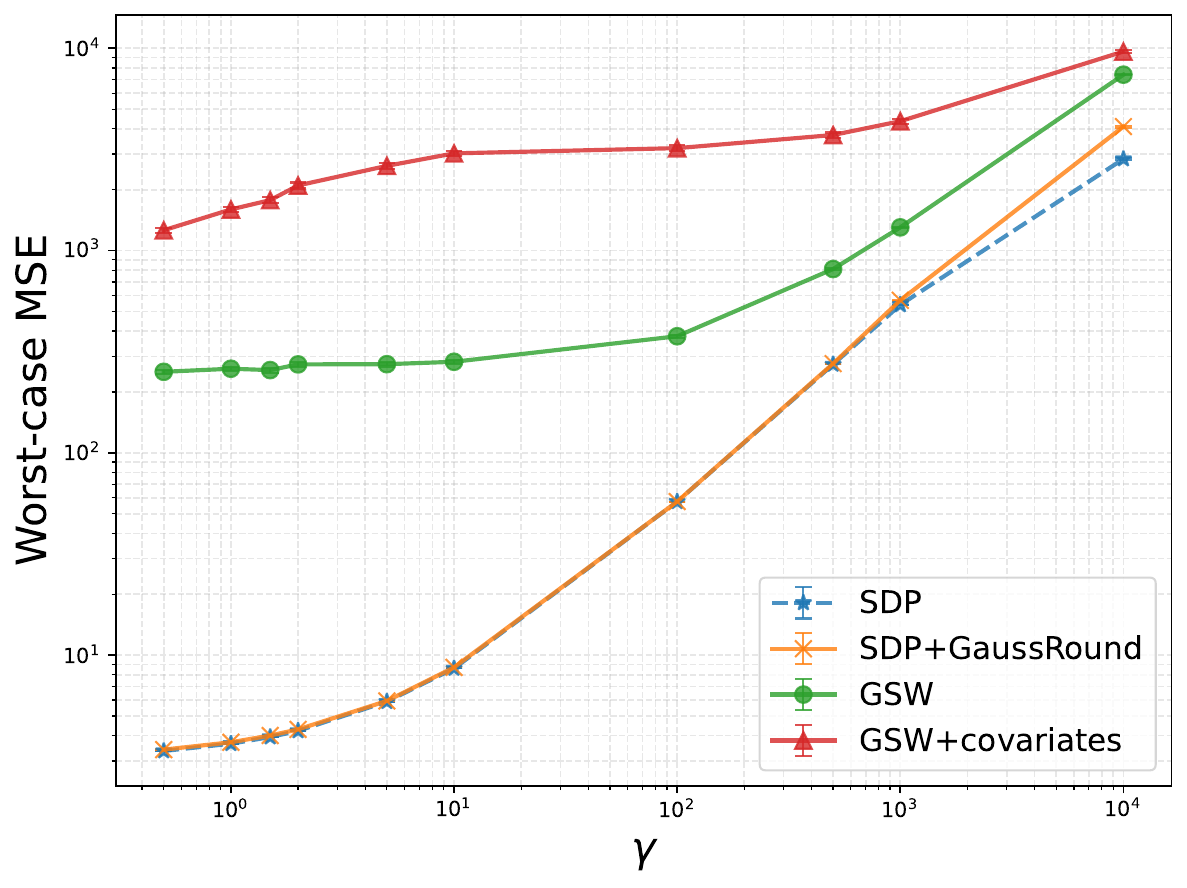}
    \end{subfigure}
    \begin{subfigure}{0.45\textwidth}
        \centering
        \includegraphics[width=\linewidth]{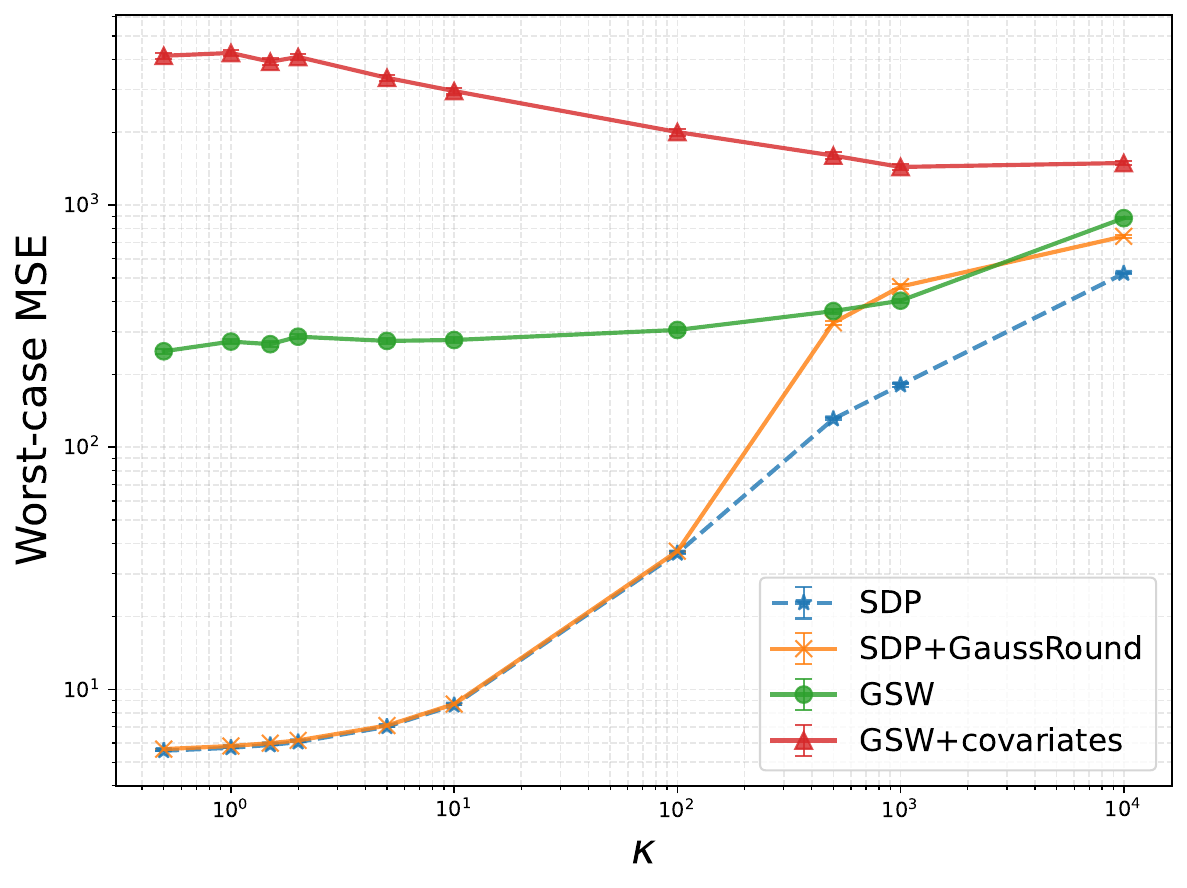}
    \end{subfigure}
    \caption{Comparisons of the worst-case MSE bounds across various designs, with the average (over 1000 trials) $\eta\approx 196$. Error bars represent the standard error.}
    \label{fig:comp_gsw_rd}
\end{figure}

In the comparisons of the worst-case MSE bounds on this network data in \Cref{fig:comp_all_rd,fig:comp_gsw_rd}, we compare the bounds derived in \Cref{thm:mse_bound_general_HT} using covariance matrices from various designs. We describe the different designs we compare in \Cref{tab:design_methods}. 
% In our comparisons, we consider also what we call ``quantile rounding'', which is equivalent to Gaussian rounding when $p=1/2$. Generally, quantile rounding for $p \leq 1/2$ is simply the procedure of selecting the threshold $t_q(i)$ for the rounding of $\xi \sim \calN(0, X^*), \, \xi \in \mathbb{R}^n$ to $ \zeta \in \{-1,1\}^n$ based on the quantiles of $\xi_i$. That is, we take $t_q(i) = \sqrt{X^*_{ii}} \cdot \Phi^{-1}(1-p)$, and let $\zeta_i = 1$ if $\xi_i \geq t_q(i)$, and $\zeta_i = -1$ otherwise. 
We note that the different cluster-randomized designs were also considered in \citep{viviano2023causal}.

% \begin{table}[ht]
% \centering
% \begin{tabular}{@{}lp{11cm}@{}}
% \toprule
% \textbf{Name} & \textbf{Description} \\
% \midrule
% SDP & SDP solution $X^*$ \\
% SDP + GaussRound & SDP followed by Gaussian rounding \\
% SDP + QuantRound & SDP followed by quantile rounding (see \Cref{sec-quantile_rounding})\\
% GSW & Adapted Gram-Schmidt Walk design \citep{harshaw2024balancing} \\
% %SDP + PG & Adapted projected gradient optimization from the optimized covariance design \citep{chen2024optimized}, initialized by our SDP solution \\
% GSW + covariates & Adapted Gram-Schmidt Walk design \citep{harshaw2024balancing} using covariate-encoding directly instead of $L^\dagger$ \\
% Louvain & Cluster-randomization with default Python implementation of the Louvain clustering algorithm \\
% $\epsilon$-net & Cluster-randomization under $(\epsilon{=}1)$-net clustering \citep{ugander2013graph} \\
% Spectral & Cluster-randomization under spectral clustering \\
% Causal & Cluster-randomization under Causal clustering \citep{viviano2023causal} with a grid-search for optimal hyperparameters \\
% Bernoulli & Uniform Bernoulli randomization \\
% % Complete & Uniform Complete randomization \\
% \bottomrule
% \end{tabular}
% \caption{Key of designs compared in this paper.}
% \label{tab:design_methods_real_data}
% \end{table}

We also compare the MSEs by combining synthetic worst-case potential outcomes with the real network data together with the caste information. We generate the homophily components in the potential outcomes, $h^\alpha$ and $h^\phi$ from a random effects model across the castes. That is, we generate potential outcomes associated with each caste $j$
\begin{align*}
    {\nu_j}_f \sim \calN(\mu_f, \sigma^2_f), 
\end{align*} for some $\mu_f$.
Then, for each node $i$ that belongs to caste $j$,
\begin{align*}
   {h_i}_f = {\nu_j}_f ,
\end{align*} for all $j$, and $f = \alpha, \phi$. In our simulations, we take $\mu_\alpha = 2$, $\mu_\phi = 3$, and $\sigma_f^2 = 3$, $f= \alpha, \phi$. Then, we take $\eta= {h_f}^T L h_f$.

% of SDP solution, our design, an adaptation of the proximal design from \citet{chen2024optimized} initialized by our SDP solution, Louvain cluster randomization, $\epsilon( = 1)$-net cluster randomization, spectral cluster randomization, causal cluster randomization, and complete randomization. Figure  compares the worst-case MSE bounds of SDP solution, our design, an adaptation of the proximal design from \citet{chen2024optimized} initialized by our SDP solution, and the adapted Gram-Schmidt Walk design.

To generate the heterogeneous variation components $\epsilon_f$, for $f = \alpha, \phi$, for $q^*=2$, we take
\begin{align*}
    \Sigma &\sim \calN(0,1)^{n \times n}, \\
    X &= \Sigma/\|\Sigma\|_2, \\
    z &\sim \calN(0, I_n), \\
    \epsilon_f &= \sqrt{\kappa} X^{1/2} z.
\end{align*}
If $q^* = \infty$ instead, we replace $X$ with $X = vv^T$, where $v$ is the leading eigenvector of $\Sigma$.

% To generate the spillover coefficients, we take $s(z) \sim \text{Unif}(- \gamma, \gamma)$ for all $i \in [n].$

To generate the interference component for each treatment assignment vector $x=2z-1$ generated from $X$ from the corresponding covariance matrix, we take 
\begin{align*}
s(z) &=  \frac{\sqrt{\gamma}L x}{\| x^T L^{1/2} \|_2}, \\
s(z) &=  s(z) - \dotp{s(z), \frac{\mathbf{1}}{n}} \mathbf{1}.
\end{align*}

We detail in \Cref{mse_comparisons_existing_designs} why these generative models appropriately capture worst-case potential outcomes.

\begin{figure}[!ht]
    \centering
    \begin{subfigure}{0.45\textwidth}
        \centering
        \includegraphics[width=\linewidth]{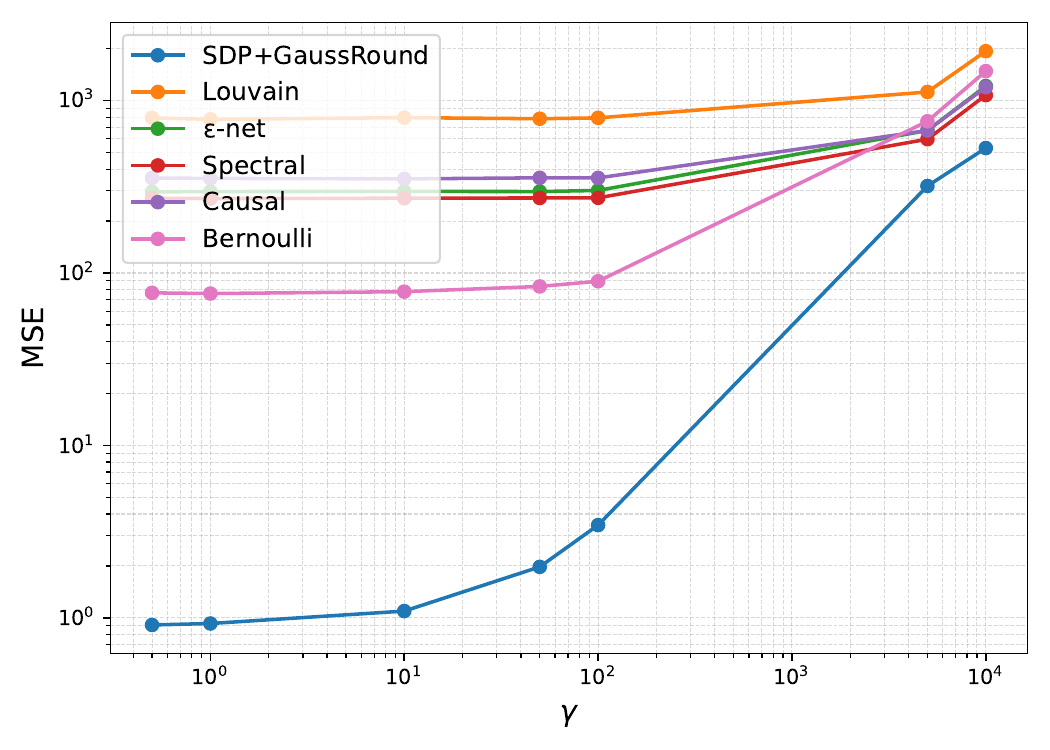}
    \end{subfigure}
    \begin{subfigure}{0.45\textwidth}
        \centering
        \includegraphics[width=\linewidth]{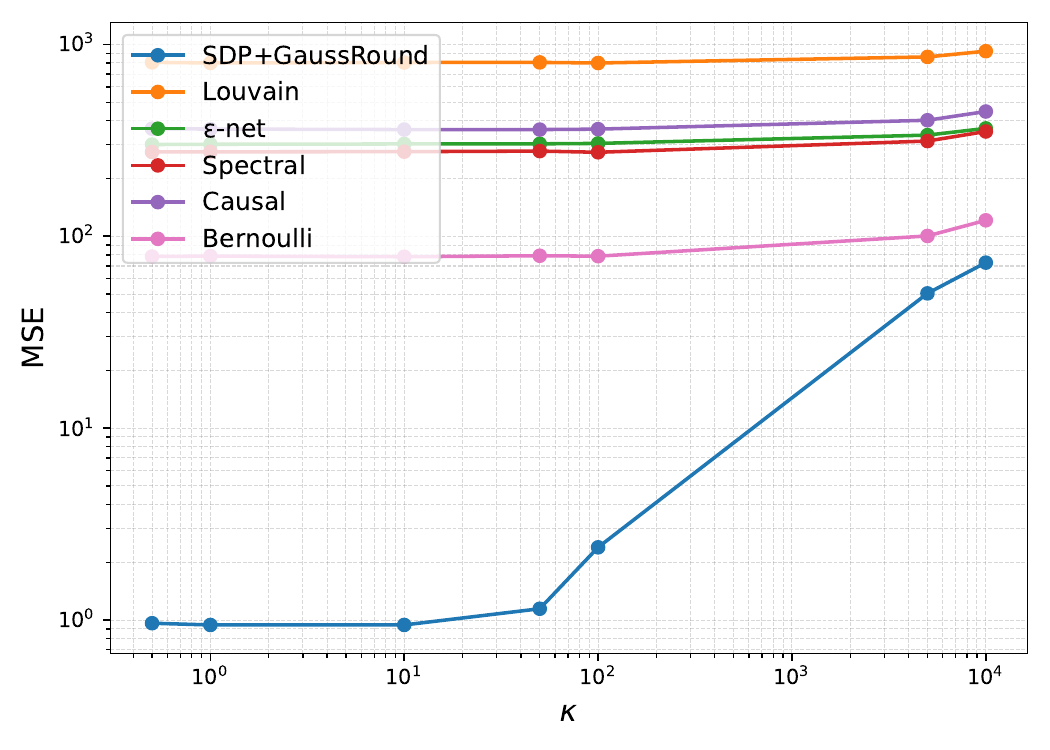}
    \end{subfigure}
    \caption{Comparisons of the MSEs across various designs, for different levels of $\gamma$ and $\kappa$, with the average (over 1000 trials) $\eta\approx 196$. Error bars represent the standard error.}
    \label{fig:comp_adversarial_rd}
\end{figure}

In \Cref{fig:comp_adversarial_rd}, we compare the MSEs across various designs, under the generated worst-case potential outcomes, for various levels of $\gamma$ and $\kappa$.
% ratios $\sqrt{\text{MSE}}/\text{GATE}$. 

In \Cref{fig:comp_gsw_rd}, in addition to the comparisons with the adapted Gram-Schmidt Walk design using the worst-case bounds with $\eta = {h_f}^TLh_f$, we also compare with the adapted Gram-Schmidt Walk design directly using a one-hot encoding of caste information (see the plotted line labeled ``GSW + covariates''), with potential outcomes generated according  to the worst-case models described above. Specifically, let $J$ be the number of different covariate groups. In our village data example, $J$ is the number of castes.
Given the random effects model above we can write the homophily component of the MSE as
\begin{align*}
    \mathbb{E}[(h^Tx)^2] &= \mathbb{E}\left[\left(\sum_{j=1}^J \left(\sum_{i: x_i = 1} \mathbf{1}\{i \in \text{Caste}_j\} - \sum_{i: x_i= - 1 } \mathbf{1}\{i \in \text{Caste}_j \}\right) \nu_j \right)^2\right] \\ 
    &= \mathbb{E}\left[\left(\sum_{j=1}^J \left(n_{1j} - n_{0j} \right)  \nu_j \right)^2\right] \\ 
    &= \sum_{j=1}^J (n_{1j} - n_{0j} )^2 \var( \nu_j) = \sigma^2 \sum_{j=1}^J (n_{1j} - n_{0j} )^2.
\end{align*}

Then, in this setting, we define $\eta= \sigma^2$, and take our pre-augmented covariate matrix $A \in \mathbb{R}^{(J + 2n) \times n }$ to be such that its $i$-th column,
\begin{align*}
    A_i^T = [ \Tilde{\eta}^{1/2} \xi_i^j \quad \Tilde{\gamma}^{1/2}L^{1/2}_i \quad {(\mathbf{1} \mathbf{1}^T)}_i^{1/2}]^T,
\end{align*} where $\xi_i^j$ is column vector with entry $1$ in row $j$ if unit $i$ belongs to caste $j$ and zero otherwise, and the parameters  $\Tilde{\eta},  \Tilde{\gamma},  \Tilde{\kappa}$ defined as before. Then, the worst case MSE bound can be written in the previous form
\begin{align*}
  \mathbb{E}[(\hat{\tau} - \tau)^2] \leq 7 \Delta [ \mathbb{E}[\| A x \|_2^2] + \Tilde{\kappa} \|X \|_{q}].
\end{align*}

